\documentclass[12pt]{article}
\PassOptionsToPackage{breaklinks}{hyperref}
\usepackage{amsmath}
\usepackage{amssymb}
\usepackage{cite}
\newcommand{\ds}{\displaystyle}
\newcommand{\vp}{\varphi}
\newcommand{\vt}{\vartheta}

\newcommand{\tl}{\tilde}
\newcommand{\om}{\omega}
\newcommand{\br}{\breve}
\begin{document}

\begin{titlepage}
\begin{flushleft}
DESY 06-209\\
Archive: quant-ph/0612032
 \end{flushleft} \vspace{0.3cm}
\begin{center}
{\Large A New Look at the \vspace{0.5cm} \\ Quantum Mechanics
of the Harmonic Oscillator}
 \vspace{1.0cm}\\

  {\large H.A.\
Kastrup\footnote{E-mail: Hans.Kastrup@desy.de}} \vspace{0.4cm}
\\ {DESY, Theory Group \\ Notkestr.\ 85, D-22603 Hamburg
\\Germany} \end{center} 
 \begin{center}{\bf Abstract}\end{center}

 In classical mechanics the harmonic oscillator (HO) provides the generic example for the use
of angle and action variables $\vp \in \mathbb{R} \bmod{2\pi}$ and $I >0\,$ which played a prominent role in the ``old'' 
Bohr-Sommerfeld quantum theory.  However, already classically there is a problem which has essential implications for the
quantum mechanics of the $(\vp,I)$-model for the HO: the transformation
$q = \sqrt{2I}\cos \vp,\, p = -\sqrt{2I}\sin \vp$ is only locally symplectic and singular for $(q,p) = (0,0)$. Globally
the phase space $\{(q,p)\}$ has the topological structure of the plane $\mathbb{R}^2$, whereas the phase space
$\{(\vp,I)\}$ corresponds globally to the punctured plane $\mathbb{R}^2 -(0,0)$ or to a simple cone  with the tip deleted.
 From the properties of the symplectic transformations on that phase space one can
derive the functions $h_0 = I,\, h_1 = I\cos \vp\,$ and $ h_2 = -I\sin \vp\,$ as the basic coordinates on $\{(\vp,I)\}\,$,
where their Poisson brackets obey the Lie algebra of the symplectic group of the plane.
This implies  a qualitative difference as
to the quantum theory of the  phase space $\{(\vp,I)\}$ compared to the usual one for $\{(q,p)\}\,$:  In the
 quantum mechanics for the $(\vp,I)$-model of the HO
 the three $h_j$ correspond to the  self-adjoint generators $K_j,\,j=0,1,2,$ of certain irreducible unitary representations
  of the symplectic group or one of its infinitely many covering groups,
the representations being parametrized by a (Bargmann) index $k>0$. This index $k$ determines the ground state energy
 $E_{k,n=0} = \hbar\,\om\,k$
of the $(\vp,I)$-Hamiltonian $H(\vec{K})= \hbar\, \om\, K_0$. For an $m$-fold covering the lowest possible value for $k$ is
$k=1/m\,$, which can be made arbitrarily small by choosing $m$ accordingly! This is not in contradiction to the usual
approach in terms of the operators $Q$ and $P$ which are now expressed as functions of the $K_j$, but keep their usual
properties. The richer structure of the $K_j$ quantum model of the HO is ``erased'' when passing to the simpler
 $(Q,P)$-model! This more refined approach to the quantum theory of the HO implies many experimental tests: Mulliken-type
experiments for isotopic diatomic molecules, experiments with harmonic traps for atoms, ions and BE-condensates, with
 charged HOs
 in  external electric fields and the (Landau) levels of charged particles in external magnetic fields,
 with the propagation of light in vacuum, passing
through strong external electric or magnetic fields. Finally it may lead to a new theoretical estimate for the
 quantum vacuum energy of fields and its relation to the cosmological constant.

\end{titlepage}
 \newpage                  

\tableofcontents 
\section{Introduction and overview}
\subsection{The issue: Quantum mechanics of the harmonic oscillator in terms of angle and action variables}
At first sight it probably appears provocative and presumptuous to present a new research paper
on the harmonic oscillator (HO), that venerable and pedagogically thoroughly squeezed simple
 model, encountered in many physics publications of all types. Despite its simplicity it has
played an important role at many instances in the history of physics, classically and quantum
 theoretically:

 It probably started  with Hooke's law 
 \begin{equation}
   \label{eq:919}
   \dot{p} = -b\,q\,,~p= M\,\dot{q}\,,~b >0\,,
 \end{equation}
in mechanics for the force exerted on a
 particle in the neighbourhood of its stable equilibrium position. Then came the HO in the plane with its two
 qualitatively different types of motion, periodical orbits (Lissajous figures) and
 quasi-periodical ones which densely fill a submanifold of the phase space, initiating the idea
of ergodic systems. Two or more linearly coupled HO with their characteristic (eigen-)
 frequencies are important for the stability analysis of many systems and play a significant role in crucial areas
of physics. By adding a friction term the model serves also as an examplary introduction to dissipative systems.

 Conceptionally important was - and still is - the locally
 canonical (symplectic) description of the position and momentum coordinates for the HO in terms of angle and action
 variables: 
 \begin{equation}
   \label{eq:920}
   q(\vp,I) = \sqrt{\frac{2\,I}{M\,\omega}}\, \cos \vp\,,~~~ p(\vp,I) = - \sqrt{2\,M\,\omega\,I}\,
 \sin \vp\,,~~~\omega = \sqrt{b/M}\,,
 \end{equation}
so that
\begin{equation}
  \label{eq:921}
  H(q,p) = \frac{1}{2M}\,p^2 + \frac{1}{2}\,M\,\omega^2\,q^2 = H(\vp,I)= \omega\,I\,.
\end{equation}
 This is the generic example for the essential concept of integrable systems, their
(non-integrable) perturbations and the associated KAM-theory \cite{arn1,arn2,thir1}.

Then there is the possible interpretation of classical  free electromagnetic standing waves in a
 cavity as a set of uncoupled harmonic oscillators. This property was essential  in Planck's
derivation of his radiation law. So the HO played an important part in 
the birth of quantum theory, too! 

In the ``old'' quantum mechanics with its Bohr-Sommerfeld framework the 
HO had the energy levels $E_n=  \hbar\, \omega\,n,\,n=0,1,\ldots $. (For a comprehensive summary
of the Bohr-Sommerfeld theory, where the angle and especially the action variables played a central role, just before
 the dawn of modern quantum mechanics see the impressive textbook by Born (and Hund)  \cite{born1}.) 

Even before Heisenberg deduced the modified energy levels
\begin{equation}
  \label{eq:826}
  E_n=\hbar\,\omega\,(n+\frac{1}{2})\,,~~n=0,1,\ldots
\end{equation}
in his famous first paper on matrix mechanics \cite{heis1}, Mulliken had concluded from his
spetroscopic analysis of the differences in the vibrational spectra of the diatomic isotopes
 $B^{10}\,O^{16}$ and
$B^{11}\,O^{16}$ \cite{mull1,herzb1} that the lowest energy state of the HO should be
\begin{equation}
  \label{eq:827}
  E_0 = \frac{1}{2}\,\hbar\, \omega\,.
\end{equation}
This has been the canonical undisputed ground state energy value of the HO ever since (for a
 comprehensive  historical overview see Ref.\ \cite{mil1}) and a standard example for the role of Heisenberg's
position - momentum  uncertainty relations. For a recent partial survey of the HO in modern physics see Ref.\ \cite{mosh1}.

Whereas angle and action variables were central ``observables'' in the old quantum mechanics,
they  disappeared almost entirely in the new quantum mechanics from 1925/26 on and the usage of the operators
$Q$ and $P$ took over nearly completely. Dirac's early attempts \cite{dir1} to use angle and action operators
also for the new framework turned out to be contradictory,
as pointed out by London \cite{lond1} and Jordan \cite{jord1} and the subject has remained controversial even up to
 now \cite{ka1}.
Before taking up that issue again, a few remarks as to the central role the ground state energy \eqref{eq:827}
started to play:

Around 1930 F.\ London deduced the van der
 Waals forces from the ground state energies of two 3-dimensional HOs \cite{lond2}.

The value \eqref{eq:827} became a nuisance (and still is!), however, when free fields were
 quantized,
 because their interpretations as a set of an infinite number of HOs implied  an
 (unobserved) infinite ground state energy. The problem has been ``swept under the rug'' 
by ignoring
the ground state energies, formally by introducing ``normal-ordering'' for the associated
 annihilation and creation operators $a$ and $a^{\dagger}$ (see below). 

Nevertheless the ground state energy \eqref{eq:827} plays a very stimulating part in the
discussions of the Casimir effect \cite{casi,mil1,milt} and also in the present attempts to understand
the dark energy in the universe and the extremely obnoxious cosmological constant problem
 \cite{wein,car,peeb,volo,cope,pad,strau,nobb}.

So the energy \eqref{eq:827} is discarded or advocated depending on the physical concepts which are being discussed.
Not a very convincing situation!

In view of the general acceptance of the value \eqref{eq:827} it  is  amazing that
 there appear to be no systematic modern experimental tests - similar to those of Mulliken - of such a conceptually
 important physical quantity! More on the experimental situation in subsec.\ 1.3 below.

It is one aim of the present paper to point out that the canonized ground state energy value
 \eqref{eq:827} may
 not be the only possible one for the HO, but that there is a canonical structure for the HO in 
terms of {\em angle and action variables} $\vp$ and $I$ the  quantum mechanics of which allows for 
ground state values
\begin{equation}
  \label{eq:828}
  E_{k,n=0} = \hbar\,\omega\,k\,,~~k >0\,,
\end{equation}
where $k$ may be any positive number, especially an arbitrary small one $> 0\,$!

I ask for a moment of patience for the justification of this seemingly outrageous claim!

The main reason for the possibility \eqref{eq:828} is the difference as to the {\em global}
 structures of the locally canonically (symplectically) equivalent phase spaces
 $\mathcal{S}_{q,p}$ and $\mathcal{S}_{\vp,I}$ of the respective canonical pairs $(q,p)$ and
$(\vp,I)$\,:
\begin{equation}
  \label{eq:829}
  \mathcal{S}_{q,p} =\{(q,p) \in \mathbb{R}^2\,\}~,
\end{equation}
\begin{equation}
  \label{eq:2158}
  \mathcal{S}_{\vp,I} =\{(\vp,I), \vp \in
 \mathbb{R} \bmod {2\pi}\,,\, I > 0\,\}\,,
\end{equation}
which shows that  $\mathcal{S}_{q,p}$ has the global topological structure of the plane
 $\mathbb{R}^2\,$, whereas  $\mathcal{S}_{\vp,I}$ has that of a simple cone with the tip deleted or
that of a punctured plane $\mathbb{R}^2 -\{0\} \cong S^1 \times \mathbb{R}^+$, where $S^1$
 denotes the unit circle and $\mathbb{R}^+$ the positive real numbers {\em without} the $0$\,.

This implies that $\mathcal{S}_{\vp,I}$ cannot be quantized in the conventional manner in terms
 of the (Born-Heisenberg-Jordan-Dirac-) Weyl group generated by the 3-dimensional Lie algebra basis 
$\{q,p,1\}$, but one has to pass to the 3-dimensional 
(proper orthochronous homogeneous Lorentz) group $SO^{\uparrow}(1,2)$ (in one ``time'' and two ``space'' dimensions) or
 to one of its (infinitely) many
 covering groups \cite{ka1}, among which the symplectic group $Sp(2,\mathbb{R})$ in the $(q,p)$-plane 
is a double covering (like the group $SU(2)$ is a double covering of the rotation group $SO(3)$).
 That symplectic group provides the key to an appropriate quantization of the phase space 
\eqref{eq:2158} and plays an essential role in what follows.

The crucial point is that both the phase space $\mathcal{S}_{\vp,I}$ and and its ``canonical group''
 $SO^{\uparrow}(1,2)$ contain the topological  ``factor'' $S^1$ which is multiply connected (with
homotopy group $\pi_1(S^1) = \mathbb{Z}$). This multi-connectedness has implications for the
infinite-dimensional irreducible unitary representations of the non-compact group 
 $SO^{\uparrow}(1,2)$ and its infinitely many covering groups because now the self-adjoint generator of the rotations
$SO(2)$ can have  more complicated spectra with a ground state like \eqref{eq:828}. And this generator is proportional
to the Hamilton operator of the HO in the $(\vp,I)\,$-framework! (For the similar case of a simple rotator see Ref.\
\cite{ka2}.)

The transformation \eqref{eq:920} from the space \eqref{eq:2158} onto the space \eqref{eq:829} with its origin deleted
is not special for the HO. It can be used for any $(1+1)$-dimnsional system with periodic motions in 
\eqref{eq:829} describable by angle and action variables in  \eqref{eq:2158}. So their quantum mechanics is
affected, too! Examples are discussed in subsec.\ 2.3. 

Quantizing the phase space $\mathcal{S}_{\vp,I}$ makes use of the positive discrete series 
$ D^{(+)}_k,\,k>0,$ of those unitary representations mentioned above \cite{bo1,ka1}. In these representations the
 self-adjoint generator $K_0$ of
the compact rotation subgroup $SO(2) \cong S^1$ constitutes the quantized counterpart of the classical action
 variable $I$ and the ``boost'' generators $K_1$ and $K_2$ correspond to the classical quantities
$I\,\cos \vp$ and $-I\,\sin \vp$, the knowledge of which allows to determine the angle $\vp \in (-\pi,\,\pi]$ uniquely. 
The choice of these basic ``observables'' on the phase space \eqref{eq:2158} can be justified systematically from the
action of the symplectic group $Sp(2,\mathbb{R})$ on the phase space \eqref{eq:829}. That action leaves the origin of
the space \eqref{eq:829} invariant!

Those basic classical observables 
\begin{equation}
  \label{eq:831}
 h_0(\vp,I) = I\,,~~ h_1(\vp,I)=I
\,\cos \vp\,,~~ h_2(\vp,I)=-I\,\sin \vp\,,
\end{equation} on  $\mathcal{S}_{\vp,I}$ 
obey the Lie algebra $\mathfrak{so}(1,2)$ of the group $SO^{\uparrow}(1,2)$ and its (infinitely many) covering groups 
in terms of Poisson brackets: 
\begin{equation}
  \label{eq:832}
 \{h_0,h_1\}_{\vp,I}=
-h_2\,,~~\{h_0,h_2\}_{\vp,I}=h_1\,,~~ \{h_1,h_2\}_{\vp,I}=h_0\,,
\end{equation} where
\begin{equation}
  \label{eq:833}
 \{h^{(1)},h^{(2)}\}_{\vp,I} \equiv
\partial_{\vp}h^{(1)}(\vp,I)\,\partial_I h^{(2)}(\vp,I)-\partial_I h^{(1)}(\vp,I)\,
\partial_{\vp}h^{(2)}(\vp,I)\,.
\end{equation}

 The corresponding quantum mechanical counterparts, the dimensionless self-adjoint operators
 \begin{equation}
   \label{eq:51}
  \tl{K}_j=K_j/\hbar  
 \end{equation}
 obey
\begin{equation}
  \label{eq:918}
  [\tl{K}_0,\,\tl{K}_1] = i\,\tl{K}_2\,,~~[\tl{K}_0,\,\tl{K}_2] = -i\,\tl{K}_1\,,~~[\tl{K}_1,\,\tl{K}_2] = -i\,\tl{K}_0\,.
\end{equation}
For the positive discrete series the operator $\tl{K}_0$ in general has the spectrum (eigenvalues)
 \begin{equation}
   \label{eq:830}
  \sigma(\tl{K}_0) = \{n+k\,,\, n=0,1,\ldots\,;~~k \in \mathbb{R}^+\,\}\,.
 \end{equation}

For the $m$th covering group $SO^{\uparrow}_{[m]}(1,2)$\,,\,$m=1,2,\ldots,\,$ of $SO^{\uparrow}(1,2)$ the allowed
 values of $k$ are
\begin{equation}
  \label{eq:2156}
  k= \frac{\mu}{m}\,,~\mu \in \mathbb{N} =\{1,2,\ldots \}\,,
\end{equation}
so that the smallest attainable value of $k$ for a corresponding irreducible unitary representation is
\begin{equation}
  \label{eq:2157}
  k=\frac{1}{m}\,.
\end{equation}
As $m$ can be an arbitrarily large natural number, $k$ can be made arbitrarily small $ >0\,$!

The quantum mechanical $(q,p)$-Hamiltonian
\begin{equation}
  \label{eq:922}
 H(q,p) \to H(Q,P) = \frac{1}{2\,M}\,P^2 + \frac{1}{2}\,M\,\omega^2\,Q^2 = -\frac{\hbar^2}{2\,M}\,
\frac{d^2}{d\,q^2}+\frac{1}{2}\,M\,\omega^2\,q^2
\end{equation}
has the unambiguous spectrum \eqref{eq:826}.
However, in view of Eq.\ \eqref{eq:830} the quantum mechanical $(\vp,I)$-Hamiltonian
\begin{equation}
  \label{eq:923}
  H(\vp,I) \to H(\vec{K}) =  \omega\,K_0\,,~~\vec{K} = \hbar\,(\tl{K}_0,\tl{K}_1,\tl{K}_2)\,,
\end{equation}
can have the spectrum
\begin{equation}
  \label{eq:924}
  E_{k,\,n}(\vp,I) = \hbar\,\omega\, (n+k)\,,\, n=0,1,\ldots\,;~~k \in \mathbb{R}^+\,.
\end{equation}

A crucial point now is the following: the spectrum \eqref{eq:826} is not just a special case of
\eqref{eq:924}, but the situation is more subtle: 

Let $|k,n\rangle\,,\,n=0,1,\ldots$ be an eigenvector of $\tl{K}_0$ with eigenvalue \eqref{eq:830}:
\begin{equation}
  \label{eq:925}
  \tl{K}_0\,|k,n\rangle = (n+k)\,|k,n\rangle\,,\,n=0,1,\ldots;~~k>0\,,
\end{equation}
then nevertheless
\begin{equation}
  \label{eq:926}
  H(Q,P)|k,n\rangle =\hbar\,\omega\,(n+1/2)\, |k,n\rangle\,,
\end{equation}
where now the operators $Q$ and $P$ are expressed as functions of the $\tl{K}_j$:
\begin{equation}
  \label{eq:927}
  Q=Q(\vec{K})= \frac{\lambda_0}{\sqrt{2}}\,(A^{\dagger}+A)\,,~~P=P(\vec{K})=
 \frac{i\,\hbar}{\sqrt{2}\,\lambda_0}\,(A^{\dagger}-A)\,,~\lambda_0 =
\sqrt{\frac{\hbar}{M\,\omega}}\,,
\end{equation}
with
\begin{equation}
  \label{eq:931}
  A = (\tl{K}_0+k)^{-1/2}\tl{K}_-\,,~~~A^{\dagger} = \tl{K}_+(\tl{K}_0+k)^{-1/2}\,,~~~\tl{K}_{\pm} = \tl{K}_1
 \pm i\,\tl{K}_2\,,
\end{equation} and
\begin{equation}
  \label{eq:2168}
 [A,\,A^{\dagger}] = {\bf 1}\,. 
\end{equation}
The non-linear relations \eqref{eq:931} are an inversion of the known Holstein-Primakoff representation of the
$\tl{K}_j$ in terms of $A$ and $A^{\dagger}$ \cite{hol1} as discussed in detail in Ref.\ \cite{ka1}.

The $k$-independent relation \eqref{eq:2168} holds in any irreducible unitary
representation $D_k^{(+)}$ and is a consequence of the commutation relations \eqref{eq:918} which imply
\begin{equation}
  \label{eq:1710}
  \tl{K}_+ |k,\,n \rangle = [(2k+n)(n+1)]^{1/2}\,|k,\,n+1 \rangle\,,~~~ \tl{K}_- |k,\,n \rangle =
 [(2k+n-1)n]^{1/2}\,|k,\,n-1 \rangle\,,
\end{equation}
so that for any $k$
\begin{equation}
  \label{eq:1711}
  A^{\dagger}\,|k,\,n \rangle = \sqrt{n+1}\,|k,\,n+1 \rangle\,,~~~
 A\,|k,\,n \rangle = \sqrt{n}\,|k,\,n-1 \rangle\,.
\end{equation}
The Eqs.\ \eqref{eq:927} and \eqref{eq:931} are just the operator versions of the classical relations 
\begin{equation}
  \label{eq:928}
  q(\vp,\,I)= \sqrt{\frac{2}{M\,\omega}}\,\frac{h_1(\vp,\,I)}{\sqrt{h_0(\vp,\,I)}}\,,~~
 p(\vp,\,I)= \sqrt{2\,M\,\omega}\,\frac{h_2(\vp,\,I)}{\sqrt{h_0(\vp,\,I)}}\,.
\end{equation}
For more details see below, here especially sec.\ 5!

The gist of the argument for allowing a possible discrepancy between the spectra \eqref{eq:925}
and \eqref{eq:926},  to be discussed in detail later on, is that - due to the multi-valuedness
 of the angle $\vp$ - the quantum version \eqref{eq:923} of the HO
Hamilton function $H(\vp,I)$  can have a richer spectrum than $H(Q,P)$ which always has the spectrum
\eqref{eq:826}, even if it acts in a Hilbert space with a representation $D_k^{(+)}\,,\,k
 \neq 1/2\,$, for which $\tl{K}_0$ has the spectrum \eqref{eq:830}!

Phrased differently: The quantities $q$ and $p$ generate global translations on the phase space
$\mathcal{S}_{q,p}$, i.e.\ no point is preferred, especially not the origin. This is different
for the global action of the generators $h_j$ which leave the origin of $\mathcal{S}_{q,p}$ 
and the corresponding point $I=0$ in $\mathcal{S}_{\vp,I}$ invariant. Thus, the operators $Q$
and $P\,$, generators of translations in momentum and position space, respectively, ``erase'' 
the topological substructure induced by the critical point $(q,p) = (0,0)$ (or $I=0$)\,. That point is, however
 ``taken care of''
by the operators $\tl{K}_j$, generators of symplectic transformations in $(q,p)$-space, which leave the point $(q=0,p=0)$
fixed!

So it makes a difference as to the choice of the primary degrees of freedom, whether one starts with $q$ and $p$ and their
topologically trivial phase space \eqref{eq:829}, or whether one starts with $\vp$ and $I$ and their topologically 
non-trivial  phase space \eqref{eq:2158}. The latter leads to a ``richer'' quantum mechanics  than that of the former which
is unable to do justice to the non-trivial topology of \eqref{eq:2158} and therefore has to ``ignore'' the additional
 structure! Whether this additional topological fine structure has indeed been ``implemented by nature'' and can be
 observed in the laboratory - or is merely a 
coordinate singularity (see subsec.\ 2.1) - has, of course, to be found out by experiments. 
\subsection{Contents overview}
The paper is organized as follows:

Sec.\ 2 collects some properties of the classical HO, with emphasis on the singular character of the transformation
\eqref{eq:920} at $(q=0,p=0)$ and on the dynamical role of the ``new'' basic coordinates $\vp$ and $I$, including the
celebrated adiabatic properties of the action variable $I$ and its role for certain 1-dimensional integrable systems with
bounded orbits.

Sec.\ 3 discusses properties of the symplectic transformation group $Sp(2,\mathbb{R})$ acting on the phase space
\eqref{eq:829}: As already mentioned above, that group transforms any two points of that space into each other, except
for the point $(0,0)$ which is left fixed. The orbits of  three independent 1-dimensional subgroups generate three
vector fields which are globally Hamiltonian. The generating Hamiltonian functions of these vector fields are essentially
 the functions
\eqref{eq:831} (expressed in terms of the variables $q$ and $p$).  The Poisson brackets of these Hamiltonian functions
 generate
the Lie algebra $\mathfrak{so}(1,2) = \mathfrak{sp}(2,\mathbb{R})$ of the groups $SO^{\uparrow}(1,2)$ and $ Sp(2,\mathbb{R})
\,$ . The quantized version of that Lie algebra belongs to
 irreducible unitary representations $D_k^{(+)}\,,\,k=1/4$ and $k=3/4$ of the so-called ``metaplectic'' group. 
These representations are implemented in the even  and  odd
parity subspaces of the usual Hilbert space $L^2(\mathbb{R},dq)$ of the HO. 

Sec.\ 4 describes the action of the group $SO^{\uparrow}(1,2) = Sp(2,\mathbb{R})/\mathbb{Z}_2$ on the $(\vp,I)\,$- phase
space \eqref{eq:2158} the points of which are ``coordinized'' by the functions \eqref{eq:831}. The action of the group
is symplectic, transitive (i.e.\ any two points may be transformed into each other), effective (i.e.\ the only group
element which leaves all points invariant is the unit element) and globally Hamiltonian, i.e.\ the functions \eqref{eq:831}
are the generating functions of the vector fields associated with three independent 1-dimensional transformation
 subgroups of $SO^{\uparrow}(1,2)$. So we have a completely satisfactory ``canonical'' structure on the phase space
\eqref{eq:831} based on the group $SO^{\uparrow}(1,2)$ and its infinitely many covering groups. This section prepares the
ground for a group theoretical quantization \cite{ish,gui,ka1a} of the phase space \eqref{eq:2158} in terms of appropriate 
irreducible unitary representations of those groups which provide the associated quantum theories. 

The central sec.\ 5 discusses the quantization of the phase space \eqref{eq:2158} in terms of the irreducible unitary
representations of the positive discrete series $D_k^{(+)}$ of the group $SO^{\uparrow}(1,2)$ and its infinitely many
 covering groups. The generator $\hbar\,\tl{K}_0$ of the rotation subgroup is the quantized version of the action
 variable $I$ and
the Hamilton function $H=\om\,I$. Its most general spectrum is given by Eq.\ \eqref{eq:830}. In physics the corresponding
 Hamilton operator \eqref{eq:923} generates time translations:
\begin{equation}
  \label{eq:2159}
  U(t) = e^{-i\,H\,t/\hbar}\,,~~H=\hbar\,\om\,\tl{K}_0\,.
\end{equation}
This means that the (dimensionless) time variable $\tl{t} = \om\,t$ mathematically represents  the angle $\vp$. As $\tl{t}$
in general does not stop at $\tl{t} = 2\pi$, it ``runs'' through several or very many coverings. As $\tl{K}_0 = 
N +k{\bf 1}$ we have 
\begin{equation}
  \label{eq:2160}
  U(\tl{t}=2\pi) =e^{-2\pi i k}{\bf 1}\,.
\end{equation}
 This shows explicitly that for an $m$-fold covering with $k$ as in Eq.\ \eqref{eq:2156} we get
 \begin{equation}
   \label{eq:2161}
   U(\tl{t} = 2\pi m) = {\bf 1}\,.
 \end{equation}
I already stressed above that in passing from the quantum theory of the Lie algebra $\mathfrak{so}(1,2)$ to that of the 
Born-Dirac-Heisenberg-Jordan-Weyl Lie algebra one loses the ``fine structure'' associated with the Bargmann index $k\,$.
This is a result the importance of which reaches probably far beyond the  simple HO\,!
It allows to avoid the celebrated  Stone-von Neumann  uniqueness theorem {\em 
without violating it}\,! The usual Heisenberg uncertainty relations for $Q$ and $P$ remain untouched,
but there are new uncertainty relations as to the operators $\tl{K}_j,\,j=0,1,2$\, \cite{ka1}.

Sec.\ 6 discusses properties and possible applications of the three types of coherent states associated
with the Lie algebra $\mathfrak{so}(1,2)$ (Schr\"{o}dinger-Glauber, Perelomov and Barut-Girardello) 
 to the HO. The last two of these coherent states are very probably of similar importance for experiments in quantum optics
as is already well-known for the Schr\"{o}dinger-Glauber coherent states. A number of interesting physical expectation
values and their dependence on the index $k$ are discussed as well as the possible experimental production of such states:
The Perelomov ones have been generated in the laboratories in the form of squeezed states, the Barut-Girardello ones to the
best of my knowledge not yet.

Sec.\ 7 describes several explicit examples of Hilbert spaces with irreducible unitary representations
of the series $D_k^{(+)}$. It starts with the conventional HO for which $k=1/2\,$ represented in the Hardy space
 $H^2_+(S^1,\vt)$
on the unit circle $S^1$. That space has the scalar product
\begin{equation}
  \label{eq:2162}
  (f_2,f_1)_+ = \frac{1}{2\pi}\,\int_{S^1}d\vt f_2^*(\vt)f_1(\vt)\,,
\end{equation}
the basis
\begin{equation}
  \label{eq:2163}
   e_n(\vt)=  e^{i\,n\,\vt}\,,~n= 0,1,2,\cdots\,,
\end{equation}
and the HO Hamilton operator
\begin{equation}
  \label{eq:2164}
  H=\hbar \,\om\,\tl{K}_0\,,~~\tl{K}_0 = \frac{1}{i}\partial_{\vt} + 1/2\,.
\end{equation}
{\em All} the well-known physical properties of the usual quantized HO can be derived in this framework, and even some 
more, because now we have {\em three} different kinds of coherent states! \\
The second part of that sec.\  deals with concrete Hilbert spaces where the index $k$ of the irreducible unitary
 representations can have any real value $>0$. One of these is the space $L^2(\mathbb{R}_+,du)$ with its orthonormal
basis of Laguerre's functions.  

Sec.\ 8 briefly recalls the description of a quantized free electromagnetic field in a cavity as an infinite set of HOs 
and the disturbing quantitative problems one encounters for the total ground state energy when using the value
\eqref{eq:827} of a single oscillator. In the $(\vp,I)$-framework one has instead $E_{k,\,n=0} = \hbar\,\om\,k\,$,
 where $k>0$, in principal,
can be arbitrarily small. This may shed new light on the notorious cosmological constant problem and the origin of the
related dark energy \cite{wein,car,peeb,volo,cope,pad,strau,nobb}. \\
If different electromagnetic modes  have different $k$ by exposing them to external electric or magnetic fields,
 the electromagnetic ``vacuum'' can  even acquire some sort of anomalous
refractive strucure. This may lead (perhaps) to an understanding of the recently observed ``dichroism'' of
 the vacuum in a strong static magnetic field \cite{zav}. \\
The sec.\ closes with a very speculative remark on the possibility of ``dark'' normal matter.

Sec.\ 9 recalls the effective HOs one has if a HO particle is charged and  an additional external electric field is
 applied or if a free charged particle is in an external magnetic field. Here, too,
 one may
introduce angle and action variables, the quantized versions of which may lead to a shift of the usual ground state levels. 

Sec.\ 10 briefly discusses the (canonical) quantum statistics of a system with the energy levels \eqref{eq:924}, in order
to see which thermodynamical quantities depend on $k$ and which not. 

 Appendix A gives the technical details for the calculation of the action variables associated with the potentials
 discussed in subsec.\ 2.3. Appendix B summarizes some essential properties of the universal covering group of
 $SO^{\uparrow}(1,2)$, its irreducible
unitary representations of the positive discrete series and those of the $m$-fold covering groups as special cases. 

\subsection{Possible experiments}
The crucial question is, of course, whether there exist HOs in nature or may be prepared in the laboratory which
 have a spectrum
of the type \eqref{eq:924}. It appears unnessary here to point out in detail the important implications this would have
 for the physics of many systems, not only for the HO!

For possible experimental setups one has to observe that the ``primary observables'' now are the operators 
$K_j,\,j=0,1,2,$ with their algebraic structure \eqref{eq:918}, {\em not} as usual the position and momentum operators
\eqref{eq:927}. Note also that $K_0$ is not the Hamiltonian, but $\om\,K_0$, so that $E_{k\,n=0}(\vp,I)$ from \eqref{eq:924}
can be the same for different $\om$ and $k$ if their product is the same, i.e.\ the energy stays the same!
 One problem for the experiments is to find
dynamical mechanisms which do not bring the usual $(q,p)$-dynamics into play, e.g.\ the dominant atomic dipole-transitions.

Following the original procedure of Mulliken and others \cite{herzb1} the value of $k$ in  the spectrum \eqref{eq:924}
 may, at least in principle, be determined as follows:
 According
 to Eqs.\ \eqref{eq:919} and
\eqref{eq:920} the frequency $\omega$ of the oscillator can be changed either by changing its
mass $M$ or by modifying the strength $b$ of the driving force. Let $\omega_1$ and $\omega_2$
be two known frequencies of the same system and let $E_a$ and $E_b$ two known fixed external energy levels
 different from
the two ground state energies $E_0(j)\,,\,j=1,2\,,$ of the two slightly different versions of the same HO. If transitions
\begin{equation}
  \label{eq:2165}
  E_a \to  E_0(1) = \hbar\,\om_1\,k\,,~~~ E_b \to  E_0(2) = \hbar\,\omega_2\,k\,,
\end{equation}
with frequencies
\begin{equation}
  \label{eq:2166}
  \om_{a,1} = [E_a-E_0(1)]/\hbar\,,~~~\om_{b,2} =[E_b-E_0(2)]/\hbar\,,
\end{equation}
  are possible and measurable, then one
can determine the value of $k$ from the difference
\begin{equation}
  \label{eq:2167}
 \omega_{a,1}-\omega_{b,2} = (E_a-E_b)/\hbar -k\,(\omega_1-\omega_2)\,. 
\end{equation}

 In the case of the vibrating diatomic molecules Mulliken investigated the
levels $E_a$ and $E_b$ where the vibrational ground states of a higher electronic level and the levels $E_0(j)$
were the vibrational ground states of a lower electronic level of the two respective isotopes 
for which the two frequencies $\omega_j$ differ because the corresponding reduced masses $\mu$ in $\om = \sqrt{b/\mu}$ 
differ \cite{herzb1}.

Note also that for $k \neq 1/2$ {\em all} energy levels of the spectrum \eqref{eq:924} are shifted compared to the usual
ones \eqref{eq:826}\,.

 More refined versions of Mulliken's experiments  with diatomic molecules using modern experimental techniques  should be
 possible and appear
highly desirable! In order to ``freeze'' the $(q,p)$-degrees of freedom when looking for $(\vp,I)$-properties 
 one should probably go to extremely low
temperatures, even below the ground state energies \eqref{eq:827}. Experiments with ultracold molecules have reached an
impressive stage of refinement \cite{mol} and the  use of Feshbach resonances \cite{fesh} has led to fascinating
 experimental results for low lying vibrational bound state levels of bosonic pairs of atoms in ultra-cold BE-condensates
\cite{kett}.

Furthermore, modern experimental techniques have provided sophisticated 1-dimensional harmonic traps \cite{qgld}, for ions
 \cite{ion}, atoms \cite{ato} and 
BE-condensates \cite{bec}, for  which the frequency $\om$ from \eqref{eq:920} can be tuned from outside, by changing the
 force strength $b$
electronically. Approximate 1-dimensional harmonic traps with ultra-cold BE-condensates mainly in the ground state
\eqref{eq:827} have been  built \cite{kett2}, the ground state energy being determined by laser light Bragg reflections
off the ``untrapped'' expanding cloud of BEC atoms. Thus, these impressive experiments appear to be associated with the
$(q,p)$-model of the HO!
Nevertheless, similar such  setups may provide
new possibilities for a search after the energy levels \eqref{eq:924}, again most likely at extremely low temperatures.

In sec.\ 6 it will be pointed out in detail that expectation values and transition probabilities involving Perelomov
coherent states are proportional to the index $k$. As these states have already been generated experimentally for
 $k=1/2$, they may perhaps also be produced for other (lower) values of $k$. 
  
Then there are possible vacuum birefringence and (or) dichroism effects of photons by strong external electric or
 magnetic fields as mentioned in sec.\ 8.

Sec.\ 9 discusses  shifts in the HO ground states  of charged particles in external electric or magnetic fields. 

Sec.\ 10 finally mentions the plans for determining the ground state energy of the HO by means of the Josephson effect
\cite{beck}!
\subsection{Generalizations}
Finally it should be remembered that the harmonic oscillator is, of course, not the only important integrable physical
system
which classically can be described by angle and action variables (e.g.\ the $const./r$ potential, see Refs.\ \cite{born1}
and \cite{arn1,arn2,thir1} for more examples).
Quantizing those systems group theoretically one has to distinguish between the cases $I \in \mathbb{R}^+$ and $ I \in
\mathbb{R}\,$. The latter has to be quantized in terms of the irreducible unitary representations of the Euclidean group
of the plane $E(2)$ and its covering groups. For details see Ref.\ \cite{ka2}. 

One has, however, to observe the following: If the group $SO(2) \cong S^1\subset SO^{\uparrow}(1,2)$ becomes a non-trivial
 subgroup of a  larger compact group (i.e.\ not just a direct abelian factor) its
topological properties can change drastically: E.g., if one passes from $SO^{\uparrow}(1,2)$ to $SO(3)$ the universal
 covering group is now
the double covering $SU(2)$. Going from   $SO^{\uparrow}(1,2)$ to $SO^{\uparrow}(1,3)$ one has the universal double
covering $SL(2,\mathbb{C})$.

 If, on the other hand,  one goes from  $SO^{\uparrow}(1,2)$ to $SO^{\uparrow}(2,3)= 
Sp(4,\mathbb{R})/\mathbb{Z}_2$, where $Sp(4,\mathbb{R})$ is the symplectic group in 4 dimensions, one again encounters
the subgroup $SO(2)\cong S^1$ as a factor in the maximal compact subgroup $SO(2) \times SO(3)$  and and also a
 positive discrete series of  irreducible unitary representations of the group $Sp(4, \mathbb{R})$ and its infinitely
 many covering groups  \cite{ka3}.
This is just another special case of  symplectic groups $Sp(2n,\mathbb{R})$ in $2n$ dimensions: They have
dimension $2n^2 +n\,$, rank $n$ (i.e.\ a maximal abelian set of $n$ commuting Lie algebra generators),
 the maximal compact subgroup $U(n) \cong
SU(n) \times U(1)$ (which has rank $n\,$, too) and (positive) discrete series of irreducible
 unitary representations \cite{barg1}, including those of their
universal covering groups  associated with the factor $U(1)\,$ (the group $SU(n)$ on the other hand is simply connected
\cite{huse}).
  This should be of interest
for the discussion of  quantum mechanical properties of higher-dimensional symplectic systems
 \cite{arn1,arn2,thir1,sour,gui,mars1}. 
\subsection{Range of the paper} As the topics of the present paper reach from experimental to mathematical
physics I shall have missed many papers relevant to the subjects mentioned. I apologize to the experts and hope to do
more justice to their work in the future. Many more related Refs.\ are contained in my paper \cite{ka1} to which I shall
refer frequently in the present one. An essential difference between this paper and Ref.\ \cite{ka1} is the almost complete
focus on the possible consequences of a consistent quantum mechanics for the angle-action variable description of 
the harmonic oscillator in different branches of physics, which is lacking in the previous paper.
\section{Some properties of the classical harmonic oscillator}
\subsection{The globally singular relationship between the canonical pairs $(q,p)$ and $(\vp,I)$}
The transformation \eqref{eq:920} is {\em locally} symplectic (``canonical''):
\begin{equation}
  \label{eq:932}
  dq\wedge dp =  d\vp\wedge dI\,, \qquad \text{or}  \qquad \frac{\partial(q,p)}{\partial(\vp,I)}=1 \,.
\end{equation}
As the angle $\vp$ is dimensionless and for other reasons it is convenient to introduce dimensionless quantities
 by means of the
unit of length $\lambda_0$ from Eqs.\ \eqref{eq:927} and Planck's constant $\hbar$ and restore the physical dimensions
when necessary:
\begin{eqnarray}
  \label{eq:936}
  \tilde{q} &=& q/\lambda_0\,,~~\lambda_0 =\sqrt{\frac{\hbar}{M\,\om}}\,, \\
\tilde{p} &=& p\,\lambda_0/\hbar\,, \label{eq:937} \\
\tilde{H}&=& H/(\hbar \omega) = \frac{1}{2}(\tl{q}^2 +\tl{p}^2)\,,\label{eq:938} \\
\tl{I}&=& I/\hbar = \tl{H} \label{eq:939}\,, \\
\tl{h}_j &=& h_j/\hbar\,,~j=0,1,2\,, \label{eq:8} \\
 \tl{t} &=& \omega\,t\,, \label{eq:944} \\
dq\wedge dp& = &\hbar\,d\tl{q}\wedge d\tl{p} = \hbar\,d\vp \wedge d\tl{I}\,,\label{eq:948}
\end{eqnarray}
Now
\begin{equation}
  \label{eq:940}
  \tl{q} = \sqrt{2\,\tl{I}}\,\cos \vp\,,~~\tl{p} = -\sqrt{2\,\tl{I}}\,\sin \vp\,.
\end{equation}

As
\begin{equation}
  \label{eq:933}
  \tl{p}\,d\tl{q} = \tl{I}\,d\vp - d(\tl{I}\,\cos \vp\,\sin \vp)\,,
\end{equation}
we have locally the four equivalent generating functions
\begin{eqnarray}
  \label{eq:934}
  dF_1(\tl{q},\vp)& =& \tl{I}\,d\vp -\tl{p}\,d\tl{q}\,,~~\partial_{\vp}F_1 = \tl{I}\,,~~\partial_{\tl{q}}F_1 = -\tl{p}\,, \\
F_1(\tl{q},\vp) &=& \frac{1}{2}\,\tl{q}^2 \tan \vp\,; \nonumber \\
dF_2(\tl{q},\tl{I}) &=& \tl{p}\,d\tl{q} + \vp\,d\tl{I}\,, \label{eq:941} \\
F_2(\tl{q},\tl{I}) &=& \tl{I}\,\arccos[\tl{q}/(\sqrt{2\,\tl{I}})]\pm \frac{1}{2} \tl{q}\,
\sqrt{2\tl{I}-\tl{q}^2}\,; \nonumber \\
dF_3(\tl{q},\tl{p}) &=& -\sqrt{2\tl{I}}\,\sin \vp\,d\tl{q}+ \sqrt{2\tl{I}}\,\cos \vp\,d\tl{p}\,,
\label{eq:942} \\ F_3(\tl{q}\,\tl{p}) &=& \tl{q}\,\tl{p}\,; \nonumber \\
dF_4(\vp,\tl{I})&=& \frac{1}{2}\,(\tl{q}^2-\tl{p}^2)\,d\vp-\frac{\tl{q}\,\tl{p}}{\tl{q}^2+
\tl{p}^2}\,d\tl{I}\,,\label{eq:943} \\
F_4(\vp,\tl{I})&=& \tl{I}\,\cos \vp \, \sin\vp\,. \nonumber
\end{eqnarray}
On $\mathcal{S}_{\vp,\tl{I}}$ we have the (trivial) equations of motion
\begin{equation}
  \label{eq:945}
  \dot{\vp} = \frac{\partial \tl{H}}{\partial\tl{I}}= \frac{\partial \tl{I}}{\partial\tl{I}}=1\,,
~~~\dot{\tl{I}} = -\frac{\partial 
\tl{I}}{\partial \vp} = 0\,,
\end{equation}
with the solutions (orbits)
\begin{equation}
  \label{eq:946}
  \vp(\tl{t}) = \tl{t} +\vp_0\,,~~\tl{I} = \text{ const.} > 0\,.
\end{equation}
Inserted into the Eqs.\ \eqref{eq:940} we get the usual orbits on $\mathcal{S}_{\tl{q},\tl{p}}$,
except for the trivial one \\ $(\tl{q}(\tl{t}),\tl{p}(\tl{t})) \equiv (0,0)$\,!

That $(\tl{q},\tl{p})=(0,0)$ or $\tl{I} =0$ is a singular point of the otherwise symplectic 
transformation \eqref{eq:940} can be seen in different ways: 
\begin{itemize} \item
The action variable appears as $\sqrt{\tl{I}}$, i.e.\ one has a branch point at $\tl{I} = 0$\,.
\item
If one introduces $\rho = \sqrt{\tl{I}}$ then the functional determinant 
  \begin{equation}
    \label{eq:947}
    \frac{\partial(\tl{q},\tl{p})}{\partial(\vp,\rho)} = \rho
  \end{equation}
becomes singular for $\rho=0$\,.
\item
The differential $d\tl{H}(\tl{q},\tl{p}) =\tl{q}\,d\tl{q} + \tl{p}\,d\tl{p}$ has a critical point
at $(\tl{q},\tl{p})= (0,0)$\,.
\item
The differentials \eqref{eq:934} -- \eqref{eq:943} of the generating functions $F_j$ become 
singular for $(\tl{q},\tl{p}) = (0,0)$ or $\tl{I}=0$\,. \end{itemize}

So one has to delete the origin of the phase space $\mathcal{S}_{\tl{q},\tl{p}}$ in order to map
it in a one-to-one manner onto $\mathcal{S}_{\vp,\tl{I}}$ and vice versa! But the punctured
 $(\tl{q},\tl{p})$\,-\,plane is no longer simply connected and topologically non-trivial
 (its first homotopy group
$\pi_1$ is $\mathbb{Z}\,$).
This non-trivial topology also manifests itself in the multi-valuedness of the angle
 $\vp$ which is mathematically represented by the unit circle $S^1 \cong \mathbb{R} \bmod{2\pi}
$. This unit circle constitutes
the multiply-connected ``configuration space'' of the phase space $\mathcal{S}_{\vp,\tl{I}}$. 
One of its here essential properties can be read off Eq.\ \eqref{eq:946}:

In the course of time the periodical motion in both phase spaces \eqref{eq:829} and \eqref{eq:2158} passes the
position $\vp_0$ a few or many times. In this way the configuration space $S^1 \subset \mathcal{S}_{\vp,\tl{I}} $
gets unwrapped onto the real axis $\mathbb{R}$ or at least a part of it, here represented by the
variable $\tl{t}$. $\mathbb{R}$ constitutes the universal covering space of $S^1$. A very
similar situation in which the same $SO(2) \cong S^1$ plays a corresponding role is discussed in Ref.\ \cite{ka2}.
The local character of the transformation \eqref{eq:940} and its singularity at $(\tl{q}=0, \tl{p}=0)$ is emphasized
in Thirring's textbook \cite{thir1}.

Note that physically the point $(\tl{q}=0, \tl{p}=0)$ is the ground state (equilibrium point) of the classical $(\tl{q},
\tl{p})$-description of the oscillator motion. In the case of the $(\vp,\tl{I})$-description the notion of an angle does
not make sense any more for $\tl{I} =0$. But $\tl{I}$ may be arbitrarily small as long as it stays positive. As $H=
\om\,I$ one can have $H \to 0$ for $I > 0$ by (formally) taking the limit $\om \to 0\,$.
\subsection{A symplectic scale transformation}
 The replacement
\begin{equation}
  \label{eq:949}
  \vp~\to~ \vp_{\beta} = \vp/\beta\,,~~~\tl{I}~\to ~\tl{I}_{\beta} = \beta\,\tl{I}\,, ~~\beta>0\,, 
\end{equation}
is symplectic ($d\vp_{\beta} \wedge d\tl{I}_{\beta}= d\vp\wedge d\tl{I}\,$).
The transformation implies (cf.\ Eq.\ \eqref{eq:946})
\begin{equation}
  \label{eq:2076}
 \tl{t} \to \tl{t}_{\beta} = \tl{t}/\beta\,.
\end{equation}
From
\begin{equation}
  \label{eq:950}
   \tl{q}_{\beta} = \sqrt{2\,\tl{I}_{\beta}}\,\cos \vp_{\beta}\,,~~\tl{p}_{\beta} = -\sqrt{2\,\tl{I}_{\beta}}\,\sin
 \vp_{\beta}\,,
\end{equation}
we get
\begin{equation}
  \label{eq:951}
  \tl{H}_{\beta} =\frac{1}{2}\,(\tl{p}_{\beta}^2 + \tl{q}_{\beta}^2\,) =\tl{I}_{\beta} = \beta\,\tl{I} =\beta\, \tl{H}\,,
\end{equation}
and 
\begin{equation}
  \label{eq:952}
  \frac{d\vp_{\beta}}{d\tl{t}_{\beta}} = \frac{\partial \tl{H}_{\beta}}{\partial \tl{I}_{\beta}}=
 \frac{\partial \tl{I}_{\beta}}{\partial \tl{I}_{\beta}} =1\,,~~ \Rightarrow
 \vp_{\beta}(\tl{t}_{\beta}) =
\tl{t}_{\beta} + \vp_{\beta}(0)\,. 
\end{equation}
Inserting this $\vp_{\beta}(\tl{t}_{\beta})$ into Eqs.\ \eqref{eq:950} yields the $\tl{t}_{\beta}$-dependence for the
variables $\tl{q}_{\beta}\,,\,\tl{p}_{\beta}\,$, analogously  to the $\tl{t}$-dependence  of the coordinates \eqref{eq:940}.

As $\tl{t} = \om\,t$ (cf.\ Eq.\ \eqref{eq:944}) the transformation of the original dimensionful quantities is ambiguous: 
\begin{enumerate} \item One can choose
  \begin{equation}
    \label{eq:2144}
    t \to t_{\beta} = t/\beta\,,~~ \om \to \om\,.
  \end{equation}
This implies (cf.\ Eq.\ \eqref{eq:920})
\begin{equation}
  \label{eq:2145}
  q \to q_{\beta} = \sqrt{\beta}\,q\,,~~ p \to p_{\beta} = \sqrt{\beta}\,p\,,~~ H \to H_{\beta} = \beta\,H =\om\,I_{\beta}\,.
\end{equation}
\item A second possibility is
  \begin{equation}
    \label{eq:2146}
    t \to t\,,~~\om \to \om_{\beta} = \om/\beta\,,
  \end{equation}
with
\begin{equation}
  \label{eq:2147}
  q \to q_{\beta} = \beta\,q\,,~~ p \to p_{\beta} = p\,,~~ H \to H_{\beta} = H =\om_{\beta}\,I_{\beta}\,.
\end{equation}
\end{enumerate}
Both possibilities are not symplectic as to $q$ and $p$.

Without further restrictions on the values of $\beta$ the transformation \eqref{eq:949} presupposes
the existence of covering spaces for $S^1$, because $\vp/\beta$ may be outside a given interval, e.g.\ $(-\pi,\,\pi]$.
\subsection{Going beyond the harmonic oscillator}

\subsubsection{Time-dependent perturbations}
 If we perturb $\tl{H}_0= \tl{I}_0$ by a time-dependent term
\begin{equation}
  \label{eq:953}
  \tilde{H}_1 = \epsilon\,\tl{I}_0\,f(\tl{t})\; \ll \tl{I}_0\,,
\end{equation}
where $f(\tl{t})$ is independent of $\vp$ and $\tl{I}_0$\,, then
\begin{equation}
  \label{eq:954}
  \dot{\vp}= \partial_{\tl{I}_0} (\tl{H}_0 + \tl{H}_1)  = 1 +\epsilon\,f(\tl{t})\,,
~\dot{\tl{I}} = -\partial_{\vp} (\tl{H}_0 + \tl{H}_1) =0\,,
\end{equation}
so that
\begin{equation}
  \label{eq:1907}
  \vp(\tl{t}) = \tl{t} + \epsilon\,\int_0^{\tl{t}}d\tau\,f(\tau) + \vp_0\,,~~\tl{I} = const.\ .
\end{equation}
Thus, only the time-dependence of $\vp$ gets modified, but not that of $\tl{I}=\tl{I}_0$\,!

 The
latter property is a special case of the famous adiabatic theorem of mechanics which says that
``small and slow'' perturbations of integrable systems leave the values of action variables
unchanged \cite{arn1,arn2,thir1,born1}. This does, of course, not mean that the energy remains
conserved! As to the important perturbation theory of integrable systems described by angle and action
variables see the Refs.\ \cite{arn1,arn2,thir1,born1}.
\subsubsection{Interactions proportional to $\tl{h}_1$ or $\tl{h}_2$}

On the phase space \eqref{eq:829} the Hamilton functions $H(\tl{q},\,\tl{p})$ depend on  the basic variables
 $\tl{q}$ and $\tl{p}$, well
beyond that of the HO. Similarly the Hamilton functions on \eqref{eq:2158} have to be expressed by the basic 
variables \eqref{eq:831}.
Simple  examples for interaction terms added to $\tl{H} = \tl{I}$ are the following ones:
\begin{equation}
  \label{eq:1}
  \tl{H} = \tl{h}_0 + \gamma\, \tl{h}_1 = \tl{I} +\gamma\, \tl{I}\,\cos \vp\,,~~|\gamma| <1\,.
\end{equation}
The eqs.\ of motion
\begin{eqnarray}
  \label{eq:2}
\dot{\vp} &=& \partial_{\tl{I}}\tl{H} = 1+\gamma\,\cos \vp\,, \\
\dot{\tl{I}} &=& -\partial_{\vp}\tl{H} = \gamma\,\tl{I}\,\sin \vp\,, \label{eq:3} 
\end{eqnarray}
have the solutions \cite{grary1}
\begin{eqnarray}
  \label{eq:4}
  \tan[(\vp(\tl{t})-\vp_0)/2] & =& \sqrt{\frac{1+\gamma}{1-\gamma}}\,\tan[\sqrt{1-\gamma^2}(\tl{t}-\tl{t}_0)/2]\,. \\
 \label{eq:5} \tl{I}(\tl{t}) &=& \tl{I}_0\,[1+\gamma\,\cos (\vp (\tl{t})-\vp_0)]^{-1}\,.
\end{eqnarray}
If we replace $\tl{h}_1$ in Eq.\ \eqref{eq:1} by $\tl{h}_2 =-\tl{I}\,\sin \vp$, we get the solutions \cite{grary2}
\begin{eqnarray}
  \label{eq:6}
  \tan[(\vp(\tl{t})-\vp_0)/2] & =& \sqrt{1-\gamma^2}\,\{\tan[\sqrt{1-\gamma^2}(\tl{t}-\tl{t}_0)/2] 
-\gamma \}\,. \\
  \tl{I}(\tl{t}) &=& \tl{I}_0\,[1-\gamma\,\sin (\vp (\tl{t})-\vp_0)]^{-1}\,. \label{eq:7}
\end{eqnarray}

According to the definitions of Refs.\ \cite{arn1,arn2} the angle $\vp(\tl{t})$ is the ``fast'' variable here and the action
variable $\tl{I}(\tl{t})$ the ``slow'' one. This language means to say that the perturbation $\gamma\,\tl{I}\,\cos \vp$
 (or $-\gamma \tl{I}\,\sin \vp$)  for small $|\gamma|$ merely leads to small oscillations of the action variable
 around its unperturbed value
 $\tl{I}_0$. This can be read off the above solutions immediately for $|\gamma | \ll 1$. Closely related to this
type of behaviour is the concept of averaging the $\vp$-dependent part of the perturbation over a period $2\pi$,
 an often powerful tool for 
estimating the influence of  perturbations on integrable systems  \cite{arn1,arn2,thir1,born1}. Such averaging is
especially discussed in Ref.\ \cite{arn2}.

On the other hand, for $|\gamma| \to 1$ the action variables $\tl{I}(\tl{t})$ in Eqs.\ \eqref{eq:5} and \eqref{eq:7}
fluctuate enormously (``resonances'')!

\subsubsection{Morse and other ``integrable''  potentials}

I briefly discuss three well-known integrable systems \cite{delang} with  potentials for which the Hamilton
 functions $H(\vp, \tl{I})$
are not just proportional to $\tl{I}$ like in the case of the HO, but are quadratic in the action variable.
This is so for the potentials
\begin{eqnarray}
  \label{eq:16}
  V_{Mo}(q) &=& V_0\,(e^{-a\,q}-1)^2\,,~~q \in \mathbb{R}\,;~~a,\,V_0\,:\, const. >0\,, \\ && 
V_{Mo}(q)\geq V_{Mo}(q=0) = 0\,, \nonumber \\
 V_{sMo}(q) &=& V_0[1-1/\cosh^2(aq)] = V_0\tanh^2(aq)\,,~~ V_{sMo}(q) \geq V_{sMo}(q = 0)=0\,, \label{eq:35} \\
&&q \in \mathbb{R}\,,~~V_0 >0\,, \nonumber \\
V_{PT}(q) &=& V_0\,\tan^2(aq)\,,~~aq \in (-\pi/2,\,\pi/2)\,,~~ V_0 > 0\,,~~V_{PT}(q=0) =0\,.\label{eq:17}
\end{eqnarray}
The first one was introduced by Morse \cite{mor} in order to describe  radial vibrations of diatomic molecules ($ q = r 
\geq 0$) somewhat better than the HO does, the second one is a sort of symmetrized Morse potential \cite{lan} and the
third one a slightly modified version of a potential discussed
 by P\"{o}schl and Teller \cite{poe}
 in order to improve upon certain properties of the Morse potential. 
The  potentials $V_{Mo}$ and $V_{sMo}$ have bound states (periodic motions) for $0< E <V_0$, the potential $V_{PT} $
 has only bound
states, for all $E >0$. 

For small $a\,q \ll 1$ the potentials reduce to the HO one:
\begin{equation}
  \label{eq:18}
  V_{Mo} (q) \approx V_{sMo}(q) \approx V_{PT}(q) \approx \frac{1}{2}\,M\om_0^2 q^2\,,~~\om_0 = a\sqrt{2\,V_0/M}\,.
\end{equation}

The ``integrable'' potential \cite{conf}
\begin{equation}
V_c(q) = V_0[aq -1/(aq)]^2\,,~~ q >0\,,~~V_c(q) \geq V_c(q=1/a) = 0\,, \label{eq:36}
\end{equation}
provides an example for which the energy is a linear function of the action variable $I$, like for the HO. For $a\,q \ll
1$ we here have  the HO approximation
\begin{equation}
  \label{eq:44}
  V_c(q) \approx \frac{1}{2}\,M\,\om_0^2 (q-1/a)^2\,,~~\om_0 = 2a\sqrt{2\,V_0/M}\,.
\end{equation}

For any potential $V(q)$ with periodic orbits on the phase space \eqref{eq:829} the action variable is defined by the
closed path integral
\begin{equation}
  \label{eq:19}
 2\pi I(E)= \oint_{C(E)}dq\,p(q,E)\,,~~p(q,E) = \pm \sqrt{2M}\,[E-V(q)]^{1/2}\,,
\end{equation}
where the  integration is to be taken $ clockwise$ along the closed path $C(E)$ determined by the energy
equation
\begin{equation}
  \label{eq:20}
  \frac{1}{2M}\,p^2 + V(q) = E\,.
\end{equation}

The factor $2\pi$ in the definition \eqref{eq:19} is due to the convention which uses the circular frequency $\om_0= 2\pi/
T$ and not $\nu = 1/T$.

The integral \eqref{eq:19} describes the area of the region the boundary of which is given by the closed curve $C(E)\,$.

If we now insert the relations \eqref{eq:920} into the integral \eqref{eq:19} we get the identity $2\pi\,I(E) = 2\pi\,I$.
This shows explicitly that the mapping \eqref{eq:920} is independent of the potential chosen.

 If $q_- < q_+$ are the inner and outer turning points of the motion we can replace 
the closed path integral in Eq.\ \eqref{eq:19} by
\begin{equation}
  \label{eq:21}
  2\pi\,I(E) = 2\sqrt{2M} \int_{q_-}^{q_+}dq\,[E-V(q)]^{1/2}\,.
\end{equation}
(Notice that $p\,dq =p\dot{q}\,dt>0$ on the path $C(E)$ in both, the upper and the lower $(q,p)$-half-planes.)

As we have three free parameters now, $M,\,a$ and $V_0\,$, we do not have to use Planck's constant in order to introduce
dimensionless quantities
\begin{equation}
  \label{eq:22}
  \tl{q} = a\,q\,,~~\tl{p} = \frac{p}{\sqrt{M\,V_0}}\,,~~\tl{E} = E/V_0\,,~~\tl{I} =I\,\om_0\,/V_0\,.
\end{equation}
\paragraph{Morse potential} \hfill

For the potential $V_M(q)$ the epression \eqref{eq:21} now takes the form
\begin{equation}
  \label{eq:23}
  \pi\,\tl{I}(\tl{E}) = 2\int_{\tl{q}_-}^{\tl{q}_+}d\tl{q}\,[\tl{E}-(e^{-\tl{q}}-1)^2]^{1/2}\,.
\end{equation}
The integral can be solved explicitly (cf.\ Appendix A) and the result is
\begin{equation}
  \label{eq:24}
 \tl{I} = 2(1-\sqrt{1-\tl{E}})\,,~~ \Rightarrow~~ \tl{E}(\tl{I}) = \tl{I}\,\left(1-\frac{1}{4}\tl{I}\right)\,.
\end{equation}
The inequality $0 <\tl{E} <1$ implies for $\tl{I}$
\begin{equation}
  \label{eq:25}
 0 < \tl{I} < 2\,.
\end{equation}
Restoring the physical dimensions we get the Hamilton function
\begin{equation}
  \label{eq:26}
  H_{Mo}(I) = \om_0\,I\,\left(1-\frac{\om_0\,I}{4\,V_0}\right)\,.
\end{equation}
It yields the eqs.\ of motion
\begin{equation}
  \label{eq:27}
  \dot{I} =0\,,~~\dot{\vp} = \om_0 -\frac{\om_0^2I}{2V_0}\,,
\end{equation}
which can be integrated immediately.

In order to quantize the system as to its  sector of bound states, we merely have to replace the action variable $I$
by the operator $\hbar\,\tl{K}_0$ (cf.\ Eq.\ \eqref{eq:923}). This leads to the Hamilton operator
\begin{equation}
  \label{eq:28}
  H_{Mo}(\vec{K}) = \hbar\,\om_0\,\tl{K}_0 - \frac{(\hbar\,\om_0)^2}{4\,V_0}\,\tl{K}_0^2\,,
\end{equation}
which, according to Eq.\ \eqref{eq:925}, yields the spectrum
\begin{equation}
  \label{eq:29}
  E_{k,\,n} = \hbar\,\om_0 (n+k)[1-\frac{\hbar\,\om_0}{4\,V_0}\,(n+k)]\,,
\end{equation}
which for $k=1/2$ is well-known \cite{mor2}.  Concrete Hilbert spaces and eigenfunctions are provided by irreducible unitary
representations as discussed in sec.\ 7. The eigenfunctions of $ H_{Mo}(\vec{K})$ do not, of course, have to be
 solutions of the Schr\"{o}dinger
eq.\ in $q$-space, as is the case in Refs.\ \cite{mor2}. But, because of the unitary equivalences, all physical predictions
are the same!

As the square bracket in Eq.\ \eqref{eq:29} should be positive one has to cut off the spectrum at a maximal $n=n_{max}$,
like it is done usually.

\paragraph{The other potentials} \hfill

For the potential \eqref{eq:35} one gets (cf.\ Appendix A) the same form for the Hamilton function 
as in Eq.\ \eqref{eq:26}, namely
\begin{equation}
  \label{eq:45}
   H_{sMo}(I) = \om_0\,I\,\left(1-\frac{\om_0\,I}{4\,V_0}\right)\,.
\end{equation}

 For the  potential \eqref{eq:17} one obtains (cf.\ Appendix A)
\begin{equation}
  \label{eq:30}
  \tl{I} = 2(\sqrt{\tl{E}+1} -1)\,,~~ \Rightarrow~~  H_{PT}(I) = \om_0\,I\,\left(1+\frac{\om_0\,I}{4\,V_0}\right)\,,
\end{equation}
which may be quantized accordingly. Again the result is well-known for $k=1/2$ \cite{wue2}.

Finally one obtains for the potential \eqref{eq:36}
\begin{equation}
  \label{eq:37}
H_c(I) = \om_0\,I\,,~~\om_0 = 2 a\sqrt{2\,V_0/M}\,.
\end{equation}
Comparison of $H_{sMo}(I)$ with $H_{Mo}(I)$  and of $H_c(I)$ with $H_{HO}(I)$ shows that the possible
orbits of motion may not depend on the details of the potentials $V(q)$, but only on some generic properties represented
by the associated $H(I)$. There is still, however, the possibility that the quantized systems have different indices $k\,$.
This is indeed the case for the solutions of the Schr\"{o}dinger eqs.\ with the Hamiltonians $H_{Mo}(Q,P)$ and $H_{sMo}(
Q,P)\,$ \cite{lan2}\,.

\subsubsection{Free non-relativistic particle}

According to the second of the Eqs.\ \eqref{eq:928} we can rewrite the Hamilton function
\begin{equation}
  \label{eq:52}
  H_0(q,p) = \frac{1}{2M}\,p^2
\end{equation}
of a free particle as
\begin{equation}
  \label{eq:53}
 H_0(\vec{h}) = \om\,h_2^2/h_0\,,~~\vec{h} = (h_0,h_1,h_2)\,. 
\end{equation}

What is remarkable is that one needs an additional time scale - here provided by $\om$ - in order to express $H_0$ in
terms of the functions \eqref{eq:831}!
\section{Action of the symplectic group  on the phase space
${\mathcal{S}_{\tl{q},\tl{p}}}$}

The transformation group $SO^{\uparrow}(1,2)$ and its double covering, the symplectic group
in 2 dimensions $Sp(2,\mathbb{R})$, play a significant role in the following discussions.
Some of their main properties have been summarized in Appendices A and B of Ref.\ \cite{ka1}. In order
to keep the present paper at least partially self-contained, some of those properties needed here are again sketched
below (secs.\ 3 -- 5) and in Appendix B  of this article. 

The present Section provides a systematic justification for the choice of the basic coordinates \eqref{eq:831} on
 the phase space
\eqref{eq:2158} in terms of the symplectic transformation group $Sp(2, \mathbb{R})$ on the phase space \eqref{eq:829},
{\em without assuming this to be the phase space of the HO}.
\subsection{Global and infinitesimal transformations, ``observables''}

The elements of the symplectic group $G_1 \equiv Sp(2,\mathbb{R})~~ (\, =SL(2,\mathbb{R})\,)$ are given by the matrices
\begin{equation}
  \label{eq:935}
  g_1= \left( \begin{array}{cc} a_{11} & a_{12}
\\ a_{21} & a_{22} \end{array} \right)\,,~ a_{jk}\in
\mathbb{R}\,,~\det g_1=1\,, 
\end{equation}
which have the (defining) property
  \begin{equation}
    \label{eq:955}
  g_1^T\cdot \left(
\begin{array}{cc} 0 & 1 \\ -1 & 0 \end{array} \right)\cdot
g_1= \left(
\begin{array}{cc} 0 & 1 \\ -1 & 0 \end{array} \right)\,.   
  \end{equation}
 If we introduce
\begin{equation}
  \label{eq:956}
   x=\begin{pmatrix}\tl{q} \\\tl{p} \end{pmatrix} \in  {\cal S}_{\tl{q},\tl{p}} \cong \mathbb{R}^2
\,,
\end{equation} then the elements $g_1$ of $Sp(2,\mathbb{R})$
 act on $x$ as
\begin{equation}
  \label{eq:957}
  x \to x^{\prime} = g_1 \cdot x\,,~~~g_1 \in G_1 \equiv Sp(2,\mathbb{R})\,,
\end{equation}
with the property 
\begin{equation}
  \label{eq:958}
  d\tl{q}^{\,\prime} \wedge d\tl{p}^{\,\prime} = d\tl{q} \wedge d\tl{p} \,,
\end{equation} 
i.e.\ the transformations \eqref{eq:957} {\em leave the symplectic form}
\begin{equation}
  \label{eq:959}
  \omega_{\tl{q},\tl{p}} = d\tl{q} \wedge d\tl{p}
\end{equation}
{\em invariant}.

The group action \eqref{eq:957} has some other remarkable
properties: 

The whole group transforms the point $x=0$ into itself and acts
 {\em transitively}
on the complement 
\begin{equation}
  \label{eq:960}
 \mathcal{ S}_{\tl{q},\tl{p};\,0} \equiv \mathcal{ S}_{\tl{q},\tl{p}} -\{x=0\} \cong \mathbb{R}^2
- \{(0,0)\}\,,
\end{equation}
i.e., if $x_1$ and $x_2$ are any two points of $\mathcal{S}_{\tl{q},\tl{p};\,0}$\,, then
they can be transformed into each other by an element of $G_1$. This can easily be seen by
considering the first two of the following 1-parameter subgroups of $G_1$:
\begin{eqnarray}\label{eq:965} R_1: && r_1= \left(
\begin{array}{cc} \cos(\theta/2) & \sin(\theta/2) 
 \\ -\sin(\theta/2) &
\cos(\theta/2) \end{array} \right)\,,~ \theta \in (-2\pi,+2\pi]\,, 
 \\ A_1: && a_1 = \left(
\begin{array}{cc} e^{\ds- \tau/2} & 0 \\ 0 &
e^{\ds \tau /2} \end{array} \right)\,,~\tau \in \mathbb{R}\,, \label{eq:964} \\
B_1: && b_1=
\left(
\begin{array}{cc} \cosh(s/2) & \sinh(s/2) \\ \sinh(s/2) &
\cosh(s/2) \end{array} \right)\,,~s \in \mathbb{R}\,;\label{eq:967} \\
N_1: &&
n_1 = \left(
\begin{array}{cc} 1 & \xi \\ 0 &
1 \end{array} \right)\,,~ \xi \in \mathbb{R}~. \label{eq:966} 
 \end{eqnarray} Each
element $g_1$ has  a (Cartan) decomposition $g_1 = k_2 \cdot a_1 \cdot k_1$ or
$g_1 = k_2 \cdot b_1 \cdot k_1$ and a unique
 (Iwasawa) decomposition $g_1=k_1\cdot a_1 \cdot
n_1$\,, where $k_1,\, k_2 \in R_1\,$. 

Now, let $x_1$ and $x_2$ be any two points of $\mathcal{S}_{\tl{q},\tl{p};\,0}$. First rotate
$x_1$ by an element of $R_1$ into $x_1^{\,\prime}$, where $\tl{p}_1^{\,\prime}=0$ and
 $\tl{q}_1^{\,\prime}$ has
the same sign as $\tl{q}_2$. Then use an element of $A_1$ so that $e^{-\tau /2}\,\tl{q}_1^{\,\prime} = 
\tl{q}_1^{\,\prime\prime}= \sqrt{\tl{q}^2_2 + \tl{p}^2_2}$\,. Finally rotate the point
 $ (\tl{q}_1^{\,\prime\prime},0)$ into $x_2$\,.

The group $G_1$ acts also {\em effectively} on  $\mathcal{ S}_{q,p;\,0}$, that is to say, if
\begin{equation}
  \label{eq:961}
  x= g_1\cdot x~~\forall x\,,
\end{equation}
  then $g_1$ is the identity element
\begin{equation}
  \label{eq:962}
  e = E_2 \equiv \begin{pmatrix} 1&0\\0&1 \end{pmatrix}\,.
\end{equation}
\subsection{Vector fields and their associated Hamiltonian functions}

The 1-parameter subgroups \eqref{eq:965} - \eqref{eq:966} generate vectorfields on 
$\mathcal{ S}_{q,p;\,0}$ in the following sense: Let $\Gamma = \{\gamma (s)\} $ be a 1-parameter
 group
such that $\gamma(s=0) = 1$ and let $f(x)$ be a smooth function. Then the $\Gamma$-associated
vectorfield $\tl{A}_{\Gamma}$ is defined by
\begin{equation}
  \label{eq:963}
  [\tl{A}_{\Gamma}f](x)=\lim_{s\to 0}\frac{1}{s}[f(\gamma(-s)\cdot x)-f(x)]\,.
\end{equation}

From the first three subgroups above  we get the following 3-dimensional basis of vectorfields associated
 with the group $G_1$:
\begin{eqnarray} \label{eq:968}\tilde{A}_{R_1} &=&\frac{1}{2}(\tl{q}\,
\partial_{\tl{p}}-\tl{p}\,\partial_{\tl{q}})\,,
\\
\label{eq:970}\tilde{A}_{A_1} &=& \frac{1}{2}(\tl{q}\,\partial_{\tl{q}} -\tl{p}\,
\partial_{\tl{p}})\,,
  \\
\label{eq:971}\tilde{A}_{B_1} &=& -\frac{1}{2}(\tl{p}\,\partial_{\tl{q}} +\tl{q}\,
\partial_{\tl{p}})\,, 
\end{eqnarray} They obey the Lie algebra $\mathfrak{sp}(2,\mathbb{R}) =
\mathfrak{so}(1,2)$:
\begin{equation}
  \label{eq:978}
  [\tl{A}_{R_1}, \tl{A}_{A_1}] = \tl{A}_{B_1}\,,~~[\tl{A}_{R_1},\tl{A}_{B_1}]=-\tl{A}_{A_1}\,,~~
[\tl{A}_{A_1},\tl{A}_{B_1}]= -\tl{A}_{R_1}\,.
\end{equation}
Notice that the vector fields \eqref{eq:968} -- \eqref{eq:971} vanish
 for $x=0$, a point to be excluded\,!

These vector fields are {\em global Hamiltonian} ones, that is to say there exist global functions
 $\check{g}(x)$ on $\mathcal{ S}_{q,p;\,0}$ such that the vector fields may be written as
 \begin{equation}
   \label{eq:972}
   -[\partial_{\tl{p}}\check{g}(x)\,\partial_{\tl{q}} - \partial_{\tl{q}}\check{g}(x)\,
\partial_{\tl{p}}]\,.
 \end{equation}
 The three Hamiltonian functions  here are
\begin{eqnarray} \label{eq:973}R_1:~~\check{g}_0(x) &=&
 \frac{1}{4}(\tl{q}^2+\tl{p}^2)\,, \\
\label{eq:974}A_1:~~\check{g}_2(x) &=& -\frac{1}{2}\,\tl{q}\,\tl{p}\,, \\ B_1:~~
\check{g}_1(x)&=& \frac{1}{4}(-\tl{q}^2+\tl{p}^2)\,. \label{eq:975}
\end{eqnarray}
(The numbering of the functions is mere convention.)

 Their Poisson brackets  obey the Lie algebra $\mathfrak{sp}(2,\mathbb{R}) =
\mathfrak{so}(1,2)\,$, too:
\begin{equation}
  \label{eq:976}
 \{\check{g}_0,\check{g}_1 \}_{\tl{q},\tl{p}} = -\check{g}_2\,,~~~~
 \{\check{g}_0,\check{g}_2 \}_{\tl{q},\tl{p}} = \check{g}_1\,,~~~~
 \{\check{g}_1,\check{g}_2 \}_{\tl{q},\tl{p}} = \check{g}_0\,.
\end{equation}
 The squares of the $\check{g}_j(x)$ fulfill the relation
\begin{equation}
  \label{eq:977}
  \check{g}_0^2-\check{g}_1^2 -\check{g}_2^2 =0\,.
\end{equation} 

On the other hand, the vector fields induced by the following translations, but now on the phase space
 $\mathcal{S}_{\tl{q},\tl{p}}$,
\begin{equation}
  \label{eq:969}
  \tl{q} \to \tl{q} +a\,,~\tl{p} \to \tl{p}\,;~~\tl{q} \to \tl{q}\,,~\tl{p} \to \tl{p} -b\,;~
a, b \in \mathbb{R}\,,
\end{equation}
are
\begin{equation}
  \label{eq:983}
 \tilde{A}_{\tl{q}}=- \partial_{\tl{q}}\,;~~\tl{A}_{\tl{p}} =\partial_{\tl{p}}\,, 
\end{equation}
with the Hamiltonian functions
\begin{equation}
  \label{eq:984}
  \check{g}_{\tl{q}}(x)=\tl{p}\,,~~\check{g}_{\tl{p}}(x) = \tl{q}\,,
\end{equation}
the Poisson brackets of which generate the usual Born-Dirac-Heisenberg-Jordan-Weyl (Lie)
 algebra \\ (called BDHJW-algebra in the following\footnote[1]{The usual terminology is ``Heisenberg-'' or 
``Weyl-'' algebra,
 but I
 think this to be unjust towards the contributions of the other authors.}) with its basis $\{\tl{q},\;\tl{p},\; 1\}$\,!

The bilinear functions \eqref{eq:973} -- \eqref{eq:975} are the generators of the infinitesimal transformations 
associated with the transformations \eqref{eq:957} of the  subgroups \eqref{eq:965} -- \eqref{eq:967}:
\begin{eqnarray}
  \label{eq:1228}
  \{\check{g}_0,\tl{q}\}=-\frac{1}{2}\;\tl{p}\;,~~~&&~~~\{\check{g}_0,\tl{p}\}=\frac{1}{2}\;\tl{q}\;, \\
\{\check{g}_1,\tl{q}\}=-\frac{1}{2}\;\tl{p}\;,~~~&&~~~\{\check{g}_1,\tl{p}\}=-\frac{1}{2}\;\tl{q}\;,\label{eq:1230}\\
\{\check{g}_2,\tl{q}\}=\frac{1}{2}\;\tl{q}\;,~~~&&~~~\{\check{g}_2,\tl{p}\}=-\frac{1}{2}\;\tl{p}\;.\label{eq:1235}
\end{eqnarray}
Integrated they give the transformations \eqref{eq:957} of the subgroups \eqref{eq:965} -- \eqref{eq:967}, except for an
unessential overall minus-sign of the group parameters, a consequence of the definition \eqref{eq:972}.

It is evident that the phase spaces \eqref{eq:956} and \eqref{eq:960} have not only quite  different topological but, as a
consequence, also essentially different
canonical structures as to the transformation groups which act transitively on them: The phase space \eqref{eq:956} has the
translations \eqref{eq:969} with their associated central extension (characterized by $\{\tl{q},\tl{p}\}= 1\,$) as its
``canonical'' group, but the phase space \eqref{eq:960} the symplectic group $Sp(2,\mathbb{R})$. This difference has
important consequences for the quantum theory as we shall see!

The Hamiltonian functions \eqref{eq:984} play a double role on the phase space $\mathcal{S}_{\tl{q},\tl{p}}\,$: They are
the generators of the (canonical) translations {\em and} at the same time they are the basic classical ``observables''
 on that phase space. Similarly one may consider the Hamiltonian functions \eqref{eq:973} -- \eqref{eq:975} as basic 
observables on $\mathcal{S}_{\tl{q},\tl{p};\,0}$. However, there is the following  ambiguity: Given a triple $(\check{g}_0 >0,
\check{g}_1,\check{g}_2)$ with the property \eqref{eq:977}, then the 2 pairs $(\tl{q},\tl{p})$ and $(-\tl{q},-\tl{p})$ are
compatible with a given triple. For further discussions of this important point see below.

The group $Sp(2,\mathbb{R})$ is not only a transformation (automorphism) group of the BDHJW-algebra but the relations
\eqref{eq:976}, \eqref{eq:1228} -- \eqref{eq:1235} and  $\{\tl{q},\tl{p}\}= 1$ show that the direct sum of the vector
 spaces of the Lie 
algebra $\mathfrak{sp}(2,\mathbb{R})$ and the BDHJW-algebra forms a 6-dimensional Lie algebra of its own. This feature
plays a major role in the harmonic (Fourier) analysis of the BDHJW-group \cite{harm}.

Whereas the coordinates of the points $x$ transform as vectors with respect to the
group $Sp(2,\mathbb{R})\,$,  (cf.\ Eq.\ \eqref{eq:957}), the Hamiltonian functions \eqref{eq:973} --
\eqref{eq:975} transform as tensors of second degree: Applying the groups \eqref{eq:965} -
\eqref{eq:967} to  $\tl{q}$ and $\tl{p}$ and inserting the results into the r.h.\ sides of the
expressions \eqref{eq:973} -- \eqref{eq:975} yields the following transformations
\begin{eqnarray}
  \label{eq:985}
  R_1:~~~ \check{g}_0(x) &\to& \check{g}_0(x^{\prime}) = \check{g}_0(x)\,,\label{eq:986} \\
\check{g}_1(x) &\to& \check{g}_1(x^{\prime}) = \cos \theta \, \check{g}_1(x) + \sin \theta\,
 \check{g}_2(x)\,,
\nonumber \\ \check{g}_2(x) &\to& \check{g}_2(x^{\prime}) = -\sin \theta\, \check{g}_1(x) + \cos \theta\,
 \check{g}_2(x)\,; \nonumber \\ \nonumber \\
A_1:~~~\check{g}_0(x) &\to& \check{g}_0(x^{\prime}) = \cosh\tau\, \check{g}_0(x) + \sinh\tau\,
 \check{g}_1(x)\,,
\label{eq:987} \\
\check{g}_1(x) &\to& \check{g}_1(x^{\prime}) = \sinh\tau\, \check{g}_0(x) + \cosh\tau\, \check{g}_1(x)\,,
\nonumber  \\
 \check{g}_2(x) &\to& \check{g}_2(x^{\prime}) = \check{g}_2(x)\,; \nonumber \\ \nonumber \\
B_1:~~~\check{g}_0(x) &\to& \check{g}_0(x^{\prime}) = \cosh s\, \check{g}_0(x) - \sinh s\,
 \check{g}_2(x)\,,
\label{eq:989} \\ 
 \check{g}_1(x) &\to& \check{g}_1(x^{\prime}) = \check{g}_1(x)\,, \nonumber \\
\check{g}_2(x) &\to& \check{g}_2(x^{\prime}) = -\sinh s\, \check{g}_0(x) + \cosh s\, \check{g}_2(x)\,.
\nonumber 
\end{eqnarray}
These formulae show that the 3 functions $\check{g}_j$ transform as a 3-vector with respect to
the ``Lorentz'' group $SO^{\uparrow}(1,2)$: The transformations \eqref{eq:986} -- \eqref{eq:989}
leave the quadratic form $ \check{g}_0^2-\check{g}_1^2 -\check{g}_2^2\,$ invariant. 
This is related to the fact that the group $Sp(2, 
\mathbb{R})$ is a double covering of the group $SO^{\uparrow}(1,2)$ with the center
 $\mathbb{Z}_2 
= \{e,-e\}$ of $Sp(2,\mathbb{R})$ as the kernel of the homomorphism $Sp(2,\mathbb{R}) \to
SO^{\uparrow}(1,2)$ (see Appendix B of Ref.\ \cite{ka1}). Notice that the kernel (center) $\mathbb{Z}_2$
leaves the bilinear expressions \eqref{eq:973} -- \eqref{eq:975} invariant.

\subsection{Space reflections and time reversal}
The center $\mathbb{Z}_2$ implements the parity operation
\begin{equation}
  \label{eq:1301}
  \Pi:~~~\tl{q} \to -\tl{q}\,,~~\tl{p} \to - \tl{p}\,,
\end{equation}
which obviously is symplectic (it leaves the 2-form $d\tl{q}\wedge d\tl{p}$ invariant). 

More subtle is the implementation of the time reversal
\begin{equation}
  \label{eq:1302}
  T:~~~\tl{t} \to -\tl{t}\,,~~\tl{q} \to \tl{q}_T = \tl{q}\,,~~\tl{p} \to \tl{p}_T = -\tl{p}\,,
\end{equation}
which is not symplectic (we have $d\tl{q}\wedge d\tl{p} \to - d\tl{q}\wedge d\tl{p}$). 
However, this can be taken care of in analogy to quantum mechanics where time reversal - according to Wigner -
 is implemented by an anti-unitary transformation in Hilbert space:
 \begin{equation}
   \label{eq:1303}
   U_T:~~~\psi_1 \to U_T\psi_1\,,~~\psi_2 \to U_T\psi_2\,,~~(U_T\psi_2,U_T\psi_1)=(\psi_1,\psi_2)=(\psi_2,\psi_1)^*\,,
 \end{equation}
where $(\psi_2,\psi_1)$ denotes the complex-valued scalar product. As $\Im (\psi_2,\psi_1)$ defines a symplectic
 form \cite{mars2}
which changes sign under the complex conjugation \eqref{eq:1303}, this suggests
 to change the order in $d\tl{q}\wedge d\tl{p}$ in the
case of the time reversal \eqref{eq:1302}:
\begin{equation}
  \label{eq:1304}
  (d\tl{q}\wedge d\tl{p})_T=d\tl{p}_T\wedge d\tl{q}_T =-d\tl{p}\wedge d\tl{q} = d\tl{q}\wedge d\tl{p}\,.
\end{equation}
This has corresponding consequences for the associated Poisson brackets: Let $f^{(j)}(\tl{q},\tl{p}),\,j=1,2,$ be two 
smooth functions on the phase space $\mathcal{S}_{\tl{q},\tl{p}}$. With
\begin{equation}
  \label{eq:1305}
  f^{(j)}_T(\tl{q},\tl{p})=f^{(j)}(\tl{q},-\tl{p}),\,j=1,2,
\end{equation}
we define the time-reversed Poisson bracket by
\begin{equation}
  \label{eq:1306}
  \{f^{(2)},f^{(1)}\}_T = \{f^{(1)}_T,f^{(2)}_T\}\,.
\end{equation}
The definition is appropriate in the following sense: The time evolution of a function $f[\tl{q}(\tl{t}),\tl{p}(\tl{t})]$
 (which does not depend explicitly on time) is given by
\begin{equation}
  \label{eq:1307}
  \dot{f}=\{f,\tl{H}\}\,,
\end{equation}
where $\tl{H}$ is the Hamilton function of the system. If $H_T(\tl{q},\tl{p}) = H(\tl{q},\tl{p})$, we have for the time-
reversed Eq.\ \eqref{eq:1307}
\begin{equation}
  \label{eq:1308}
  \frac{d\,f_T}{d(-\tl{t})} = \{f,\tl{H}\}_T = \{\tl{H},f_T\}\,,~~ \Rightarrow \dot{f}_T = \{f_T,\tl{H}\}\,;
\end{equation}
which is what one wants!
\subsection{The space $\mathcal{S}_{\tl{q},\tl{p};\,0}$ as a ``homogeneous'' one}

The phase space $\mathcal{S}_{\tl{q},\tl{p};\,0}$ can be interpreted as a {\em homogeneous} one as follows:

The subgroup \eqref{eq:966} leaves the points of the line $\{(\tl{q},\tl{p}=0)\}$ invariant, i.e.\ it is the
``isotropy'' or ``little'' group of these points. We have already seen that the group $Sp(2,\mathbb{R})$ acts transitively
on $\mathcal{S}_{\tl{q},\tl{p};\,0}$. Both properties imply that we can represent $\mathcal{S}_{\tl{q},\tl{p};\,0}$  as a
homogeneous space, namely
\begin{equation}
  \label{eq:1309}
 \mathcal{S}_{\tl{q},\tl{p};\,0} \cong  Sp(2,\mathbb{R})/N_1,
\end{equation}
  i.e.\ the points $x \in \mathcal{S}_{\tl{q},\tl{p};\,0}$ are in one-to-one
 correspondence with the rest classes $g_1\cdot N_1, \text{where}\, g_1 \in Sp(2,\mathbb{R})$. This is immediately
 plausible:
The group $G_1= Sp(2,\mathbb{R})$ has the unique (Iwasawa) subgroup decomposition $R_1\cdot A_1\cdot N_1\,$, with the
 topological product structure $S^1 \times \mathbb{R}^+ \times \mathbb{R}$\,. ``Dividing out the subgroup $N_1$'' means 
dividing out the topological factor $\mathbb{R}$. The remaining pro\-duct $S^1 \times \mathbb{R}^+$ corresponds to the polar
 coordinates of the punctured plane $\mathbb{R}^2 -\{(0,0)\} \cong \mathcal{S}_{\tl{q},\tl{p};\,0}$\,.

\subsection{Some quantum aspects}
The present subsec.\ is intended to illustrate the role the symplectic group $Sp(2,\mathbb{R})$ from above plays in the
conventional quantum mechanics of the HO, a role which remains unmentioned in the usual QM textbook discussions.
 
Let us  apply the conventional quantization procedure to the functions \eqref{eq:973} --
\eqref{eq:975} by replacing $\tl{q}$ and $\tl{p}$ by the operators $\tl{Q}$ and $\tl{P}$ and
(Weyl) symmetrizing where necessary. We then get
\begin{eqnarray} \label{eq:988}\check{g}_0(x) &\to & \frac{1}{4}\,(\tl{P}^2+\tl{Q}^2) = \tl{K}_0\,,
  \\
\label{eq:990}\check{g}_1(x) &\to & \frac{1}{4}(\tl{P}^2-\tl{Q}^2)= -\tl{K}_1\,.\\ \label{eq:991}
\check{g}_2(x)&\to &  
 - \frac{1}{4}\,(\tl{Q}\,\tl{P}+ \tl{P}\,\tl{Q}) = \tl{K}_2\,.
\end{eqnarray} With
\begin{equation}
  \label{eq:992}
  \tl{Q} =\frac{1}{\sqrt{2}}\,(a^{\dagger}+a)\,,~~\tl{P} =\frac{i}{\sqrt{2}}\,(a^{\dagger}-a)\,,~~
[a,a^{\dagger}]= {\bf 1}\,,
\end{equation}
we have
\begin{equation}
  \label{eq:993}
 \tl{K}_0= \frac{1}{4}\,(2\,a^{\dagger}a +1)\,,~~\tl{K}_1 = \frac{1}{4}\,({a^{\dagger}}^2+a^2)\,,~~\tl{K}_2 =
 -\frac{i}{4}\,({a^{\dagger}}^2-a^2)\,,
\end{equation}
and
\begin{equation}
  \label{eq:995}
  \tl{K}_+=\tl{K}_1+i\,\tl{K}_2 = \frac{1}{2}\,{a^{\dagger}}^2\,,
 ~\tl{K}_-=\tl{K}_1-i\,\tl{K}_2 = \frac{1}{2}\,a^2\,. 
\end{equation}
The associated Lie algebra is
\begin{equation}
  \label{eq:1100}
  [\tl{K}_0,\tl{K}_1]=i\,\tl{K}_2\,,~~[\tl{K}_0,\tl{K}_2]=-i\,\tl{K}_1\,,~~[\tl{K}_1,\tl{K}_2]=-i\,\tl{K}_0\,,
\end{equation}
or
\begin{equation}
  \label{eq:1128}
  [\tl{K}_0,\tl{K}_{\pm}]=\pm \tl{K}_{\pm}\,,~~[\tl{K}_+,\tl{K}_-]=-2\tl{K}_0\,.
\end{equation}
The relations \eqref{eq:993} and \eqref{eq:995} constitute a well-known realization of the
 Lie algebra $\mathfrak{sp}(2,\mathbb{
R}) = \mathfrak{so}(1,2)$ which yields two irreducible  
positive discrete series unitary representations of a twofold covering group
of $Sp(2, \mathbb{R})$ \cite{ka4}:

 Let $|n_{osc} \rangle$ be a number eigenstate of the
harmonic oscillator Fock space:
\begin{eqnarray}
  \label{eq:994}
  a^{\dagger}\,|n_{osc} \rangle = \sqrt{n_{osc} +1}\,|n_{osc}+1 \rangle,&&~
 a\,|n_{osc} \rangle = \sqrt{n_{osc}}\,|n_{osc}-1 \rangle,~ \\a^{\dagger}a|n_{osc}\rangle = n_{osc}\,
|n_{osc}\rangle,&&~n_{osc}=0,1,2,\ldots\,. \nonumber
\end{eqnarray}
As $\tl{K}_-$ annihilates the states
$|n_{osc} =0\rangle$ {\em and} $|n_{osc} =1\rangle$,
  \begin{equation}
    \label{eq:996}
  \tl{K}_-|n_{osc}=0 \rangle =0\,,~~  \tl{K}_-|n_{osc}=1 \rangle =0\,,
\end{equation}
we get two different irreducible unitary
representations associated with the Lie algebra $\mathfrak{sp}(2,\mathbb{R})$ $=\mathfrak{so}(1,2)$,
 one which is given by
states with even numbers of Fock space quanta and one with odd
numbers, both generated by the creation operator $\tl{K}_+$\,:
 Because
\begin{equation}
  \label{eq:262}
 \tl{K}_0 |n_{osc}\rangle
=\frac{1}{2}\,(n_{osc}+1/2)|n_{osc}\rangle\,,~n_{osc} =0,1,2,\ldots,
\end{equation}
we see that $\tl{K}_0$
has the eigenvalues
\begin{equation}
  \label{eq:259}
   (2\,n_{osc} + 1/2)/2= n+\frac{1}{4}~~\text{and}~~ (2\,n_{osc}+1 +
1/2)/2= n+\frac{3}{4}\,,~n=0,1,\ldots,\,
\end{equation}
 in the cases of even and odd numbers of quanta, respectively. 
That is to say, we get one irreducible unitary representation
with $k=1/4$ and one with $k=3/4$. 

As to the related groups these are true
representations of a 2-fold covering $Mp(2,\mathbb{R})$ of $Sp(2,\mathbb{R})=
SL(2,\mathbb{R}) \cong SU(1,1)
$ and a 4-fold covering
of $SO^{\uparrow}(1,2)$. These 2-fold covering groups of the symplectic
groups $Sp(2n,\mathbb{R})$ in $2n$ dimensions are called ``metaplectic''
 \cite{wei,meta}  ones (for more
 details see below). 

As the operators \eqref{eq:988} -- \eqref{eq:991} commute with the parity transformation
\begin{equation}
  \label{eq:997}
  \Pi\,:~~~\tl{Q} \to -\tl{Q}\,,~~~\tl{P}\to -\tl{P}\,,~~~\Pi^2 = {\bf 1}\,,
\end{equation}
the two irreducible representations may be associated with the eigenvalues $\pm 1$ of $\Pi$, respectively.
 
The two representations with $k=1/4$ and $k=3/4$  may, of course, be realized
 in the 2 subspaces ${\cal H}_+$ and ${\cal H}_-$ of
the conventional Hilbert space $L^2(\mathbb{R},d\tl{q})$ of the harmonic oscillator with the orthonormal basis
\begin{equation}
  \label{eq:998}
  u_{n_{osc}}(\tl{q})= \frac{e^{\ds -\tl{q}^2}}{\sqrt{2^{n_{osc}}\sqrt{\pi}\,n_{osc}!}}\,H_{n_{osc}}(\tl{q})\,,
~~H_{n_{osc}}(-\tl{q})= (-1)^{n_{osc}}H_{n_{osc}}(\tl{q})\,,
\end{equation}
where $H_n(\tl{q})$ is the $n$th Hermite polynomial.

The subspace ${\cal H}_+$ for the unitary representation with $k=1/4$
 is spanned by 
the Hermite functions with even Hermite polynomials $H_{ n_{osc}}$ 
 and the subspace ${\cal H}_-$ for the
 representation
with $k=3/4$ is spanned by the Hermite functions with odd Hermite polynomials.

In the ``even'' subspace ${\cal H}_+$ the Hamiltonian
\begin{equation}
  \label{eq:1000}
  \tl{H}_{osc}=2\tl{K}_0
\end{equation}
has the eigenvalues 
\begin{equation}
  \label{eq:1001}
  (n_{osc}+ 1/2)\,,~~n_{osc}=2n\,,~n=0,1,2,\dots\,,
\end{equation} and in the ``odd'' subspace ${\cal H}_-$ its eigenvalues are
\begin{equation}
  \label{eq:1002}
  (n_{osc} +1/2)\,,~~n_{osc}=2n+1\,,~n=0,1,2,\dots\,.
\end{equation}

Notice that the operators \eqref{eq:992} map $\mathcal{H}_+$ onto $\mathcal{H}_-$ and vice versa!

The operators \eqref{eq:931} for the two irreducible representations are
\begin{equation}
  \label{eq:999}
  A_{(1/4)}= \frac{1}{\sqrt{2}}\,(
N_{\Phi} +1)^{-1/2}\,a^2\,,~~~~ A_{(1/4)}^{\dagger}= \frac{1}{\sqrt{2}}\,
{a^{\dagger}}^2(N_{\Phi}+1)^{-1/2}\,,~~N_{\Phi}=a^{\dagger}a\,,
\end{equation}
and
\begin{equation}
  \label{eq:1014}
   A_{(3/4)}= \frac{1}{\sqrt{2}}\,(N_{\Phi} +2)^{-1/2}\,a^2\,,~~~~ A_{(3/4)}^{\dagger}= \frac{1}{\sqrt{2}}\,
{a^{\dagger}}^2(N_{\Phi}+2)^{-1/2}\,.
\end{equation}
(The index $\Phi$ stands for ``Fock''.)

It follows from the properties of $a$ and $a^{\dagger}$ that
\begin{equation}
  \label{eq:1015}
  A_{(1/4)}|n_{osc}=2n \rangle= \sqrt{n}\,|2n-2\rangle\,,~~~
 A_{(1/4)}^{\dagger}|n_{osc}=2n\rangle= \sqrt{n+1}\,|2n+2\rangle\,,
\end{equation}
and
\begin{equation}
  \label{eq:1016}
  N_{(1/4)}= A^{\dagger}_{(1/4)}A_{(1/4)}=\frac{1}{2}\,N_{\Phi}\,,~~~ [A_{(1/4)},
 A^{\dagger}_{(1/4)}]={\bf 1}\,.
\end{equation}
Analogously we get
\begin{equation}
  \label{eq:1048}
 A_{(3/4)}|n_{osc}=2n +1\rangle= \sqrt{n}\,|2n-1\rangle\,,~~~
 A_{(3/4)}^{\dagger}|n_{osc}=2n+1\rangle= \sqrt{n+1}\,|2n+3\rangle\,,
\end{equation}
and
 \begin{equation}
  \label{eq:1049}
N_{(3/4)}= A^{\dagger}_{(3/4)}A_{(3/4)}=\frac{1}{2}\,(N_{\Phi}-{\bf 1})\,,~~~ [A_{(3/4)},
A^{\dagger}_{(3/4)}]={\bf 1}\,.  
\end{equation}
This means
\begin{equation}
  \label{eq:1050}
  N_{1/4}\,|n_{osc}=2n \rangle = n\,|n_{osc}=2n\rangle\,,~~  N_{3/4}\,|n_{osc}=2n+1 \rangle =
 n\,|n_{osc}=2n+1\rangle\,.
\end{equation}
According to Eqs.\ \eqref{eq:927}, \eqref{eq:1016} and \eqref{eq:1049} we may define in $\mathcal{H}_+$ and $\mathcal{H}_-$
the position and momentum operators
\begin{equation}
  \label{eq:1131}
  \tl{Q}_{(k)}= \frac{1}{\sqrt{2}}\,(A_{(k)}+A_{(k)}^{\dagger})\,,~~\tl{P}_{(k)}= \frac{i}{\sqrt{2}}\,(A_{(k)}^{\dagger}-
A_{(k)})\,,~~[\tl{Q}_{(k)},\tl{P}_{(k)}] =i\,{\bf 1}\,,~~k= 1/4\,,\,3/4\,.
\end{equation}
The operators $\tl{Q}_{(1/4)}$ and $\tl{P}_{(1/4)}$ or $\tl{Q}_{(3/4)}$ and $\tl{P}_{(3/4)}$ have on the subspace
 $\mathcal{H}_+$ or $\mathcal{H}_-$, respectively, the same matrix elements  the operators \eqref{eq:992} have on
 $\mathcal{H} = \mathcal{H}_+ \oplus \mathcal{H}_-$\,! This is possible because in an infinite dimensional linear
 (Hilbert) space 
a genuine subspace may be isomorphic to the space itself. Here such a  correspondence can be implemented by $\mathcal{H}
\ni |n\rangle \leftrightarrow |2n\rangle \in \mathcal{H}_+$ or  $\mathcal{H}
\ni |n\rangle \leftrightarrow |2n+1 \rangle \in \mathcal{H}_-$\,, respectively.

There is a crucial difference, however, between the ``elementary'' operators \eqref{eq:992} and the ``composite''
ones \eqref{eq:1131}: 
Using the general operator formula
\begin{equation}
  \label{eq:1189}
  e^{\ds C}B\,e^{\ds -C}= B+[C,B] +\frac{1}{2!}\,[C,[C,B]]+\frac{1}{3!}\,[C,[C,[C,B]]]+\cdots\,,
\end{equation}
we get from Eqs.\ \eqref{eq:992} and \eqref{eq:993}
\begin{equation}
  \label{eq:1213}
  U(\theta)\,a\,U(-\theta)= e^{i\theta/2}\,a\,,~~ U(\theta)\,a^{\dagger}\,U(-\theta)= e^{-i\theta/2}\,a^{\dagger}\,
,~~U(\theta)= e^{-i\,\tl{K}_0\,\theta}\,,~~\tl{K}_0= \frac{1}{4}\,(2\,a^{\dagger}a +1)\,,
\end{equation}
so that
\begin{equation}
  \label{eq:1219}
  U(\theta)\,\tl{Q}\,U(-\theta) = \cos (\theta/2)\,\tl{Q}-\sin (\theta/2)\,\tl{P}\,,~~
U(\theta)\,\tl{P}\,U(-\theta) = \sin (\theta/2)\,\tl{Q}+\cos (\theta/2)\,\tl{P}\,.
\end{equation}
Especially for $\theta = 2\pi$ we get the reflection
\begin{equation}
  \label{eq:1221}
   U(\theta = 2\pi)\,\tl{Q}\,U[-(\theta= 2\pi)]= -\tl{Q}\,,~~ U(\theta = 2\pi)\,\tl{P}\,U[-(\theta= 2\pi)] = -\tl{P}\,.
\end{equation}
This shows that the operators \eqref{eq:992} transform according to the subgroup \eqref{eq:965} of $Sp(2,\mathbb{R})$\,.

For $\theta = 4\pi$ the transformations \eqref{eq:1219} act as the identity on the pair $\tl{Q},\, \tl{P}$, but we have
\begin{equation}
  \label{eq:1709}
  U(\theta = 4\pi) = e^{-i\,4\pi(2N_{osc}+1)/4}=e^{-i\pi}{\bf 1} = -{\bf 1}\,,~~N_{osc} = a^{\dagger}a\,.
\end{equation}
This shows again that $U(\theta),\, \theta \in [0,4\pi)\,,$ is not a true representation of the group $Sp(2,\mathbb{R})$,
but that it is one of its 2-fold covering $Mp(2,\mathbb{R})$ for which $U(\theta = 8\pi)= {\bf 1}$\,.

The reflections \eqref{eq:1221} may also be implemented by the simplified operator $\Pi$\,:
\begin{equation}
  \label{eq:1255}
  \Pi\tl{Q}\Pi^{-1}= - \tl{Q}\,,~~\Pi\tl{P}\Pi^{-1}= - \tl{P}\,,~~\Pi = e^{i\pi N_{osc}}\,,
\end{equation}
where the phase has been choosen such that
\begin{equation}
  \label{eq:1343}
  \Pi|n_{osc}\rangle = (-1)^{n_{osc}}|n_{osc} \rangle\,.
\end{equation}

Contrary to the relations \eqref{eq:1213} we have on the other hand 
\begin{equation}
  \label{eq:1225}
 U(\theta)\,A_{(k)}\,U(-\theta)= e^{i\theta}\,A_{(k)}\,,~~ U(\theta)\,A^{\dagger}_{(k)}\,U(-\theta)=
 e^{-i\theta}\,A^{\dagger}_{(k)}
\,,~~U(\theta)= e^{-i\,\tl{K}_0\,\theta}\,, 
\end{equation}
where $\tl{K}_0$ is now given by $A_{(k)}^{\dagger}A_{(k)} +k{\bf 1}$. \\
Thus, the operators \eqref{eq:1131} transform according to the group $Sp(2,\mathbb{R})/\mathbb{Z}_2 \cong
SO^{\uparrow}(1,2)$. This reflects the fact that the operators $\tl{K}_0,\,\tl{K}_+$ and $\tl{K}_-$ transform 
according to the
 adjoint representation
of $Sp(2,\mathbb{R})$ \cite{ka5}.

The transformation properties of the operators \eqref{eq:1131} under the subgroups generated by the operators $\tl{K}_1$
and $\tl{K}_2$ are  more complicated than those of Eqs.\ \eqref{eq:992}. For the latter we have, e.g.
\begin{equation}
  \label{eq:1243}
  e^{-i\tau \tl{K}_2}\tl{Q}e^{i\tau \tl{K}_2} = e^{\tau /2} \tl{Q}\,,~~ e^{-i\tau \tl{K}_2}\tl{P}e^{i\tau \tl{K}_2} = 
e^{-\tau /2} \tl{P}\,,
\end{equation}
where $\tl{K}_2$ is given by Eq.\ \eqref{eq:993}. The transformation \eqref{eq:1243} is the usual ``squeezing'' 
transformation
of quantum optics \cite{ka6}. The corresponding transformation properties of the $\tl{Q}_{(k)}$ and $\tl{P}_{(k)}$ are
more complicated (see sec.\ 4.4 below).

It is evident that by replacing the operators $a$ and $a^{\dagger}$ in Eqs.\ \eqref{eq:993} and \eqref{eq:995}
by the operators $A_{(k)}$ and $A^{\dagger}_{(k)}$ of Eqs.\ \eqref{eq:999} and \eqref{eq:1014} one may repeat the whole
procedure indicated above, thereby splitting the subspaces $\mathcal{H}_+$ and $\mathcal{H}_-$ again into two subspaces
and so on.

What is important for us at the present state of the discussion is that the quantized version of the
``$(\tl{q},\tl{p})$-model'' of the HO carries two different irreducible unitary representations of a 2-fold
covering of the symplectic  group $Sp(2,\mathbb{R})$.

There is much more to come with the quantized version of the $(\vp,\tl{I})$-model of the HO:

\section{Action of the proper orthochronous homogeneous Lorentz group in 1+2 dimensions on the phase
space ${\mathcal{S}_{\vp,\tl{I}}}$}
\subsection{The basic canonical ``observables'' on $\mathcal{S}_{\vp,\tl{I}}$}
If we insert the relations
\begin{equation}
  \label{eq:979}
  \tl{q} = \sqrt{2\tl{I}}\cos\vp\,,~\tl{p} =-\sqrt{2\tl{I}} \sin\vp\,,
\end{equation}
into the expressions \eqref{eq:973} -- \eqref{eq:975}
we get another set of functions $\check{h}_j(\vp,\tl{I})\,,\,\,j=0,1,2,$ which again
obey the Lie algebra $\mathfrak{sp}(2,\mathbb{R})=\mathfrak{so}(1,2)$ with respect to the Poisson
brackets \eqref{eq:833}:
\begin{eqnarray} \label{eq:1347}\check{h}_0(\vp,\tl{I}) &=& \frac{1}{2}\tl{I}\,,
 \\
\label{eq:1349} \check{h}_1(\vp,\tl{I})&=& -\frac{1}{2}\tl{I}\cos(2\vp)\,,
 \\ \label{eq:1348}
\check{h}_2(\vp,\tl{I}) &=& \frac{1}{2}\tl{I} \sin(2\vp)\,, 
\end{eqnarray}
with
\begin{equation}
  \label{eq:1359}
 \{\check{h}_0,\check{h}_1 \}_{\vp,\tl{I}} = -\check{h}_2\,,~~
 \{\check{h}_0,\check{h}_2 \}_{\vp,\tl{I}} = \check{h}_1\,,~~
 \{\check{h}_1,\check{h}_2 \}_{\vp,\tl{I}} = \check{h}_0\,.  
\end{equation}
This is not yet quite the form \eqref{eq:831} we would like to have. But implementing the scaling \eqref{eq:949}
 with $\beta=2$ yields
the functions \eqref{eq:831}, except for the signs of $h_1$ and $h_2$ which may be reversed without affecting their
 properties and the Lie algebra structure \eqref{eq:832}. \\
Thus, we obtain on the phase space
\begin{equation}
  \label{eq:1404}
  \mathcal{S}_{\vp,\tl{I}} = \{\sigma =(\vp,\tl{I}); \vp \in \mathbb{R} \bmod {2\pi}, \tl{I} > 0\}
\end{equation}
 the basic dimensionless functions
\begin{equation}
  \label{eq:1372}
  \tl{h}_0(\vp,\tl{I})=\tl{I}>0\,,~~\tl{h}_1(\vp,\tl{I})=\tl{I}\cos \vp\,,~~\tl{h}_2(\vp,\tl{I})=-\tl{I} \sin \vp\,,
\end{equation}
 which obey the Lie algebra
\begin{equation}
  \label{eq:1373}
  \{\tl{h}_0,\tl{h}_1 \}_{\vp,\tl{I}} = -\tl{h}_2\,,~~
 \{\tl{h}_0,\tl{h}_2 \}_{\vp,\tl{I}} = \tl{h}_1\,,~~
 \{\tl{h}_1,\tl{h}_2 \}_{\vp,\tl{I}} = \tl{h}_0\,. 
\end{equation}
The two obvious main reasons to pass from the functions \eqref{eq:1347} -- \eqref{eq:1348} to the functions \eqref{eq:1372}
are the following ones:

First, one would like  $\tilde{h}_0$ to be equal to the Hamiltonian $\tl{H}=\tl{I}$ and, secondly, the basic periodic
functions on $S^1$ are $\cos \vp$ and $\sin \vp$ from which all the higher ones, $\cos n \vp,\, \sin n \vp,\, n=2,3,
\ldots,$ can be constructed. The functions $\cos 2 \vp$ and $\sin 2\vp$ cannot serve that purpose! For  related
 discussions of this point see Ref.\ \cite{bo2}.

A given  triple $(\tl{h}_0,\tl{h}_1,\tl{h}_2)$ with the property
\begin{equation}
  \label{eq:980}
  \tl{h}_0^2-\tl{h}_1^2-\tl{h}_2^2 = 0\,,~~\tl{h}_0 > 0\,,
\end{equation}
determines a point $\sigma \in \mathcal{S}_{\vp,\tl{I}}$ uniquely.
Eq.\ \eqref{eq:980} shows that the phase space $\mathcal{S}_{\vp,\tl{I}}$ is diffeomorphic to a (light) cone with the
tip deleted, i.e.\ it is topologically equivalent to $S^1 \times \mathbb{R}^+$. Thus, $\mathcal{S}_{\vp,\tl{I}}$ has the
same topological structure as $\mathcal{S}_{\tl{q},\tl{p};\,0}$ from above! It is, therefore, not surprising that
 the canonical
group $Sp(2,\mathbb{R})$ is intimately related to the coresponding one of  $\mathcal{S}_{\vp,\tl{I}}\,$, namely the ``proper
orthochronous homogeneous Lorentz'' group $SO^{\uparrow}(1,2) \cong Sp(2,\mathbb{R})/\mathbb{Z}_2$ which has the 
symplectic group
$Sp(2,\mathbb{R})$ as a double covering. $SO^{\uparrow}(1,2)$ is that connected subgroup of the four ``pieces'' of the
group $O(1,2)$ which contains the unit element and is time-direction preserving \cite{ka17}. More on this in subsec.\ 
4.5 below.

The transformations of $SO^{\uparrow}(1,2)$ on  $\mathcal{S}_{\vp,\tl{I}}$ are conveniently implemented by passing to the
group $G_0 \equiv SU(1,1)$ which is isomorphic to the group $Sp(2, \mathbb{R})$:
The elements $g_0 \in G_0$ 
  are given by
 \begin{equation}
   \label{eq:981}
 g_0= \left( \begin{array}{ll} \alpha & \beta \\
 \beta^* & \alpha^*
\end{array} \right)\,,~~ \det{g_0} =|\alpha|^2-|\beta|^2=1\,.  
 \end{equation} They
 act on a 2-dimensional complex vector
space $\mathbb{C}^2$ as 
\begin{equation}
  \label{eq:1757}
  g_0 \cdot \left(
\begin{array}{c} z_1 \\ z_2 \end{array} \right)= \left(
\begin{array}{c} z_1^{\prime} \\ z_2^{\prime} \end{array}
\right)\,,~~\text{with}~|z_1^{\prime}|^2 -|z_2^{\prime}|^2 = |z_1|^2-|z_2|^2~.
\end{equation}
The isomorphism between the two groups $G_0$ and $G_1$ can be realized by the unitary matrix
\begin{equation}
 \label{eq:982}
 C_0 =
\frac{1}{\sqrt{2}}\left( \begin{array}{cc} 1 & -i \\ -i & 1
\end{array} \right)\,,~\mbox{det}C_0=1\,,~C_0^{-1}=
\frac{1}{\sqrt{2}}\left( \begin{array}{cc} 1 & i \\ i & 1
\end{array} \right)=C_0^{\dagger}~, 
\end{equation}
  which yields
\begin{equation}
  \label{eq:1421}
  C_0\cdot g_1\cdot C_0^{-1}=g_0~. 
\end{equation}

 The hermitian
matrices 
\begin{equation}
  \label{eq:1422}
 \underline{\sigma}  =
    \left( \begin{array}{cc} \tl{h}_0 = \tl{I} & \tl{h}_1+i\,\tl{h}_2=\tl{I}e^{-i\vp} \\ \tl{h}_1-i\,\tl{h}_2=
\tl{I}e^{i\vp} & \tl{h}_0=\tl{I}
\end{array} \right)\,,~~ \mbox{det}\underline{\sigma} = \tl{h}_0^2- \tl{h}_1^2-\tl{h}_2^2=0\,,
\end{equation}
are in 1-1 correspondence to the points $\sigma \in \mathcal{S}_{\vp,\tl{I}}$. 
The transformations $ \underline{\sigma}\rightarrow  \underline{\sigma}^{\prime}$ under $SO^{\uparrow}(1,2)$ are
 implemented by
\begin{equation}\label{eq:1423}
   \underline{\sigma} \rightarrow  \underline{\sigma}^{\prime}= g_0\cdot  \underline{\sigma}\cdot g_0^{\dagger}\,,~~ \det
 \underline{\sigma}^{\prime} =
\det  \underline{\sigma}\,, 
\end{equation}
  where $g_0^{\dagger}$ denotes the hermitian conjugate of
the matrix $g_0$.

The last equality in Eq.\ \eqref{eq:1423} follows from the property $\mbox{det}g_0 = \mbox{det}g_0^{\dagger}= 1$. Because
 $ \mbox{det}\underline{
\sigma} = \tl{h}_0^2- \tl{h}_1^2-\tl{h}_2^2\,,$ the transformations \eqref{eq:1423} are indeed Lorentz transformations!

 One sees
immediately that $g_0$ and $-g_0$ lead to the same transformations
of  the 3-vectors $(\tl{h}_0,\tl{h}_1,\tl{h}_2)$ and therefore of $\tl{I}$ and $\varphi$. Thus, the group $SU(1,1)$
 acts on  the space $\mathcal{S}_{\vp,\tl{I}}$ 
only  almost effectively with the kernel $Z_2$ representing the
center of the twofold covering groups $SU(1,1)$ or $Sp(2, \mathbb{R})$ of $SO^{\uparrow}(1,2)$. It
is well-known that the latter group acts effectively and
transitively on the forward light cone\cite{bo3} and thus on $\mathcal{S}_{\vp,\tl{I}}$. 

 Applying a general  $g_0$ to the
matrix \eqref{eq:1422}
  yields the mapping:, 
 \begin{eqnarray}
\sigma = (\varphi,\tl{I})
&\rightarrow& \sigma^{\prime} = (\varphi^{\prime},\tl{I}^{\prime})\,: \nonumber \\
 e^{\ds i\varphi^{\prime}}&=& \frac{\alpha^*\,e^{\ds i\varphi} 
 + \beta^*}{\alpha+e^{\ds i\varphi}\,\beta}~. \label{eq:1429}\\
 \tl{I}^{\prime}&=& |\alpha +e^{\ds i\varphi}\, \beta|^2 \, \tl{I}\,,\label{eq:1426}  
\end{eqnarray}
As 
\begin{equation}
  \label{eq:1428}
 \frac{\partial \varphi^{\prime}}{\partial \varphi} = |\alpha +
e^{\ds i \varphi} \beta |^{-2}  ~, 
\end{equation}
 we have the equality
 \begin{equation}
   \label{eq:1431}
  d\varphi^{\prime}\wedge d\tl{I}^{\prime} = d\varphi \wedge d\tl{I}~, 
 \end{equation}
  that
is, the transformations \eqref{eq:1429} and \eqref{eq:1426} are symplectic.

It is, however, more instructive to look at the actions of 1-parameter subgroups of $SU(1,1)$:
The unitary transformation \eqref{eq:982} maps the  subgroups \eqref{eq:965}-\eqref{eq:966} of $G_1$ 
onto the following subgroups of $G_0$:
\begin{eqnarray} R_0: && r_0= \left(
\begin{array}{cc}  e^{\ds i\theta/2} & 0\\  0 &
e^{\ds -i\theta/2} \end{array} \right)\,,~ \theta \in
(-2\pi,+2\pi]\,, \label{eq:1437} \\ A_0: && a_0 = \left(
\begin{array}{cc} \cosh(\tau/2) & i\sinh(\tau /2) \\ -i\sinh(\tau/2) &
\cosh(\tau /2) \end{array} \right)\,,~\tau \in \mathbb{R}\,,\label{eq:1438} \\
B_0: && b_0 = \left(
\begin{array}{cc} \cosh(s/2) & \sinh(s/2) \\ \sinh(s/2) &
\cosh(s/2)
 \end{array} \right)\,, ~s \in \mathbb{R}\,, \label{eq:1436} \\ 
 N_0: && n_0 = \left(
\begin{array}{cc} 1+i\xi/2 & \xi/2 \\ \xi/2 &
1-i\xi/2 \end{array} \right)\,,~ \xi \in \mathbb{R}~~. \label{eq:1453}
\end{eqnarray} 
(I here, too, list four - not independent - subgroups of $G_0$ because we shall need $N_0$ for the representation of 
$\mathcal{S}_{\vp,\tl{I}}$ as a homogeneous space below.)
Their actions \eqref{eq:1423} on the 3-vector $(\tl{h}_0,\tl{h}_1,\tl{h}_2)$ are given by
\begin{eqnarray}
 \label{eq:1457}
  R_0:~~~  \tl{h}_0 &\to& \tl{h}_0^{\prime} = \tl{h}_0\,, \\
\tl{h}_1 &\to& \tl{h}_1^{\prime} = \cos\theta\,\tl{h}_1 - \sin\theta\,\tl{h}_2\,,\nonumber \\
\tl{h}_2 &\to& \tl{h}_2^{\prime} = \sin\theta\,\tl{h}_1 + \cos\theta\,\tl{h}_2\,,\nonumber \\
A_0:~~~ \tl{h}_0 &\to & \tl{h}_0^{\prime} = \cosh \tau\,\tl{h}_0 + \sinh \tau\,\tl{h}_2\,,\label{eq:1459} \\
\tl{h}_1 &\to& \tl{h}_1^{\prime} =\tl{h}_1\,, \nonumber \\
 \tl{h}_2 &\to & \tl{h}_2^{\prime} = \sinh \tau \,\tl{h}_0 + \cosh \tau\,\tl{h}_2\,, \nonumber \\
&& \nonumber \\
B_0:~~~  \tl{h}_0 &\to & \tl{h}_0^{\prime} = \cosh s\,\tl{h}_0 + \sinh s\,\tl{h}_1\,,\label{eq:1489} \\
 \tl{h}_1 &\to & \tl{h}_1^{\prime} = \sinh s\,\tl{h}_0 + \cosh s\,\tl{h}_1\,, \nonumber \\
\tl{h}_2 &\to& \tl{h}_2^{\prime}=\tl{h}_2\,, \nonumber \\
&& \nonumber \\
N_0:~~~ \tl{h}_0 &\to & \tl{h}_0^{\prime} = (1+\xi^2/2)\,\tl{h}_0 + \xi\, \tl{h}_1 -(\xi^2/2)\,\tl{h}_2\,,\label{eq:1492} \\
\tl{h}_1 &\to& \tl{h}_1^{\prime} = \xi\,\tl{h}_0 +\tl{h}_1 -\xi\,\tl{h}_2\,, \nonumber \\
 \tl{h}_2 &\to & \tl{h}_2^{\prime} = (\xi^2/2)\,\tl{h}_0 + \xi\,\tl{h}_1 +(1-\xi^2/2)\,\tl{h}_2\,. \nonumber
\end{eqnarray}
So we have  rotations in the $\tl{h}_1-\tl{h}_2$ plane and two Lorentz ``boosts'',
one in the $\tl{h}_0-\tl{h}_2$ plane and the other in the $\tl{h}_0-\tl{h}_1$ plane!
All transformations leave the form $ \tl{h}_0^2-\tl{h}_1^2-\tl{h}_2^2 $ invariant.

For the variables $\vp$ and $\tl{I}$ these transformations mean
\begin{eqnarray} R_0:~&& \tl{I}^{\prime}=\tl{I}\,, \label{eq:1495} \\
&&  e^{ i\varphi^{\,\prime}}= e^{ i(\varphi-
\theta)}\,, \nonumber \\
   \label{eq:1496}A_0:~ && \tl{I}^{\prime}= \rho_a(\tau,\varphi)\,\tl{I}\,,~~
\rho_a(\tau,\varphi)=\cosh \tau -\sinh \tau \,\sin\varphi\,,  \\
&&\cos \varphi^{\prime}=
\cos\varphi/\rho_a(\tau,\varphi)\,,\nonumber \\ \nonumber
&&\sin \varphi^{\prime}=(\cosh
\tau\sin\varphi - \sinh \tau) /\rho_a(\tau,\varphi)\,, \\
B_0:~&&\tl{I}^{\prime}=\rho_b(s,\vp)
\,\tl{I}\,,~ \rho_b(s,\vp)
= \cosh \tau + \sinh \tau\,\cos \vp\,,\label{eq:1499} \\ \nonumber&&
 \cos\vp^{\prime} =
 (\cosh s \cos \vp +\sinh s)/\rho_b(s,\vp)\,, \\
\nonumber&& \sin\vp^{\prime} = \sin\vp/\rho_b(s,\vp)\,, \\
N_0:~ && \tl{I}^{\prime}=
\rho_n(\xi,\varphi)\,\tl{I}~,~\rho_n(\xi,\varphi)=1+\xi \cos\varphi +
\xi^2(1+\sin\varphi)/2\,,~~~~ \label{eq:1586} \\ \nonumber
&&\cos\varphi^{\prime}=
[\cos\varphi+\xi(1+\sin\varphi)]/\rho_n(\xi,\varphi)\,,  \\
&&\sin\varphi^{\prime}= [\sin\varphi-\xi \cos\varphi -
\xi^2(1+\sin\varphi)/2]/\rho_n(\xi,\varphi)\,. \nonumber
\end{eqnarray}

As the center $\mathbb{Z}_2$ of $SU(1,1)$ acts as the identity in the transformations \eqref{eq:1457} -- \eqref{eq:1586}
the above transformation subgroups are actually those of $SO^{\uparrow}(1,2) = SU(1,1)/\mathbb{Z}_2$ which we shall
denote by $R=R_0/\mathbb{Z}_2$, $A=A_0/\mathbb{Z}_2$ etc.\ in the following.

 Transitivity of the $SO^{\uparrow}(1,2)$ group action on $\mathcal{S}_{\vp,\tl{I}}$
can be seen as follows: 
 Any point $\sigma_1=(\varphi_1,\tl{I}_1)$ may be transformed into any
other point $\sigma_2=(\varphi_2, \tl{I}_2)$: first
transform $(\varphi_1,\tl{I}_1)$ into $(0,\tl{I}_1)$ by
$r_0(\theta=\varphi_1)$, then map this point into $(\varphi_0
= -\arctan(\sinh \tau_0\,),\tl{I}_2)$ by $a_0(\tau_0;\,
\cosh \tau_0=\tl{I}_2/\tl{I}_1)$ and finally transform $(\varphi_0,\tl{I}_2)$
by $r_0(\theta = \varphi_0-\vp_2)$ into
$\sigma_2=(\varphi_2,\tl{I}_2)$. 

 For infinitesimal values of the
  parameters $\theta,\, \tau \,$ and $ s $ the transformations
 \eqref{eq:1495} -- \eqref{eq:1586}
 take the
  form
 \begin{eqnarray}R~:&& \delta \varphi =
 -\theta,~|\theta|\ll 1,~~~ \delta \tl{I}
 =0\,, \label{eq:1633}\\ A~:&& \delta \varphi =-(\cos\varphi)\,\tau,
~~\delta \tl{I} =
 -\tl{I}\,(\sin\varphi)\,\tau,~~
 |\tau|\ll 1\,,\label{eq:1652} \\ B~:&& \delta\varphi=-(\sin\varphi)
\,s,~~\delta \tl{I} =
 \tl{I}\,(\cos\varphi)\,s,~~|s|\ll 1~~. \label{eq:1695} \end{eqnarray} According to Eq.\
\eqref{eq:963} they induce on $\mathcal{S}_{\vp,\tl{I}}$
 the vector fields \begin{eqnarray} \label{eq:1696}\tilde{A}_{R} &=&
 \partial_{\varphi}\,,
\\ \tilde{A}_{A} &=&
 \cos\varphi\,\partial_{\varphi}+\tl{I}\,\sin\varphi\,\partial_{\tl{I}}\,,
\label{eq:1697} \\
 \tilde{A}_{B}& =& \sin\varphi\,
 \partial_{\varphi}-\tl{I}\,\cos\varphi\,\partial_{\tl{I}}\,. \label{eq:1698}\end{eqnarray}
  It is easy to check that the Lie algebra of these vector fields is isomorphic to the
 Lie algebra of $SO^{\uparrow}(1,2)$,
 and all its covering groups, of course. 

The vector fields \eqref{eq:1696} -- \eqref{eq:1698} are (global) Hamiltonian ones in the sense of Eq.\
\eqref{eq:972}. The corresponding Hamiltonian functions $f(\vp,\tl{I})$ are:

\begin{eqnarray} \label{eq:1699}\tilde{A}_{R}: ~f_{R}(\vp,\tl{I})
 &=& -\tl{I}\,, \\
\tilde{A}_{A}: ~f_{A}(\vp,\tl{I}) &=&- \tl{I}\,\cos\vp
\,,\label{eq:1700}\\
\tilde{A}_{B}:~f_{B}(\vp,\tl{I})& =&- \tl{I}\,\sin\vp\,.
 \label{eq:1701}
\end{eqnarray}
 The Hamiltonian functions $f_{R},f_{A}$ and $f_{B}$ obey the Lie algebra
$\mathfrak{so}(1,2)$ with respect to the Poisson brackets \eqref{eq:833}:
\begin{equation}
  \label{eq:1702}
 \{f_{R},f_{A}\} =-f_{B}\,,~~~\{f_{R},f_{B}\} = f_{A}\,,~~~\{f_{A},f_{B}\}=f_{R}\,. 
\end{equation}
Reversing the   minus signs on the right-hand side of Eqs.\ \eqref{eq:1699} and \eqref{eq:1700} we finally arrive again
at our three  basic classical observables  introduced before:
\begin{equation}
 \label{eq:1703}
 \tl{h}_0(\vp,\tl{I}) \equiv -f_{R} =\tl{I}\,,~~~ \tl{h}_1(\vp,\tl{I}) \equiv -f_{A}=
\tl{I}\,\cos\varphi\,,~~~\tl{h}_2(\vp,\tl{I}) \equiv f_{B} = - \tl{I}\,\sin\varphi\,.  
\end{equation}
 Thus, the canonical
group $SO^{\uparrow}(1,2)$ of the symplectic space $\mathcal{S}_{\vp,\tl{I}}$ 
determines
the basic ``observables'' \eqref{eq:1372} of that classical space. 
\subsection{$\mathcal{S}_{\vp,\tl{I}}$ as a homogeneous space}
The transformation formulae \eqref{eq:1586} show that
 the subgroup $N_0$
leaves the half-line $\varphi = -\pi/2,\, \tl{I}>0,$ pointwise invariant, that is,
$N_0$ is the stability group of those points. This implies
that the symplectic space  $\mathcal{S}_{\vp,\tl{I}}$ is diffeomorphic to the coset space
$SO^{\uparrow}(1,2)/N_0\,$:
\begin{equation}
  \label{eq:1704}
  \mathcal{S}_{\vp,\tl{I}} \cong 
SO^{\uparrow}(1,2)/N_0\,.
\end{equation}
 Notice that the subgroups $N_0$ and
$A_0$  do not contain the second center element
$-e$ of $SU(1,1)$. The center $\mathbb{Z}_2$ is a subgroup of
$R_0$. 

In the language of the $\tl{h}_j$ the transformations \eqref{eq:1492} leave the points
$(\tl{I},\tl{h}_1=0,\tl{h}_2=\tl{I})$ invariant.

\subsection{On the relationship between the phase spaces $\mathcal{S}_{\tl{q},\tl{p};\,0}$ and 
$\mathcal{S}_{\vp,\tl{I}}$}
The two phase spaces  $\mathcal{S}_{\tl{q},\tl{p};\,0}$ and  $\mathcal{S}_{\vp,\tl{I}}$ have the same topological structure
$S^1 \times \mathbb{R}^+$, but their correspondence is nevertheless not one-to-one. The term ``topological'' is somewhat 
imprecise here: the positive real numbers $\mathbb{R}^+$ can be mapped in a one-to-one fashion onto the full real line
$\mathbb{R}$ by $\tl{I}= e^a,\,a = \ln \tl{I},\, \tl{I} \in \mathbb{R}^+,\,a \in \mathbb{R},\,$. This mapping is, however,
not symplectic because $d\vp \wedge d\tl{I} = e^ad\vp \wedge da\,$. But as we are interested in preserving the symplectic
structures, we consider only (usually local) ``symplectomorphisms'' \cite{sour}.

In order to see the essential difference between the spaces $\mathcal{S}_{\tl{q},\tl{p};\,0}$ and  $\mathcal{S}_{\vp,\tl{I}}$
let us look at the orbits of the transformation group $R_1$ (cf.\ Eq.\ \eqref{eq:965}) on the former (cf.\ Eq.\ 
\eqref{eq:957})
 and those of $R_0$ (cf.\ Eq.\ \eqref{eq:1437}) on the latter (cf.\ Eqs. \eqref{eq:1457}): 

If we start with $\theta =0$ and increase
$\theta$ to $\pi$ then - according to Eq.\ \eqref{eq:965} - the positive $\tl{q}$-axis is rotated by $90^{\circ}$ onto
 the positive $\tl{p}$-axis and the latter is rotated onto the negative $\tl{p}$-axis. On the other hand - according to
Eq.\ \eqref{eq:1457} - the $\tl{h}_1$- and $\tl{h}_2$-axis are both rotated by $180^{\circ}$, i.e.\ they change sign.
For $\theta = 2\pi$ the transformation \eqref{eq:1457} becomes the identity, whereas now the transformation \eqref{eq:965}
changes the sign of $x=(\tl{q},\tl{p})^T$ (``T'' here means ``transpose''): $x\to -x$. Finally, for $\theta =4\pi$ both
 groups act as the identity.

 All this is, of course, a consequence of the
fact that the effective transformation group on $\mathcal{S}_{\tl{q},\tl{p};\,0}$ is $Sp(2,\mathbb{R})\,$, whereas the
 corresponding transformation group on $\mathcal{S}_{\vp,\tl{I}}$ is $SO^{\uparrow}(1,2) = Sp(2,\mathbb{R})/\mathbb{Z}_2$!
The situation is completely similar to the well-known transformations of the group $SU(2)$ on a 2-dimensional
 (complex spinor) vector space which induces a corresponding transformation   of the group $SO(3)
 =SU(2)/\mathbb{Z}_2$ on a
3-dimensional vector space. Here, too, a given element of $SO(3)$ corresponds to {\em two} elements $\pm u \in SU(2)$\,!
In our case the $x \in \mathcal{S}_{\tl{q},\tl{p};\,0}$ are the ``spinors'' and the $\sigma \in  \mathcal{S}_{\vp,\tl{I}}$
are the ``vectors''.

The remarks show in which way the two spaces $\mathcal{S}_{\tl{q},\tl{p};\,0}$ and $\mathcal{S}_{\vp,\tl{I}}$ differ 
{\em globally} despite the {\em local} equality $d\vp \wedge d\tl{I} = d\tl{q} \wedge d\tl{p}$ which is obviously
invariant under the center $\mathbb{Z}_2$ (it sends $x$ to $-x$ and $\sigma$ to $\sigma$).

One may characterize the situation also in the following way \cite{ka7}: If we identify the points $x$ and $-x$ on
$\mathcal{S}_{\tl{q},\tl{p};\,0}$ then the group $Sp(2,\mathbb{R})$ acts on this quotient space in the same way as the group
$SO^{\uparrow}(1,2)$ on $\mathcal{S}_{\vp,\tl{I}}$ and we have the correspondence
\begin{equation}
  \label{eq:1456}
  \mathcal{S}_{\vp,\tl{I}} \cong \mathcal{S}_{\tl{q},\tl{p};0}/\mathbb{Z}_2\,,
\end{equation}
which follows also from comparing the homogeneous spaces \eqref{eq:1309} and \eqref{eq:1704}.

 A quotient space like \eqref{eq:1456} is called an ``orbifold''.
An orbifold may be generated
from a manifold $\mathtt{M}$ by identifying points which are connected by
a finite discontinuous group $D_n$ of $n$ elements so that the orbifold
is given by the quotient space $\mathtt{M}/D_n$.
 An orbifold generally has
additional singularities as compared to the mani\-fold from which it is
constructed, as we shall see now: 

In our case the orbifold $ \mathcal{S}_{\tl{q},\tl{p};\,0}/\mathbb{Z}_2$ is a cone: Take the lower half
of the $(\tl{q},\tl{p})$-plane and rotate it around the $\tl{q}$-axis till it coincides
with the upper half of the plane such that the negative $\tl{p}$-axis lies
on the positive one. Then rotate the left half of the upper half plane
around the positive $\tl{p}$-axis till the negative $\tl{q}$-axis coincides with
the positive one. Finally glue the two $\tl{q}$-half-axis together. The
resulting space is a cone with its ``tip'' (vertex) at $x=0$.
The tip  constitutes a singularity to be deleted. It is a fixed point of the action of $\mathbb{Z}_2$.
 We thus arrive at the cone structure of the symplectic space \eqref{eq:1456} by a different route. 

\subsection{Relationships between the coordinates $\tl{q},\, \tl{p}$ and $\tl{h}_0,\, \tl{h}_1, 
\, \tl{h}_2 $}
Further below we shall encounter several important relations between the quantum operator versions $\tl{Q},\,\tl{P},\,
\tl{K}_0,\,\tl{K}_1\,$ and $\tl{K}_2$ of the corresponding classical basic quantities  $\tl{q},\,\tl{p},\,\tl{h}_0,\,
\tl{h}_1\,$
 and $\tl{h}_2$. So it
 is useful to list a number of relations between the latter:
 \begin{eqnarray}
   \label{eq:1458}
   \tl{q}(\vp,\tl{I})&=& \sqrt{2\tl{I}}\cos \vp = \sqrt{\frac{2}{\tl{h}_0}}\,\tl{h}_1\,;~~~ 
 \tl{h}_0 = \tl{I},~~\tl{h}_1 = \tl{I}\cos \vp\,,  \\
\tl{p}(\vp,\tl{I}) &=& -\sqrt{2\tl{I}}\sin \vp = \sqrt{\frac{2}{\tl{h}_0}}\,\tl{h}_2\,;~~~\label{eq:1460} 
 \tl{h}_2 = -\tl{I}\sin \vp\,,  \\
\alpha &=& \frac{1}{\sqrt{2}}(\tl{q}+i\tl{p}) = \sqrt{\tl{I}}\,e^{\ds -i\vp}= \tl{h}_+/\sqrt{\tl{h}_0}\,,\label{eq:1461} \\
&& \tl{h}_+ = \tl{h}_1 +i\tl{h}_2 = \tl{I}\,e^{\ds -i\vp}\,,~ 
\tl{h}_- = \tl{h}_1 -i\tl{h}_2 = \tl{I}\,e^{\ds i\vp}\,. \label{eq:1493}
 \end{eqnarray}

Also of interest are a number of Poisson brackets:
\begin{eqnarray}
  \label{eq:1494}
  \{\tl{h}_0,\tl{q}\}_{\vp,\tl{I}}&=& -\tl{p}\,, \\
 \{\tl{h}_0,\tl{p}\}_{\vp,\tl{I}}&=& \tl{q}\,, \label{eq:1497}\\
 \{\tl{h}_0,\alpha \}_{\vp,\tl{I}}&=& i\,\alpha \,.\label{eq:1498}
\end{eqnarray}
These are  just the canonical eqs.\ of motion for the HO.

 More complicated are the following Poisson brackets
\begin{eqnarray}
  \label{eq:1500}
  \{\tl{h}_1,\tl{q}\}_{\vp,\tl{I}}&=& -\frac{1}{2}\,\sin\vp\, \tl{q}-\cos\vp\,\tl{p}\,, \\
  \{\tl{h}_1,\tl{p}\}_{\vp,\tl{I}}&=& -\frac{1}{2}\,\sin\vp\, \tl{p}+\cos\vp\,\tl{q}\,, \label{eq:1576}\\
 \{\tl{h}_2,\tl{q}\}_{\vp,\tl{I}}&=& -\frac{1}{2}\,\cos\vp\, \tl{q}+\sin\vp\,\tl{p}\,,\label{eq:1577} \\
 \{\tl{h}_2,\tl{p}\}_{\vp,\tl{I}}&=& -\frac{1}{2}\,\cos\vp\, \tl{p}-\sin\vp\,\tl{q}\,, \label{eq:1593}\\
\end{eqnarray}
where the right-hand sides may be expressed in different ways by using the quantities defined in Eqs.\ 
\eqref{eq:1458} -- \eqref{eq:1493}. Examples are
\begin{eqnarray}
  \label{eq:1607}
  \{\tl{h}_+,\alpha \}_{\vp,\tl{I}}&=& \frac{1}{\sqrt{\tl{I}}}\,(-\frac{1}{2}\,\alpha^* +i\alpha)\alpha =
-\frac{1}{2}\sqrt{\tl{I}}+ \frac{i}{\tl{I}^{3/2}}(\tl{h}_+)^2\,, \\
 \{\tl{h}_-,\alpha \}_{\vp,\tl{I}}&=& \frac{1}{\sqrt{\tl{I}}}\,(-\frac{1}{2}\,\alpha +i\alpha^*)\alpha =
\frac{i}{2}\sqrt{\tl{I}} -\frac{1}{2\tl{I}^{3/2}}(\tl{h}_-)^2\,.\label{eq:1608}
\end{eqnarray}
The brackets  $\{\tl{h}_+,\alpha^* \}_{\vp,\tl{I}}$ and $ \{\tl{h}_-,\alpha^* \}_{\vp,\tl{I}}$ follow from complex
conjugation of the relations \eqref{eq:1608} and \eqref{eq:1607}.

\subsection{Space reflections and time reversal}

The space reflections \eqref{eq:1301} may be implemented on $ \mathcal{S}_{\vp,\tl{I}}$ by
\begin{equation}
  \label{eq:1490}
  \Pi\,: ~~~\vp \to \vp \pm \pi\,,~~~\tl{I} \to \tl{I}\,.
\end{equation}
The reflection $\Pi$ leaves the symplectic form $d\vp \wedge d\tl{I}$ invariant (locally) and implies
\begin{equation}
  \label{eq:1491}
  \Pi\,:~~~\tl{h}_0 \to \tl{h}_0\,,~~~\tl{h}_1 \to -\tl{h}_1\,,~~~\tl{h}_2 \to -\tl{h}_2\,.
\end{equation}
The time reversal \eqref{eq:1302} can be implemented by
\begin{equation}
  \label{eq:1705}
  T\,: \tl{t} \to -\tl{t}\,,~~~\vp \to -\vp\,,~~~\tl{I} \to \tl{I}\,.
\end{equation}
In order to make this transformation into a symplectic one, we also have to change the order of the factors in 
$d\vp\wedge d\tl{I}$ as discussed  after Eq.\ \eqref{eq:1302} above. We now have
\begin{equation}
  \label{eq:1706}
  T\,:~~~\tl{h}_0 \to \tl{h}_0\,,~~~\tl{h}_1 \to \tl{h}_1\,,~~~\tl{h}_2 \to -\tl{h}_2\,.
\end{equation}
Notice that the space reflection properties \eqref{eq:1491} of the $\tl{h}_j$ are different from those of the
$\check{h}_j$ of Eqs.\ \eqref{eq:1347} -- \eqref{eq:1348}. The T-reversal properties are the same.

The relationship of the above $\Pi$- and $T$-transformations to the different ``pieces'' of the homogeneous
Lorentz group $O(1,2)$ is as follows: It follows from
\begin{equation}
  \label{eq:1707}
  \tl{h}_j \to \tl{h}_j^{\prime}= \sum_{k=0}^2\Lambda_j^{\,k}\,\tl{h}_k\,,~ (\tl{h}_0^{\prime})^2-(\tl{h}_1^{\prime})^2-
(\tl{h}_2^{\prime})^2 = (\tl{h}_0)^2-(\tl{h}_1)^2-(\tl{h}_2)^2\,,
\end{equation}
that
\begin{equation}
  \label{eq:1708}
  \det (\Lambda_j^{\,k})= \pm 1\,,~~~\mbox{sgn}\Lambda_0^{\,0} = \pm 1\,.
\end{equation}
The group $SO^{\uparrow}(1,2)$ which contains the identity transformation is characterized by 
$ \det (\Lambda_j^{\,k})= 1, \,\mbox{sgn}\Lambda_0^{\,0} = 1\,.$ The above transformations $\Pi$ and $T$ have
both $\mbox{sgn}\Lambda_0^{\,0} = 1\,$, but $ \det (\Lambda_j^{\,k})= 1\,$ and $ \det (\Lambda_j^{\,k})=- 1\,,$
respectively.
\section{Quantizing  the angle - action variables  phase space $ \mathbf{\mathcal{S}_{\vp,\tl{I}}}$ of the 
harmonic oscillator}
\subsection{Lie algebra of the self-adjoint observables $\tl{K}_j$ and the structure of their irreducible 
representations}
The quantum theory of the HO described on the phase space $\mathcal{S}_{\tl{q},\tl{p}}$ is a settled affair, due to the
Stone-von Neumann uniqueness theorem for the irreducible unitary representations of the BDHJW-group \cite{ka8}!

The situation is different, however, for the quantum theory of the HO described by the phase space $\mathcal{S}_{\vp,
 \tl{I}}$ of its angle and action variables. We have seen that the ``canonical'' group of that phase space is the group
$SO^{\uparrow}(1,2)$ which has an infinite number of covering groups, due to its maximal compact rotation subgroup $SO(2)$.
The group $SO^{\uparrow}(1,2)$  - and its covering groups - has 3 classes of irreducible unitary representations
 \cite{irrep}:
 the ``principal'', the ``supplementary'' or ``complementary'' series and two ``discrete'' series. In the principal and
 supplementary series
the spectra of the generator $\tl{K}_0$ are unbounded from below and above. One of the discrete series has a strictly 
positive
spectrum of $\tl{K}_0$ and the other a strictly negative one. In our case $\tl{K}_0$ corresponds to the positive
 action variable
$\tl{I}$ and, therefore, ought to be a positive definite operator. This leaves only the positive discrete series of the
irreducible unitary representations. These may be - formally - constructed as follows: \\
 As the group $SO^{\uparrow}(1,2)$
is noncompact, its irreducible unitary representations are
infinite-dimensional. Different concrete representation Hilbert spaces will be discussed later in sec.\ 7. 

In an irreducible unitary
 representation of the group  the
classical functions 
$\tl{h}_0,\,\tl{h}_1,\,\tl{h}_2$ with their Lie algebra \eqref{eq:832} correspond to
self-adjoint operators $\tl{K}_0,\tl{K}_1,\tl{K}_2$ which obey the commutation
relations 
\begin{equation}
  \label{eq:1712}
 [\tl{K}_0,\tl{K}_1]=i\tl{K}_2~,~~[\tl{K}_0,\tl{K}_2]=-i\tl{K}_1~,~~[\tl{K}_1,\tl{K}_2]=-i\tl{K}_0~, 
\end{equation}
 or,
with the definitions 
\begin{equation}
  \label{eq:1713}
   \tl{K}_+ =\tl{K}_1+i\tl{K}_2\,,~~\tl{K}_-=\tl{K}_1-i\tl{K}_2\,,
\end{equation} we have
\begin{equation}
  \label{eq:1714}
 [\tl{K}_0,\tl{K}_+]=\tl{K}_+~,~~[\tl{K}_0,\tl{K}_-]=-\tl{K}_-~,~~[\tl{K}_+,\tl{K}_-]=-2\tl{K}_0~. 
\end{equation}
  The
relations \eqref{eq:1712} are invariant under the replacement $\tl{K}_1
\rightarrow -\tl{K}_1, \tl{K}_2 \rightarrow -\tl{K}_2$ and  
 $\tl{K}_1
\rightarrow -\tl{K}_2, \tl{K}_2 \rightarrow \tl{K}_1$.
The relations \eqref{eq:1714} are
invariant under $\tl{K}_+ \rightarrow \mu \tl{K}_+,\, \tl{K}_- \rightarrow
\mu^* \tl{K}_-,\, |\mu|=1\,$, and under the transformations $ \tl{K}_+ \leftrightarrow \tl{K}_-, \tl{K}_0
\rightarrow -\tl{K}_0$.
 In irreducible unitary representations with a scalar product $(f_1,f_2)$ the
operator $\tl{K}_-$ is the adjoint operator of $\tl{K}_+:\,(f_1,\tl{K}_+f_2)=(\tl{K}_-f_1,f_2)$,
 and vice versa.

The  (Casimir) operator 
\begin{equation}  
 \mathfrak{C}=\tl{K}_1^2+\tl{K}_2^2-\tl{K}_0^2 
\end{equation}
commutes with all three $\tl{K}_j$ and therefore is a multiple of the identity operator in an irreducible representation.
  It obeys the  relations 
  \begin{equation}
    \label{eq:1715}
  \tl{K}_+\tl{K}_-=\mathfrak{C}+\tl{K}_0(\tl{K}_0-1)\,,~~\tl{K}_-\tl{K}_+= \mathfrak{C}+\tl{K}_0(\tl{K}_0+1)~.  
  \end{equation}
  Most unitary
representations make use of the fact that $\tl{K}_0$ is the generator
of a compact group and that its eigenfunctions $g_m$ are
normalizable elements of the associated Hilbert space $\mathcal{H}\,$ \cite{muk2}.

The relations \eqref{eq:1714} show that the $K_{\pm}$ act es creation and annihilation operators and they imply
 \begin{eqnarray}
  \label{eq:1716} \tl{K}_0\,g_m&=&m\,g_m\,,~~m \in \mathbb{R}\,,~~(g_m,g_m) = 1\,,
  \\
\tl{K}_0\,\tl{K}_+g_m&=&(m+1)\tl{K}_+g_m\,, \label{eq:1717}\\ 
\tl{K}_0\,\tl{K}_-g_m &=& (m-1)\tl{K}_-g_m\,, \label{eq:1718}
\end{eqnarray} which, combined with \eqref{eq:1715}, lead to
\begin{eqnarray} (g_m,\tl{K}_+\tl{K}_-g_m)&=&(\tl{K}_-g_m,\tl{K}_-g_m)=\mathfrak{c}+m(m-1) \ge
0\,, \label{eq:1719} \\ (g_m, \tl{K}_-\tl{K}_+g_m)&=&\mathfrak{c}+m(m+1)\ge 0 \,,
 ~\mathfrak{c}=(g_m,\mathfrak{C}g_m).
\label{eq:1720}
\end{eqnarray} It follows that
\begin{equation}
  \label{eq:1721}
  (\tl{K}_+g_m,\tl{K}_+g_m)= 2m+(\tl{K}_-g_m,\tl{K}_-g_m)
\ge 0. 
\end{equation}
  As we assume that we have an irreducible
representation the functions $g_m$ are eigenfunctions of
the Casimir operator $\mathfrak{C}\,$: 
\begin{equation}
  \label{eq:929}
 \mathfrak{C}\,g_m=\mathfrak{c}\,g_m\,. 
\end{equation}
 The relations
\eqref{eq:1716} -- \eqref{eq:1721}  show that the eigenvalues $m$ of $\tl{K}_0$ in principle can
be any real number, where, however, different eigenvalues
 differ by an integer.  

As already said above: For the ``principle'' and the ``complementary''
 series the spectrum of $\tl{K}_0$ is  unbounded from below and above \cite{irrep}, but
as $\tl{K}_0$ corresponds to the classical positive definite quantity $\tl{I}$,
these unitary representations are of no interest here. \\
 Here the
positive discrete series $D^{(+)}_k$ of irreducible  unitary
representations are important. These are
 characterized by the property that there exists a lowest eigenvalue
 $m=k$ such that 
 \begin{equation}
   \label{eq:1722}
   \tl{K}_0\,g_k =k\,g_k\,,~~ \tl{K}_-\,g_k=0\,.
 \end{equation}
  Now the relations \eqref{eq:1719} and \eqref{eq:1721} imply
  \begin{equation}
   \label{eq:1723}
    \mathfrak{c}=k(1-k)~,~~~ k>0~,~~m=k+n,~ n=0,1,2,\ldots . 
  \end{equation}
That $k>0$ follows from Eq.\ \eqref{eq:1721} with $m=k,\,\tl{K}_-g_k=0\,$, but $(\tl{K}_+g_k,\tl{K}_+g_k) >0\,$,
 because the scalar
product is positive definite!
  Exploiting  the relations \eqref{eq:1716}-\eqref{eq:1718} yields
  \begin{eqnarray}
 \tl{K}_0g_{k,n}&=&(k+n)\,g_{k,n}\,,~ k > 0\,,~ n=0,1,\ldots,\,,~~(g_{k,n},g_{k,n}) =1\,,\label{eq:1724}\\
 \tl{K}_+g_{k,n}&=&\mu_n\,[(2k+n)(n+1)]^{1/2}\,g_{k,n+1}\,,~~|\mu_n|=1~,
\label{eq:1725} \\
 \tl{K}_-g_{k,n}&=&\frac{1}{\mu_{n-1}}[(2k+n-1)n]^{1/2}\,g_{k,n-1}\,.
\label{eq:1726}\end{eqnarray}
 The phases $\mu_n$ guarantee that $(f_1,\tl{K}_+f_2)=(\tl{K}_-f_1,f_2)$. In
most cases of interest $\mu_n$ is independent of $n$. Then one can absorb it
into the definition of $K_{\pm}$ and forget about the phases $\mu_n$!

Up to now $k$ may be any positive real number. A detailed analysis (see Appendix B) shows \cite{irrep} that $k=1,2,\ldots,$
 for the group
$SO^{\uparrow}(1,2)$ itself, $k=1/2,1,3/2,\ldots,$ for the isomorphic groups $Sp(2,\mathbb{R}) \cong SL(2,\mathbb{R})
 \cong SU(1,1)\,$
and  $k=1/4,1/2,3/4,1,\ldots,$ for the metaplectic group $Mp(2, \mathbb{R})$ we encountered above. 

For the universal covering
group $\tl{G} \equiv SO^{\uparrow}_{[\infty]}(1,2)$ the ``Bargmann index''  $k$ may have any positive value $> 0$.
 Further below we shall see that for an m-fold covering $SO^{\uparrow}_{[m]}(1,2)$ the index $k$ can take the rational values
 \begin{equation}
   \label{eq:2142}
   k =\frac{\mu}{m}\,,~~\mu = 1,2,\ldots\,.
 \end{equation}
Here the natural number $m$ may be arbitrary large, i.e.\ the lowest value  $k= 1/m $ can be made arbitrary small $>0\,$!

As long as I do not specify the concrete Hilbert space used I shall employ Dirac's bracket notation and write
 $g_{k,n} = |k,n \rangle\,.$ 
It follows from Eq.\ \eqref{eq:1725} that
\begin{eqnarray}
  \label{eq:1754}
  |k,n \rangle &=& \frac{1}{\sqrt{(2k)_n\,n!}}(\tl{K}_+)^n|k,0 \rangle\,,\\ 
(2k)_n &\equiv& 2k\,(2k+1)\cdots (2k+n-1) =
\frac{\Gamma (2k +n)}{\Gamma (2k)}\,,\label{eq:2143} \\
&& (2k)_{n=0}=1\,,~~(1)_n = n!\,,~~(-2k)_n = (-1)^n\,n!\,{2k \choose n}\,. \label{eq:9}
\end{eqnarray}

The Casimir operator relation
\begin{equation}
  \label{eq:1814}
  \tl{K}_1^2+ \tl{K}_2^2 = \tl{K}_0^2 +k(1-k)\,{\bf 1}
\end{equation}
modifies the corresponding classical Pythagorean relation
\begin{equation}
  \label{eq:1815}
  \tl{h}_1^2 + \tl{h}_2^2 = \tl{h}_0^2\,,
\end{equation}
unless $k=1$! So for a HO with $k=1/2$ ``Pythagoras'' is ``violated'' by quantum effects!
\subsection{The operators $\tl{Q}$ and $\tl{P}$ as functions of the operators $\tl{K}_j$}
The relations \eqref{eq:920}\,, \eqref{eq:928}\,, \eqref{eq:1458} and \eqref{eq:1460} as well show the dependence of
the canonical coordinates $\tl{q}$ and $\tl{p}$ on the canonical coordinates $\vp$ and $\tl{I}$. It is important that
a corresponding operator relation expresses the position operator $\tl{Q}$ and the momentum operator $\tl{P}$ in terms of
the operators $\tl{K}_j,\,j=0,1,2\,$. That this is indeed possible was already stated in the introduction. 
The relationship can be read off the Eqs.\ \eqref{eq:1724} -- \eqref{eq:1726} as follows:

 If we have annihilation and creation operators $a$ and $a^{\dagger}$
in a (Fock) Hilbert space with a number state basis $|n \rangle$ such that
\begin{equation}
  \label{eq:1727}
  a\,|n \rangle = \sqrt{n}\,|n-1 \rangle\,,~~~a^{\dagger}\,|n \rangle = \sqrt{n+1}\,|n+1 \rangle\,,~~~[a,\,a^{\dagger}] =
{\bf 1}\,,
\end{equation}
we can define
\begin{equation}
  \label{eq:1728}
  \tl{Q} = \frac{1}{\sqrt{2}}\,(a + a^{\dagger})\,,~~~\tl{P} = \frac{i}{\sqrt{2}}\,( a^{\dagger}-a)\,,~~~[\tl{Q}, \tl{P} ]
= i{\bf 1}\,.
\end{equation}

The operators \eqref{eq:1727} have been used to construct non-linear realizations of the generators $\tl{K}_j$ \cite{hol1}:
\begin{equation}
  \label{eq:74}
  \tl{K}_0 = N+k{\bf 1}\,,~~ \tl{K}_+ = a^{\dagger}\sqrt{N+2k{\bf 1}}\,,~~\tl{K}_- = \sqrt{N+2k{\bf{1}}}\,a\,,~~N =
 a^{\dagger}a\,.
\end{equation}

However, as I pointed out in Ref.\ \cite{ka1}, it is much more interesting to invert these relations:
\subsubsection{Operator version of the polar coordinates in the plane}

Now, as $k > 0$ and the operator $\tl{K}_0$ is positive definite in any irreducible unitary representation of the positive
discrete series $D_k^{(+)}$, the operator
\begin{equation}
  \label{eq:1729}
  B_k = (\tl{K}_0+k)^{-1/2}
\end{equation}
is well-defined and self-adjoint. As
\begin{equation}
  \label{eq:1730}
  B_k |k,n \rangle = (2k +n)^{-1/2}\,|k,n \rangle\,,
\end{equation}
then according to the relations \eqref{eq:1724} -- \eqref{eq:1726} (with $\mu_n = 1$) the operators
\begin{equation}
  \label{eq:1731}
  A_{(k)}(\vec{\tl{K}}) = B_k\,\tl{K}_-\,, \;\;\; A_{(k)}^{\dagger}(\vec{\tl{K}}) = \tl{K}_+\,B_k
\end{equation}
have the properties
\begin{eqnarray}
  \label{eq:1732}
  A_{(k)}\,|k,n \rangle = \sqrt{n}\,|k,n-1 \rangle\,,&&~~~A_{(k)}^{\dagger}\,|k,n \rangle = \sqrt{n+1}\,|k,n+1 \rangle\,,\\
~~~A^{\dagger}_{(k)}A_{(k)}|k,n
\rangle = n\,|k,n\rangle\,,&&~~~[A_{(k)},A^{\dagger}_{(k)}] = {\bf 1}\,.\label{eq:1733}
\end{eqnarray}
The actions of the operators \eqref{eq:1731} are independent of the (Bargmann) index $k$ which characterizes the
 irreducible representation of
the group $SO^{\uparrow}(1,2)$ or one of its covering groups. So we may drop their index $(k)\,$.

This $k$-independence is another manifestation of the Stone-von Neumann uniqueness theorem which says that - provided
certain regularity conditions are fulfilled - all the irreducible representations of $\tl{Q}$ and $\tl{P}$ with the
property \eqref{eq:1728} are unitarily equivalent, i.e.\ have the same matrix element whatever Hilbert space is employed! \\
Before drawing consequences let me derive the relation
\begin{equation}
  \label{eq:1734}
  N= A^{\dagger}(\vec{\tl{K}})\,A(\vec{\tl{K}}) = \tl{K}_0 -k{\bf 1}
\end{equation}
in a different way:
If $f(\tl{K}_0)$ is a ``suitable'' function of the operator $\tl{K}_0$, then a repeated application of the relations 
\eqref{eq:1714}
yields
\begin{equation}
  \label{eq:1735}
  \tl{K}_-\,f(\tl{K}_0) = f(\tl{K}_0+{\bf 1})\,\tl{K}_-\,,~~~f(\tl{K}_0)\,\tl{K}_+ = \tl{K}_+\,f(\tl{K}_0+{\bf 1})\,,
\end{equation}
where ``suitable'' means that $f(\tl{K}_0)$ and $f(\tl{K}_0+{\bf 1})$ are both well-defined operators;
We then have
\begin{equation}
  \label{eq:1736}
  A^{\dagger}A = \tl{K}_+(\tl{K}_0+k)^{-1}\tl{K}_- = (\tl{K}_0+k-1)^{-1}\tl{K}_+\tl{K}_- =
 (\tl{K}_0+k-1)^{-1}[k(1-k)+\tl{K}_0(\tl{K}_0-k)]\,,
\end{equation}
where the first of the relations \eqref{eq:1715} has been used. \\ As $k(1-k)+\tl{K}_0(\tl{K}_0-k) = 
(\tl{K}_0+k-1)(\tl{K}_0-k)$ the Eq.\
\eqref{eq:1734} follows immediately.

Explicitly written in terms of the operators $\tl{K}_0$,\,$\tl{K}_1$ and $\tl{K}_2$ we have
\begin{eqnarray}
  \label{eq:1748}
  \tl{Q}(\vec{\tl{K}})&=& \frac{1}{\sqrt{2}}\,(A+A^{\dagger})=\frac{1}{\sqrt{2}}\,(\tl{K}_1B_k+ B_k\tl{K}_1)+
\frac{i}{\sqrt{2}}\,(\tl{K}_2B_k-B_k\tl{K}_2)\,,
 \\
\tl{P}(\vec{\tl{K}}) &=& \frac{i}{\sqrt{2}}\,(A^{\dagger}-A) = \frac{i}{\sqrt{2}}\,(\tl{K}_1B_k-B_k\tl{K}_1)-
\frac{1}{\sqrt{2}}\,(\tl{K}_2B_k+B_k\tl{K}_2)\,.
 \label{eq:1749}
\end{eqnarray}
These relations show that - contrary to the classical case (cf.\ Eqs.\ \eqref{eq:1458} and \eqref{eq:1460}) - the operators
$\tl{Q}$ and $\tl{P}$ are not just proportional to $\tl{K}_1$ and $\tl{K}_2$, but contain mixtures of both!
\subsubsection{Two kinds of energy spectra for the quantum mechanical HO}
We now come to the crucial point of the whole paper: \\
Obviously the (dimensionless) $(\tl{q},\tl{p})$-Hamiltonian
\begin{equation}
  \label{eq:1737}
  \tl{H}[\tl{Q}(\vec{\tl{K}}),\tl{P}(\vec{\tl{K}})] = \frac{1}{2}\,\tl{Q}^2 + \frac{1}{2}\,\tl{P}^2 =
 A^{\dagger}A + \frac{1}{2}
\end{equation}
obeys the eigenvalue equation
\begin{equation}
  \label{eq:1738}
  \tl{H}(\tl{Q},\tl{P})\,|k, n \rangle = (n + 1/2)\,|k, n\rangle\,.
\end{equation}
On the other hand we have for the $(\vp,\tl{I})$-Hamiltonian
\begin{equation}
  \label{eq:1739}
  \tl{H}(\vec{K}) = \tl{K}_0\,,~~~\vec{K} = \hbar\,(\tl{K}_0, \tl{K}_1, \tl{K}_2)\,,~~H(\vec{K})= \hbar\,\om\,\tl{K}_0\,,
\end{equation}
that 
\begin{equation}
  \label{eq:1740}
  \tl{H}(\vec{K})\,|k, n\rangle = (n + k)|k,n \rangle \,,~~ k > 0\,.
\end{equation}
The last equation shows that the ground state energies of the Hamiltonian \eqref{eq:1739} in principle may take any
real positive value! 

 In sec.\ 3 we encountered the values $k=1$ (for $SO^{\uparrow}(1,2)\,$), $k= 1/2$ (for $Sp(2,
 \mathbb{R})$\,)
and $k=1/4,\,3/4$ (for the 4-fold covering group $Mp(2, \mathbb{R})$ of $SO^{\uparrow}(1,2)\,$). One can show (see below 
and  Appendix B) that for an $m$-fold covering ($ m \in \mathbb{N}$) the lowest possible value for $k$ is $k = 1/m$.
 Thus, we can make $k>0 $ as small as we like by going to higher and higher coverings.

These surprising new possibilities come, of course, from the non-trivial topological structure $\mathbb{R}^2 - \{(0,0)\}
\cong S^1 \times \mathbb{R}^+$ of the phase space $\mathcal{S}_{\vp,\tl{I}}\,$, a structure which is being ``erased''
 when going over to the phase space $\mathcal{S}_{\tl{q},\tl{p}}$ with its trivial topology $\mathbb{R}^2\,$! \\ 
Actually, the more general eigenvalues of Eq.\ \eqref{eq:1740} are a consequence of the ``richer'' quantum theory of
symplectic group $Sp(2, \mathbb{R})$ of the plane which constitutes the ``canonical'' group of the phase space
$\mathcal{S}_{\vp,\tl{I}}\,$.

{\em It is, of course, of crucial importance,  to look for this additional structure experimentally
 (see subsec.\ 1.3 of the Introduction)!!}

If $k \neq 1/2$ then the two energy spectra
\begin{equation}
  \label{eq:1252}
  E_n^{(q,p)} =  \hbar\,\omega\,(n+1/2)\, ,~~~  E_{k,n}^{(\vp,I)} = \hbar \,\omega\,(n+k) \,,
\end{equation}
are different and transitions between different levels should (in principle) be possible if the $E_{k,n}^{(\vp,I)}$ - levels 
do appear at all in nature or can be produced in the laboratory!  Of special interest here is the case where $0 < k
< 1/2$ because then transitions from the $(q,p)$-ground state   to a lower lying $(\vp,I)$-level
are in principle possible provided an appropriate dynamical mechanism is available. An obvious challenge is that for 
$k \neq 1/2$ the same states $|k,n \rangle$ belong to different energy eigenvalues of the operators $H(Q,P)$ and
 $H(\vec{K})$! Notice, however, that for $\tl{H}(\vec{K})$ the ``observables'' $\tl{K}_0$, $\,\tl{K}_1$ and $\tl{K}_2$
 are the primary ones, whereas
$\tl{Q}$ and $\tl{P}$ are ``derived'' or ``composite''  quantities, at least in the present context!

It may also happen, perhaps, that transitions between levels of the two different spectra are more or less strongly impeded
 and that, therefore, certain levels remain ``in the dark''! (See also sec.\ 8). 

\subsubsection{Time evolution and the ground states for different covering groups} 
Let us look at the provoking situation from a slightly different point of view:

The unitary time evolution operator for the $(\vp,\tl{I})$-model of the HO is
\begin{equation}
  \label{eq:1741}
  U(\tl{t}) = e^{-i\,\tl{H}\,\tl{t}}\,,~~\tl{H} = \tl{K}_0\,,~~\tl{t} = \theta\,.
\end{equation}

This equation shows that  the rotation angle $\theta$ can be identified with the time variable $\tl{t}$ which
 - in principle - represents the universal covering space of the circle $S^1$.

From the commutation relations \eqref{eq:1712} and the formula \eqref{eq:1189} we get the  (Heisenberg) eqs.\ of
motion
\begin{eqnarray}
  \label{eq:1223}
  U(-\tl{t})\,\tl{K}_1\,U(\tl{t}) &=& \cos\tl{t}\,\tl{K}_1 - \sin\tl{t}\,\tl{K}_2\,, \\
 U(-\tl{t})\,\tl{K}_2\,U(\tl{t}) &=& \sin\tl{t}\,\tl{K}_1 + \cos\tl{t}\,\tl{K}_2\,,\label{eq:1245}\\
 U(-\tl{t})\,\tl{K}_+\,U(\tl{t}) &=& e^{i\,\tl{t}}\,\tl{K}_+\,,\label{eq:1246} \\
 U(-\tl{t})\,\tl{K}_-\,U(\tl{t}) &=& e^{-i\,\tl{t}}\,\tl{K}_-\,.\label{eq:1248}
\end{eqnarray}
As the operator \eqref{eq:1729} commutes with $U(\tl{t})$ the creation and annihilation operators $A^{\dagger}$ and
$A$ from Eq.\ \eqref{eq:1731} transform as $\tl{K}_+$ and $\tl{K}_-$ in Eqs.\ \eqref{eq:1246} and \eqref{eq:1248}. 
This means that
the position and momentum operators \eqref{eq:1748} and \eqref{eq:1749}
have the usual time evolution:
\begin{eqnarray}
  \label{eq:1254}
   U(-\tl{t})\,\tl{Q}\,U(\tl{t}) &=& \cos\tl{t}\,\tl{Q} + \sin\tl{t}\,\tl{P}\,, \\
 U(-\tl{t})\,\tl{P}\,U(\tl{t}) &=& -\sin\tl{t}\,\tl{Q} + \cos\tl{t}\,\tl{P}\,.\label{eq:1742}
\end{eqnarray}
Here, all the explicit $k$-dependence has dropped out! \\
 However, because of the relation \eqref{eq:1734} we have
\begin{equation}
  \label{eq:1743}
  U(\tl{t} = 2\pi) = e^{-2\pi i k}{\bf 1}\,.
\end{equation}
If $k$ is a positive rational number, $k = n/m\,,\,n,m \in \mathbb{N}$, then the unitary operator \eqref{eq:1743}
belongs to the center of a unitary representation of a $m$-fold covering of $SO^{\uparrow}(1,2)$, the ``lowest''
 representation
 of which is
given by $k=1/m$. Only for $k=1,2,\ldots, $ the operator \eqref{eq:1743} is the identity operator, representing the
identity of $SO^{\uparrow}(1,2)$. If $k= n/m$ then $U(\tl{t} = m\,2\pi)$ is the corresponding identity operator.

Here we see, why the values of $k$ in the interval $(0,\,1]$ may be generically the most important ones in the context
of the HO (see also the related discussions in Ref.\ \cite{bo2}).
 The center
\begin{equation}
  \label{eq:1792}
  \mathbb{Z}_m = \{e^{2\pi\,i\, \mu/m}\,,~\mu =  1 ,\,\cdots,\, m  \}
\end{equation}
of the m-fold covering may be generated by the single element
\begin{equation}
  \label{eq:1816}
  e^{2\pi\,i/m}\,.
\end{equation}
For $\mu = m+1$ we obviously get the same element. Corresponding arguments apply to the unitary operator \eqref{eq:1743}.

The relation \eqref{eq:1743} may also be interpreted  in the following way: 
Applying the operator \eqref{eq:1741} to the ground state yields
\begin{equation}
  \label{eq:2064}
  U(\tl{t})\,|k,0\rangle = e^{-i\,k\,\tl{t}}\,|k,0 \rangle \,.
\end{equation}
As $\tl{t} = \om\,t$ can be used as an angle parametrizing one of the covering groups of the subgroup $SO(2)$,
the interval
\begin{equation}
  \label{eq:2065}
  T_{2\pi,\,k} = \frac{2\pi}{\om_k}\,,~~\om_k \equiv  k\,\om
\end{equation}
is the time the system needs in order to ``run'' through that group. So in a heuristic sense the index $k$ and the
``angle'' $\om\,T_{2\pi\,,k}$ are complementary! The larger the latter the smaller the former! I repeat: The index $k$
can principally be extremely small as long as it stays positive!

\subsubsection{The index $k$ in number states matrix elements}
The index $k$ plays a significant role in matrix elements of the operators $\tl{K}_j,\,j=0,1,2,\,$ with respect to 
the number states $|k,n\rangle$: 

It follows from
\begin{equation}
  \label{eq:1744}
\tl{K}_1 = \frac{1}{2}(\tl{K}_+ +\tl{K}_-)\,,~~~~\tl{K}_2 = \frac{1}{2i}(\tl{K}_+ -\tl{K}_-)\,,
  \end{equation}
that
\begin{equation}
  \label{eq:1745}
  \langle k,n|\tl{K}_j|k,n \rangle = 0\,,~j=1,2,
\end{equation}
and
\begin{equation}
  \label{eq:1746}
  (\Delta \tl{K}_j)^2_{k,n}= \langle k,n|\tl{K}_j^2|k,n \rangle -\langle k,n|\tl{K}_j|k,n \rangle^2 = 
\frac{1}{2}(n^2+2nk+k)\,,~j=1,2,
\end{equation}
so that
\begin{equation}
  \label{eq:1747}
 (\Delta \tl{K}_1)_{k,n}\,(\Delta \tl{K}_2)_{k,n} = \frac{1}{2}(n^2+2kn+k)\,,~~~(\Delta \tl{K}_1)_{k,n=0}\,
(\Delta \tl{K}_2)_{k,n=0} =
 \frac{k}{2}\,,
\end{equation}
Thus, $\tl{K}_1$ and $\tl{K}_2$ have the same standard deviations (``uncertainties'') and  the
 product of these
uncertainties in the ground state is given by $k/2$, i.e.\ the smaller $k$ the smaller the minimal standard deviations!

For the operators $\tl{Q}(\vec{K})$ and $\tl{P}(\vec{K})$ we have 
\begin{equation}
  \label{eq:1758}
\langle k,n|\tl{Q}|k,n \rangle = 0\,,~~\langle k,n|\tl{P}|k,n \rangle = 0\,, 
\end{equation}
and 
\begin{equation}
  \label{eq:1759}
  (\Delta \tl{Q})_{k,n}^2 =\langle k,n|\tl{Q}^2|k,n \rangle = n+1/2\,,~~(\Delta \tl{P})_{k,n}^2 = \langle k,n|\tl{P}^2|k,n
 \rangle = n + 1/2\,,
\end{equation}
which are the usual $k$-independent relations, implying
\begin{equation}
  \label{eq:1760}
  (\Delta \tl{Q})_{k,n}\,(\Delta \tl{P})_{k,n} = n + 1/2\,.
\end{equation}
\subsubsection{Space reflection and time reversal}

From Eqs.\ \eqref{eq:1223} and Eqs.\ \eqref{eq:1245}, or Eqs.\ \eqref{eq:1254} and \eqref{eq:1742} we can infer the
space reflection operator
\begin{equation}
  \label{eq:1750}
  \Pi:~~~\Pi\,\tl{Q}\,\Pi^{\dagger} = -\tl{Q}\,,~~\Pi\,\tl{P}\,\Pi^{\dagger} = -\tl{P}\,,~~\Pi = U(\tl{t} =-\pi)=
e^{i\pi\,(N+k)}\,.
\end{equation}
Now
\begin{equation}
  \label{eq:1751}
  \Pi^2 = e^{2\pi i k}\,,~~\Pi\,|k,n\rangle = (-1)^n\,e^{i \pi k}\,|k,n \rangle\,,
\end{equation}
which shows the $k$-dependence of the phases associated with the so defined operator $\Pi$. 

The antiunitary time reversal transformation $T$ (cf.\  Eq.\ \eqref{eq:1706}) may be implemented by the substitutions
\begin{equation}
  \label{eq:1752}
  T:~~~\tl{K}_0 \to \tl{K}_0\,,~~\tl{K}_1 \to \tl{K}_1\,,~~\tl{K}_2 \to - \tl{K}_2\,,~~i \to -i\,,
\end{equation}
which imply
\begin{equation}
  \label{eq:1753}
 T:~~~ K_{\pm} \to K_{\pm}
\end{equation}
and leave the commutation relations \eqref{eq:1712} and \eqref{eq:1714} invariant. The transformations \eqref{eq:1752}
imply the correct ones for the operators \eqref{eq:1748} and \eqref{eq:1749}.

Contrary to what happens in the case of the canonical pair angle and orbital angular momentum where reflection and time
reversal invariance are generally in conflict with fractional orbital angular momenta \cite{ka2} this is not so for
 fractional ground state energies $ \propto k$ of the HO!

Like in the case of the corresponding Poisson brackets \eqref{eq:1607} and \eqref{eq:1608} the commutators $[K_{\pm},
A]$ etc.\ are rather complicated and will not be listed here.
 One can nevertheless define the following ``squeezing'' operator \cite{dod} ``by hand'':
\begin{equation}
  \label{eq:1755}
  S= e^{-iV\gamma}\,,~~V=\frac{i}{2}(A^2 - (A^{\dagger})^2)\,,~~\gamma \in \mathbb{R}\,,
\end{equation}
which has the property
\begin{equation}
  \label{eq:1756}
  S\,\tl{Q}\,S^{\dagger} = e^{\gamma}\,\tl{Q}\,,~~~ S\,\tl{P}\,S^{\dagger} = e^{-\gamma}\,\tl{P}\,,
\end{equation}
\subsection{Restoring the physical dimensions}
Up to now I have used dimensionless quantities, classical and quantum ones, as introduced in subsec.\ 2.1.
 Here I briefly summarize the main physical quantities with their dimensions restored.
For the classical quantities the procedure is obvious from subsec.\ 2.1. So I confine myself to the operators and their
eigenvalues:

The primary operators with the dimension of an action are
\begin{equation}
  \label{eq:2141}
  K_j =\hbar\, \tl{K}_j\,,~~j=0,1,2\,,~~K_{\pm}=\hbar\,\tl{K}_{\pm}\,;
\end{equation}
they have the commutation relations (cf.\ Eqs.\ \eqref{eq:1712} and \eqref{eq:1714})
\begin{equation}
  \label{eq:2148}
  [K_0,\,K_1] = i\,\hbar\,K_2\,,~~[K_0,\,K_2] = -i\,\hbar\,K_1\,,~~[K_1,\,K_2] = -i\,\hbar\,K_0\,,
\end{equation}
and
\begin{equation}
  \label{eq:2149}
  [K_0,\,K_+] = \hbar\,K_+\,,~~[K_0,\,K_-] = -\hbar\,K_-\,,~~[K_+,\,K_-] = -2\hbar\,K_0\,.
\end{equation}
We have, e.g.\
\begin{equation}
  \label{eq:2150}
  K_0|k,n\rangle =  \hbar\,(n+k)\, |k,n\rangle\,.
\end{equation}
The Hamilton operator is given by
\begin{equation}
  \label{eq:2151}
  H(\vec{K})=\om\,K_0\,,~~~H|k,n\rangle = \hbar\,\om\,(n+k)\, |k,n\rangle\,.
\end{equation}
The number operator remains dimensionless:
\begin{equation}
  \label{eq:2152}
  N=\tl{K}_0-k{\bf 1}\,.
\end{equation}
The conventional annihilation and creation operators \eqref{eq:1731} should also remain dimensionless:
\begin{equation}
  \label{eq:2153}
  A(\vec{\tl{K}})=B_k\,\tl{K}_-\,,~~A^{\dagger}(\vec{\tl{K}})=\tl{K}_+\,B_k\,,~~B_k = (\tl{K}_0+k)^{-1/2}=(N+2k)^{-1/2}\,,
\end{equation}
so that
\begin{equation}
  \label{eq:2154}
  [A,\,A^{\dagger}] = {\bf 1}\,.
\end{equation}
The physical position and momentum operators are then given by (cf.\ Eqs.\ \eqref{eq:936} and \eqref{eq:937})
\begin{equation}
  \label{eq:2155}
  Q = \frac{\lambda_0}{\sqrt{2}}(A^{\dagger}+A)\,,~~P= \frac{i\,\hbar}{\sqrt{2}\,\lambda_0}(A^{\dagger}-A)\,,
~~[Q,\,P]=i\,\hbar\,,~~\lambda_0 =
\sqrt{\frac{\hbar}{m\,\om}}\,.
\end{equation}

\section{Three types of coherent states}
\subsection{Definition and physical interpretation}
It is well-known \cite{ka9} that one can associate three different types of coherent states (CS) with the Lie algebra
of the $\tl{K}_j,\,j=0,1,2,$ in a representation $D_k^{(+)}$: Barut-Girardello, Perelomov  and the conventional
 Schr\"odinger-Glauber
coherent states.
The three kinds of CS may be defined by the relations
 \begin{eqnarray}\label{eq:1761} \tl{K}_-|k,z\rangle
 &=& z\,|k,z\rangle\,,~~ z=|z|\,e^{-i\phi} \in \mathbb{C}\,,
  \\ \label{eq:1762} E_{k,-}|k,\lambda \rangle &=& \lambda\,
 |k,\lambda \rangle\,,~~ \lambda = |\lambda|\,e^{-i\,\theta} \in
 \mathbb{D}~, \\ E_{k,-}&=& (\tl{K}_0+k)^{-1}\tl{K}_-\,,~~ \mathbb{D} = \{\lambda
\in \mathbb{C}\,,~ |\lambda| < 1 \}\,,\label{eq:1763}  \\A|k,
\alpha
\rangle &=& \alpha\,|k,\alpha
 \rangle\,,~~A= B_k\,\tl{K}_-\,,~~ \alpha = |\alpha|\,e^{-i\,\beta} \in \mathbb{C}\,. \label{eq:1764}
  \end{eqnarray}
The minus-sign for the phases of the complex numbers is mere convenience\footnote[2]{Ref.\ \cite{ka1} has the opposite sign
convention.}.

Expanding with respect to a number basis $|k,n \rangle$ yields \cite{ka9}
\begin{eqnarray}
  \label{eq:1765}
 |k,z\rangle&=& \frac{1}{\sqrt{g_k(|z|^2)}}\sum_{n=0}^{\infty}
\frac{z^n}{\sqrt{(2k)_n\,n!}}\,|k,n\rangle\,, \\
&& g_k(|z|^2) = \sum_{n=0}^{\infty}\frac{|z|^{2n}}{ (2k)_n\,n!}
=\Gamma(2k) |z|^{1-2k}\,I_{2k-1}(2|z|)\,; \label{eq:1766} \\
|k,\lambda
\rangle &=& (1-|\lambda|^2)^k\, \sum_{n=0}^{\infty} \left (
\frac{(2k)_n}{n!}\right )^{1/2}\,\lambda^n\,|k,n\rangle\,,~~|\lambda| < 1\,; \label{eq:1767} \\
 |k,\alpha \rangle & =& e^{-|\alpha|^2/2}\, \sum_{n=0}^{\infty}
 \frac{\alpha^n}{\sqrt{n!}}\,|k,n\rangle\,. \label{eq:1768}
\end{eqnarray}
The function $I_{\nu}(x)$ in Eq.\ \eqref{eq:1766} is the usual modified Bessel function of the first kind:
\begin{equation}
  \label{eq:1769}
 I_{\nu}(x) = \left
(\frac{x}{2}\right)^{\nu}\sum_{n=0}^{\infty}
\frac{1}{n!\,\Gamma(\nu+n+1)}\left (\frac{x}{2}\right)^{2n}\,. 
\end{equation}
The series \eqref{eq:1765} - \eqref{eq:1768} are formal ones the convergence properties of which can be specified once
the number states and their Hilbert space are given explicitly.

The physical interpretation of the complex numbers $z\,$,\,$\lambda$ and $\alpha$ can be deduced from the following
expectation values:
\subsubsection{Barut-Girardello coherent states}
\begin{eqnarray}
  \label{eq:1770}
 \langle \tl{K}_0\rangle_{k,z} &\equiv& \langle
 k,z|\tl{K}_0|k,z\rangle=
k+|z|\, \rho_k(|z|)\,, \\ && \rho_k(|z|)=
\frac{I_{2k}(2|z|)}{I_{2k-1}(2|z|)} <1\;,~~ k \geq 1/4\,, \label{eq:1771} \\
(\Delta \tl{K}_0)^2_{k,z} &=& |z|^2\,[1-\rho^2_k(|z|)] +(1-2k)\,|z|\,\rho_k(|z|)\,,\label{eq:1865} \\
\langle N\rangle_{k,z} &\equiv& \bar{n}_{k,z} =
|z|\, \rho_k(|z|)\,,~~N=\tl{K}_0-k{\bf 1}\,, \label{eq:1790} \\
\langle N^2\rangle_{k,z} &=& |z|^2 + (1-2k)|z|\,\rho_k(|z|)\,, \label{eq:1833} \\
\label{eq:1772} \langle \tl{K}_1\rangle_{k,z}
&=&\frac{1}{2}(z^*+z)= \Re (z)= |z|\,\cos\phi\,,\\ \label{eq:1773}\langle
\tl{K}_2\rangle_{k,z} &=&\frac{1}{2i}(z^*-z)= -\Im (z)= |z|\,\sin\phi\,,\\
(\Delta \tl{K}_1)^2_{k,z}= (\Delta \tl{K}_2)^2_{k,z} &=& \frac{1}{2}\,\langle \tl{K}_0 \rangle_{k,z}\,, \label{eq:1877} \\
\tan \phi &=& \langle \tl{K}_2\rangle_{k,z}/\langle \tl{K}_1\rangle_{k,z}\,; \label{eq:1809}\end{eqnarray}
The behaviour of the ratio $\rho_k$  from Eq.\ \eqref{eq:1771} for all $ k>0$ is discussed in Appendix C.
\subsubsection{Perelomov coherent states}
  \begin{eqnarray}
\langle \tl{K}_0 \rangle_{k, \lambda} &\equiv&
 \langle
k, \lambda|\tl{K}_0| k, \lambda \rangle =
k\,\frac{1+|\lambda|^2}{1-|\lambda|^2}= k\, \cosh|w|\,,\label{eq:1774} \\
&& w=|w|\,e^{-i\theta} \in \mathbb{C}\,,~\lambda = \tanh(|w|/2)\,e^{-i\theta}\,,~|w|= \ln\left(\frac{1+|\lambda|}{1- 
|\lambda|}\right)\,,
\label{eq:1775} \end{eqnarray} \begin{eqnarray}
\langle N \rangle_{k, \lambda} &\equiv& \bar{n}_{k,\lambda} = k\,(\cosh |w| -1)\,, \label{eq:1791} \\
\Rightarrow~~~|\lambda|^2 &=& \frac{\bar{n}_{k,\lambda}}{\bar{n}_{k,\lambda} +2k}\,,~~k\,\sinh |w|=
 \sqrt{\bar{n}_{k,\lambda}\,(\bar{n}_{k,\lambda} +2k)} \,,\label{eq:1834}\\
(\Delta \tl{K}_0)^2_{k,\lambda}&=& \frac{k}{2}\,\sinh^2|w| = \frac{1}{2k}\,\bar{n}_{k,\lambda}\,(\bar{n}_{k,\lambda} +2k)\,,
\label{eq:1835} \\
\langle \tl{K}_1 \rangle_{k, \lambda}& = &
2\,k\, \frac{|\lambda|}{1-|\lambda|^2}\,\cos\theta  =
k\,\sinh|w| \,\cos\theta\,,\label{eq:1777} \\ (\Delta \tl{K}_1)^2_{k,\lambda} &=& \frac{k}{2}\,\frac{1+2\cos 2\theta\,
 |\lambda|^2
+ |\lambda|^4}{(1-|\lambda|^2)^2}\,,\label{eq:1878} 
\\ \label{eq:1776}\langle \tl{K}_2 \rangle_{k, \lambda} & =&  2\,k\, 
\frac{|\lambda|}{1-|\lambda|^2}\,\sin\theta  =  k \,\sinh|w|\,\sin\theta\,, \\
(\Delta \tl{K}_2)^2_{k,\lambda} &=& \frac{k}{2}\,\frac{1-2\cos 2\theta\, |\lambda|^2
+ |\lambda|^4}{(1-|\lambda|^2)^2}\,,\label{eq:1879} \\
\langle \tl{K}_0\rangle_{k,\lambda}^2 &=&\langle \tl{K}_1\rangle_{k,\lambda}^2 +\langle \tl{K}_2\rangle_{k,\lambda}^2 +k^2\,;
 \label{eq:1785} \\
\tan \theta &=& \langle \tl{K}_2 \rangle_{k, \lambda}/\langle \tl{K}_1 \rangle_{k, \lambda}\,. \label{eq:1810} 
\end{eqnarray}
\subsubsection{Schr\"{o}dinger-Glauber coherent states}
\begin{eqnarray}
  \label{eq:1778}
  \langle \tl{Q} 
\rangle_{k, \alpha} &=& \sqrt{2}\,\Re (\alpha) = \tl{q} = \sqrt{2}|\alpha| \cos \beta\,, \\
 \langle \tl{P} \rangle_{k, \alpha} &=& \sqrt{2}\,\Im (\alpha) = \tl{p} = -\sqrt{2}|\alpha| \sin \beta \,, \label{eq:1779}\\
 \langle \tl{H}(\tl{Q},\tl{P}) \rangle_{k, \alpha} &=& |\alpha|^2 + 1/2\,,\label{eq:1780} \\ \nonumber \\
\langle \tl{K}_0\rangle_{k,\alpha} &=& \langle N\rangle_{k,\alpha}+k
=|\alpha|^2 +k\,,~~N=\tl{K}_0-k{\bf 1}\label{eq:1781}\\
\langle \tl{K}_1\rangle_{k,\alpha} &=&
 |\alpha|\,\cos\beta\, \langle  k, \alpha|\sqrt{N+2k}|k, \alpha\rangle
 \,,\label{eq:1782} \\
\langle \tl{K}_2\rangle_{k,\alpha} &=& |\alpha|\, \sin\beta\,\langle k,  \alpha|\sqrt{N+2k}|k, \alpha\rangle\,,
\label{eq:1783} \\
&& \langle k, \alpha|\sqrt{N+2k}|k, \alpha\rangle =
e^{-|\alpha|^2}\sum_{n=0}^{\infty}\sqrt{2k+n}\,\frac{|\alpha|^{2n}}{n!} \equiv h_1^{k}(|\alpha|)\, \label{eq:1784}\,, \\
\tan \beta &=& \langle \tl{K}_2\rangle_{k,\alpha}/\langle \tl{K}_1\rangle_{k,\alpha}\,. \label{eq:1811}
\end{eqnarray}
\subsubsection{Physical interpretation of the complex variables}
\paragraph{Barut-Girardello states} \hfill

Eqs.\ \eqref{eq:1772} and \eqref{eq:1773} show that we can interpret $\Re (z)$ as the classical variable $\tl{h}_1$ and
$\Im (z)$ as $\tl{h}_2$, i.e.\ we have
\begin{equation}
  \label{eq:1786}
  z= \tl{h}_1 +i\,\tl{h}_2= \tl{h}_+ = |z|\,e^{-i\phi}\,,~|z|=\tl{I} > 0\,,~\phi=\vp\,.
\end{equation}

Deviations from the classical value $|z|$ etc.\ in the relations \eqref{eq:1770} and \eqref{eq:1865} -- \eqref{eq:1833}
are controlled by the ratio $\rho_k$. It has the limiting values \cite{ka9}
\begin{equation}
 \label{eq:54}
 \rho_k(|z|)
\to
\frac{|z|}{2\,k}\,\left(1-\frac{|z|^2}{2\,k\,(2k+1)}\right)~~\mbox{for}~|z|\to
0\,,
\end{equation}
and for very large $|z|$, the {\em correspondence limit\,}, we get

\begin{eqnarray} \label{eq:55} \rho_k(|z|)&\asymp& 1 -\frac{4k-1}{4|z|}+
\frac{16\,(k^2-k)+3}{32|z|^2}+O(|z|^{-3})~, \\
\label{eq:56} \rho_k^2(|z|)&\asymp& 1 -\frac{4k-1}{2|z|}+
\frac{8\,k^2-6\,k+1}{4\,|z|^2}+ O(|z|^{-3})~ \mbox{ for } |z|
\to \infty~.  \label{eq:1068}  \end{eqnarray}
 
 The last two relations imply that for large $|z|$ 
 \begin{eqnarray} \label{eq:57} \langle K_0\rangle_{k,z}& \asymp &|z|
+\frac{1}{4}+O(|z|^{-1})~,\\ \label{eq:58}(\Delta K_0)^2_{k,z} &\asymp&
\frac{1}{2}|z| + O(|z|^{-1})~, \\
\label{eq:59} \bar{n}_{k,z}&\asymp& |z| +\frac{1}{4}-k+ O(|z|^{-1})~, \\
\label{eq:60} (\Delta N)^2_{k,z}&\asymp&\frac{1}{2}|z| + O(|z|^{-1})\asymp
\frac{1}{2}\bar{n}_{k,z}~,
\end{eqnarray}

\paragraph{Perelomov states} \hfill

Here the situation is different from the previous one: 
The expectation values \eqref{eq:1777} and \eqref{eq:1776} are proportional to the index $k$, a completely non-classical
 quantity.
This suggests to divide out the factor $k$ and make the ``classical''  interpretations
\begin{equation}
  \label{eq:1787}
  \tl{h}_1 = \tl{I}\,\cos\theta\,,~~\tl{h}_2 = -\tl{I}\,\sin\theta\,,~~\tl{I} = \sinh |w|\,,~~|w| >0\,,~~ \theta = \vp\,.
\end{equation}
It means that
\begin{equation}
  \label{eq:1789}
  |w|= \ln \left(\tl{I} +\sqrt{\tl{I}^2 +1}\right)\,,\;\;\; |\lambda|= \tanh(|w|/2)= \frac{\tl{I}}{1+\sqrt{\tl{I}^2 
+1}}\,,
\end{equation}
so that
\begin{equation}
  \label{eq:1788}
  \lambda = \frac{\tl{h}_1+i\,\tl{h}_2}{1+\sqrt{\tl{I}^2+1}}\,.
\end{equation}
It follows that the expectation value \eqref{eq:1774} of $\tl{K}_0$ approaches the value $k\,\tl{I}$ in the classical limit
 for which  $|w| \to \infty$ or $|\lambda| \to 1^-\,$.

It is remarkable that the above expectation values with respect to the states $|k,\lambda \rangle$ are all proportional to
$k$, i.e.\ they have a sensitive $k$-dependence. This may be of interest for experimental tests. 
\paragraph{Schr\"{o}dinger-Glauber states} \hfill

The first three of the expectation values \eqref{eq:1778} - \eqref{eq:1783} are well-known. They show that
\begin{equation}
  \label{eq:1807}
  |\alpha|^2= \tl{I}\,,~~\beta = \vp.
\end{equation}
 The others have been discussed
in subsec.\ 3.3 of Ref.\ \cite{ka1}.
\paragraph{Measuring the phases} \hfill

The three relations \eqref{eq:1809}, \eqref{eq:1810} and \eqref{eq:1811} show that the operators $\tl{K}_1$ and $\tl{K}_2$
 can be
used in order to ``measure'' phases of complex amplitudes.
\subsection{Generation from the ground state}
The coherent states \eqref{eq:1765} - \eqref{eq:1768} may be generated from the ground state $|k,0 \rangle$ by 
 unitary or similar operators. The unitary operators are also  useful for the experimental generation of those states
(see subsec.\ 6.5). Another  problem is the appropriate experimental preparation of the ground state $|k, 0\rangle $
 on which
 the unitary operators act.
\subsubsection{Schr\"{o}dinger-Glauber states}
 The  coherent states \eqref{eq:1768} can be generated from the groundstate $|k,0 \rangle $ by
 the unitary operator
\begin{equation}
  \label{eq:1802}
  U_{SG}= e^{ \alpha\,A^{\dagger} - \alpha^*\,A}= e^{ -|\alpha|^2/2}\,e^{ \alpha\,A^{\dagger}}\,
e^{-\alpha^*\,A}\,,~~U_{SG}\,|k,0 \rangle = |k,\alpha \rangle\,,
\end{equation}
which is well-known for the case $k= 1/2$.
The operator \eqref{eq:1802} has the ``displacement'' (translation) properties
\begin{equation}
  \label{eq:2080}
  U_{SG}^{\dagger}A\,U_{SG}= A + \alpha\,,~~~ U_{SG}^{\dagger}A^{\dagger}\,U_{SG}= A^{\dagger} + \alpha^*\,,
\end{equation}
so that
\begin{equation}
  \label{eq:2081}
   U_{SG}^{\dagger}A^{\dagger}A\,U_{SG}= A^{\dagger}A + \alpha\,A^{\dagger}+ \alpha^*\,A + |\alpha|^2\,,
\end{equation} with 
\begin{equation}
  \label{eq:75}
  \langle k,0| U_{SG}^{\dagger}A^{\dagger}A\,U_{SG}|k,0 \rangle = |\alpha|^2\,.
\end{equation}
If $\alpha$ becomes time-dependent, the transformed number operator \eqref{eq:2081} corresponds to  a driven harmonic
 oscillator,
i.e.\ an oscillator coupled to an external source \cite{exts}. Such external sources are actually used in order to
generate these coherent states  experimentally \cite{sarg}. In textbooks and articles laser light is frequently
 mentioned as being
in a coherent state. The characteristic Poisson distribution of the associated photons is, however, only reached for
lasers well above threshhold \cite{walls1}.
\subsubsection{Perelomov states}
The states \eqref{eq:1767} can be generated from $|k,0 \rangle$ by the unitary operator \cite{perem}
\begin{equation}
  \label{eq:1800}
  U_P =e^{ (w/2)\tl{K}_+ - (w^*/2)\tl{K}_-} = e^{ \lambda\,\tl{K}_+}\,e^{ \ln(1-|\lambda|^2)\tl{K}_0}\,
e^{ -\lambda^*\,\tl{K}_-}\,,~~U_P\,|k,0\rangle = |k, \lambda \rangle \,,
\end{equation}
where the complex number $w$ is the same as in Eq.\ \eqref{eq:1775}.

Instead of the displacements \eqref{eq:2080} we here have the Lorentz transformations \cite{ka10}
\begin{eqnarray}
U^{\dagger}_P\,\tl{K}_0\,U_P &=& \cosh|w|\,\tl{K}_0 + \label{eq:2082}\\ &&
+\frac{1}{2}\,\sinh|w|(e^{-i\,\theta}\, \tl{K}_+ +
e^{i\,\theta}\,\tl{K}_-)\,,\nonumber \\
U^{\dagger}_P\,\tl{K}_+\,U_P &=& \frac{1}{2}(\cosh|w|+1)\,\tl{K}_+ +\label{eq:2083} \\
&& +\frac{1}{2}e^{2\,i\,\theta}(\cosh|w| -1)\, \tl{K}_- +
e^{i\,\theta}\,\sinh|w|\,\tl{K}_0\,, \nonumber \\
U^{\dagger}_P\,\tl{K}_-\,U_P &=& \frac{1}{2}(\cosh|w|+1)\,\tl{K}_- + \label{eq:2084}\\
&& +\frac{1}{2}e^{-2\,i\,\theta}(\cosh|w| -1)\, \tl{K}_+
 + e^{-i\,\theta}\,\sinh|w|\,\tl{K}_0\,, \nonumber
\end{eqnarray}
The relation corresponding to Eq.\ \eqref{eq:75} here is
\begin{equation}
  \label{eq:76}
  \langle k,0| U_{P}^{\dagger}\tl{K}_0 U_{P}|k,0 \rangle = k\, \cosh|w|\,.
\end{equation}

In terms of the vectors
\begin{equation}
  \label{eq:2085}
  \vec{K}_{\perp} =(\tl{K}_1,\,\tl{K}_2)\,,~~\vec{n} = (\cos \theta,\,\sin \theta)\,,
\end{equation}
these relations may be written as
\begin{eqnarray}
U^{\dagger}_P\,\tl{K}_0\,U_P& = &\cosh|w|\,\tl{K}_0 +\sinh|w|\,(\vec{n}\cdot
\vec{K}_{\perp})~, \label{eq:2086} \\
 U^{\dagger}_P\vec{K}_{\perp}\,U_P &=& \vec{K}_{\perp}+(\cosh|w|-1)
(\vec{n}\cdot\vec{K}_{\perp})\,\vec{n}+ \sinh|w|\,\vec{n}\,\tl{K}_0\,. \label{eq:2087}
\end{eqnarray}

The operator \eqref{eq:1800} now generates interaction terms for the original $\tl{K}_0$ which are proportional
 to $\tl{K}_+$
 and $\tl{K}_-\,$, or to $\tl{K}_1$ and (or) $\tl{K}_2\,$. (Their classical counterparts for $\theta = 0$ and $\theta =
\pi/2$ were briefly discussed in subsec.\ 2.3.)
The use of the induced interaction term in Eq.\ \eqref{eq:2086} in theoretical descriptions of experiments will be
 discussed in subsec.\ 6.5\,.
\subsubsection{Barut-Girardello states}
Here the situation appears to be more complicated, because no corresponding unitary operator has been derived by now.
The present situation is as follows \cite{aga1}:

Because of the relation \eqref{eq:1792} we can write
\begin{equation}
  \label{eq:1793}
  \sum_{n=0}^{\infty}\frac{z^n}{\sqrt{(2k)_n\,n!}}|k,n\rangle = \sum_{n=0}^{\infty}\frac{z^n}{(2k)_n\,n!}
(\tl{K}_+)^n|k,0\rangle\,.
\end{equation}
As
\begin{equation}
  \label{eq:1794}
  \frac{1}{(2k)_n}(\tl{K}_+)^n|k,0\rangle = (E_{k,+})^n|k,0\rangle\,,~~E_{k,+}= \tl{K}_+(\tl{K}_0+k)^{-1}=
(E_{k,-})^{\dagger}\,,
\end{equation}
where $E_{k,-}$ as in Eq.\ \eqref{eq:1763}, we have 
\begin{equation}
  \label{eq:1795}
  \sum_{n=0}^{\infty}\frac{z^n}{\sqrt{(2k)_n\,n!}}|k,n\rangle = F_k(z)\,|k,0\rangle\,,~~F_k(z)=e^{ z\,E_{k,+}}\,.
\end{equation}
The non-unitary operators $F_k(z)$ and $F^{\dagger}_k(z) = \exp (z^*\,E_{k,-})\,$    have the following properties:
\begin{equation}
  \label{eq:1796}
  \langle k,0|F^{\dagger}_k(z)F_k(z)|k,0 \rangle = g_k(|z|^2) >0\,,~~F^{\dagger}_k(z) \,|k,0\rangle = |k,0 \rangle\,,~~
 F^{\dagger}_k(z)
\,|k,\lambda \rangle = e^{ z^*\,\lambda}\,|k, \lambda \rangle\,,
\end{equation}
where $g_k(|z|^2)$ is defined in Eq.\ \eqref{eq:1766} and $|k,\lambda \rangle$ in Eq.\ \eqref{eq:1762}. \\ Thus, we have
\begin{equation}
  \label{eq:2088}
  F_k(z)\,|k,0 \rangle = \sqrt{g_k(|z|^2)}\,|k,z \rangle\,,
\end{equation}
i.e.\ $F_k(z)$ generates the unnormalized Barut-Girardello states. 
It corresponds to the similar generating parts
 \begin{equation}
    \label{eq:930}
   e^{ \alpha\,A^{\dagger}}\,, ~~~~e^{ \lambda\,\tl{K}_+} 
  \end{equation}
of the unitary operators \eqref{eq:1802} and \eqref{eq:1800} for the unnormalized Schr\"{o}dinger-Glauber and Perelomov
states. But, contrary to $A^{\dagger}$ and $\tl{K}_+$ the operators $E_{k,+}$ and $E_{k,-}$ are not elements of a Lie 
algebra. They have - among others more complicated ones - the commutators
 \begin{equation}
  \label{eq:2090}
  [E_{k,-},\,E_{k,+}] = \frac{2k-1}{(\tl{K}_0+k)(\tl{K}_0+k-1)}\,,~~~[\tl{K}_-,\,E_{k,+}] = 1\,,~~~[E_{k,-},\,\tl{K}_+] =1\,.
\end{equation}
It follows from the completeness relation \eqref{eq:1840} and the last of the relations \eqref{eq:1796} that one has
for $F^{\dagger}_k(z)F_k(z)$ the ``spectral representation''
\begin{equation}
  \label{eq:2095}
  F^{\dagger}_k(z)F_k(z) =\int_{\mathbb{D}}d\mu_k(\lambda)\, e^{ z^*\,\lambda + z\,\lambda^*}\,|k,\lambda\rangle 
\langle k, \lambda |\,.
\end{equation}
\subsubsection{Transitions between Perelomov and Barut-Girardello coherent states}

Notice that, according to Eqs.\ \eqref{eq:1765} and \eqref{eq:1767}, 
\begin{eqnarray}
 \langle k, \lambda |k, z \rangle & =&
 \frac{(1-|\lambda|^2)^k}{\sqrt{g_k(|z|^2)}}\,e^{ \lambda^*\,z}\,, \nonumber \\ p_k(\lambda \leftrightarrow z) &=&
 |\langle k, \lambda |k, z \rangle|^2 =
 \frac{(1-|\lambda|^2)^{2k}}{g_k(|z|^2)}\,e^{ 2|\lambda|\,|z|\,\cos (\phi-\theta)}\,. \label{eq:1801}
\end{eqnarray}
As \cite{ka9}
\begin{equation}
  \label{eq:10}
  g(|z|^2) \asymp \frac{\Gamma (2k)}{2\sqrt{\pi}}\,|z|^{1/2-2k}\,e^{ 2|z|}\,[1+O(1/|z|)]~ \text{ for large } |z|\,,
\end{equation}
we get for the transition probability in the (classical) limit of large $|z|\,$:
\begin{equation}
  \label{eq:11} p_k(\lambda \leftrightarrow z) \asymp
  \frac{2\sqrt{\pi}}{\Gamma (2k)\,\sqrt{|z|}}\,[|z|(1-|\lambda|^2)]^{2k}\,
e^{-2|z|[1-|\lambda|\cos (\phi-\theta)]}~ \text{ for large } |z|\,.
\end{equation}
According to Eqs.\ \eqref{eq:1786} and \eqref{eq:1786} we have
\begin{equation}
  \label{eq:12}
  |\lambda| \asymp 1-\frac{k}{|z|}~\text{ for large } |z|\,.
\end{equation}
Inserting this approximation for $|\lambda|$ into the relation \eqref{eq:11} yields in leading order for large $|z|$
\begin{equation}
  \label{eq:13}
   p_k(\lambda \leftrightarrow z) \asymp \frac{2\sqrt{\pi}\,(2k)^{2k}}{\Gamma (2k)\,\sqrt{|z|}}\,
e^{ -2|z|[1-\cos (\phi-\theta)]}~ \text{ for } |z| \to \infty\,.
\end{equation}
Expanding  $\cos (\phi-\theta)$ around $(\phi-\theta)=0$ gives an approximate Gaussian distribution for $p_k(\lambda
 \leftrightarrow z) $:
 \begin{equation}
   \label{eq:14}
  p_k(\lambda \leftrightarrow z) \asymp \frac{2\sqrt{\pi}\,(2k)^{2k}}{\Gamma (2k)\,\sqrt{|z|}}\,
e^{ -|z| (\phi-\theta)^2}~ \text{ for } |z| \to \infty\,.  
 \end{equation}
This shows that for a given large $|z|$ the transition probability is maximal for $\phi = \theta$. \\
On the other hand, it follows from
\begin{equation}
  \label{eq:15}
  \lim_{k \to 0^+} (2k)^{2k} =1\,,~~~ \Gamma (2k) \to \frac{1}{2k} \text{ for } k \to 0^+\,,
\end{equation}
that $p_k$ becomes very small for very small $k$.

Properties of the matrix elements $\langle k, \alpha | k, z \rangle $ and $ \langle k, \alpha | k, \lambda \rangle $
are discussed in chap.\ 3 of Ref.\ \cite{ka1}. In the special case $k= 1/2$ they are described in subsec.\ 7.1 below.

\subsection{Time evolution}

It follows from 
\begin{equation}
  \label{eq:1803}
  U(\tl{t})|k,n \rangle = e^{-i(n+k)\,\tl{t}}\,|k,n \rangle\,,~~U(\tl{t}) = e^{-i\,\tl{K}_0\,\tl{t}}\,,
\end{equation}
that
\begin{eqnarray}
  \label{eq:1804}
  U(\tl{t})|k,z\rangle &=& e^{-i\,k\,\tl{t}}|k,z(\tl{t}) \rangle\,,~~z(\tl{t})=z\,e^{-i\,\tl{t}}\,, \\
U(\tl{t})|k, \lambda \rangle &=& e^{-i\,k\,\tl{t}}|k, \lambda(\tl{t}) \rangle\,,~~\lambda(\tl{t})=\lambda\,e^{-i\,\tl{t}}\,,
\label{eq:1805} \\
U(\tl{t})|k,\alpha \rangle &=& e^{-i\,k\,\tl{t}}|k, \alpha(\tl{t}) \rangle\,,~~\alpha(\tl{t})=\alpha\,e^{-i\,\tl{t}}\,.
\label{eq:1806}
\end{eqnarray}
These equations show that the time evolution does not change the form of the coherent states. It essentially shifts
only the phases of the complex numbers $z$, $\lambda$ and $\alpha$ linearly in time:
\begin{equation}
  \label{eq:1808}
  \phi \to \phi + \tl{t}\,,~~\theta \to \theta + \tl{t}\,,~~\beta \to \beta + \tl{t}\,.
\end{equation}
\subsection{Some general properties}
I finally list some general properties of the above coherent states which are very useful for applications:

\subsubsection{Scalar products}
Using the orthonormality of the number states $|k,n \rangle $ two different states within one of the types listed in
Eqs.\ \eqref{eq:1765} -- \eqref{eq:1768} have the scalar product
\begin{eqnarray}
  \label{eq:1836}
  \langle k, z_2|k,z_1 \rangle &=&
\sum_{n=0}^{\infty}\langle k, z_2|k,n \rangle  \langle k,n|k,z_1
\rangle =  \frac{g_k(
z^*_2\,z_1)}{\sqrt{g_k(|z_2|^2)\,g_k(|z_1|^2)}}\,; \\
\langle k, \lambda_2 | k, \lambda_1
\rangle &=& (1-|\lambda_1|^2)^k\,
(1-|\lambda_2|^2)^k\,(1-\lambda^*_2\, \lambda_1)^{-2\,k}\,; \label{eq:1837} \\
\langle \alpha_2|\alpha_1 \rangle &=&
e^{-(|\alpha_2|^2+|\alpha_1|^2)/2}\,e^{ \alpha^*_2\,\alpha_1}\,. \label{eq:1838}
\end{eqnarray}
Different states are not orthogonal, but they are ``complete'' in the sense that they provide a resolution of the
identity as follows \cite{ka9}:
\subsubsection{Completeness}
\begin{eqnarray}
  \label{eq:1839}
  \int_{\mathbb{C}} d\mu_k(z)
\,|k,z\rangle \langle k, z| &=& {\bf 1}\,, \\
d\mu_k(z) &=&
\frac{2}{\pi\,\Gamma(2k)}\,|z|^{2k}\,
K_{2k-1}(2|z|)\,g_k(|z|^2)\,d|z|d\phi\,,~k>0\,\label{eq:1841} ; \\
\int_{\mathbb{D}} d\,\mu_k(\lambda)\, |k,\lambda \rangle \langle k,
\lambda | &=& {\bf 1}\,, \label{eq:1840} \\
d\mu_k(\lambda)& =& \frac{2k-1}{\pi}(1-|\lambda|^2)^{-2}
|\lambda|\,d|\lambda| \, d\theta \,,~k > 1/2\,;\label{eq:1842} \\
\frac{1}{\pi}\int_{\mathbb{C}}d^2\alpha |\alpha\rangle \langle
\alpha|& =& {\bf 1}\,, \label{eq:1843} \\
d^2\alpha & =&
d\,\Re(\alpha)\,d\,\Im (\alpha) \,. \label{eq:1844}
\end{eqnarray}
The modified Bessel function of the third kind $K_{\nu}(2|z|)$ (cf.\ Ref.\ \cite{knu1}) in the measure \eqref{eq:1841}
 has the property
$K_{-\nu}(2|z|) = K_{\nu}(2|z|)$ which makes the measure well-defined for $k>0$, because in the limit $|z| \to 0$
one has  
\begin{equation}
  \label{eq:1845}
  \tl{K}_0(2|z|) \to \ln(1/|z|)\,,~K_{\nu}(2|z|) \to \frac{\Gamma (\nu)\,|z|^{-\nu}}{2}+\frac{\Gamma (-\nu)\,|z|^{\nu}}{2}
\text{ for } 0 < |\nu| <1\,, 
\end{equation}
and
\begin{equation}
  \label{eq:1846}
 \tl{K}_1(2|z|) \to 1/(2|z|)+ |z|\ln |z|\,,~K_{\nu}(2|z|) \to \Gamma (|\nu|)\,|z|^{-|\nu|} \text{ for } |\nu| >1\,. 
\end{equation}
The extension of Hilbert spaces with the measure \eqref{eq:1842} for states $|k,\lambda \rangle$ with
 $0 < k \leq 1/2$ will be
discussed below.

The relation \eqref{eq:1843} holds for all $k>0$.

\subsubsection{Hilbert spaces of holomorphic functions associated with the three  types of coherent states} 
It is well-known that the three types of coherent states \eqref{eq:1765} -- \eqref{eq:1768} can be associated with Hilbert
spaces of holomorphic functions \cite{ka11}, the (normalized!) basis elements of which are given by the coefficients under
the sums of the expensions with respect to the states $|k,n \rangle$ \cite{ka9}:
\paragraph{Barut-Girardello holomorphic functions} \hfill 

\begin{eqnarray}
  \label{eq:1847}
  (f_2,f_1)_{k,z} &\equiv& \int_{\mathbb{C}}d\hat{\mu}_k(z)\,f_2^*(z)f_1(z)\,, \\
d\hat{\mu}_k(z) &=& \frac{2}{\pi\,\Gamma(2k)}
\,|z|^{2k}\,
K_{2k-1}(2|z|)\,d|z|d\phi\,,~k>0\,, \nonumber\\
\hat{f}_{k,n}(z)& =& \frac{z^n}{\sqrt{(2k)_n\,n!}}\,,~~(\hat{f}_{k,n_2},\hat{f}_{k,n_1})_{k,z} = \delta_{n_2\,n_1}\,, 
\label{eq:1848} \\
\Delta_k(z_2^*,z_1) = \sum_{n=0}^{\infty} \hat{f}_{k,n}^*(z_2)\,\hat{f}_{k,n}(z_1) &=& g_k(z_2^*\,z_1)\,, \label{eq:1849} \\
\int_{\mathbb{C}}d\hat{\mu}_k(z_2)\,\Delta_k(z_2^*,z_1)\,\hat{f}_{k,n}(z_2) &=& \hat{f}_{k,n}(z_1)\,, \label{eq:1850} \\
\int_{\mathbb{C}}d\hat{\mu}_k(z_2)\,\Delta_k(z_2^*,z_1)\,f(z_2) &=& f(z_1)\,,~f(z) = \sum_{n=0}^{\infty} a_n\,z^n\,,
\label{eq:1851} \\
\int_{\mathbb{C}}d\hat{\mu}_k(z)\,\Delta_k(z_2^*,z)\Delta_k(z^*,z_1)&=& \Delta_k(z_2^*,z_1)\,, \label{eq:1852} \\
(f_2,f_1)_{k,z} = \sum_{n=0}^{\infty} (2k)_n\,n!\,a_{n,2}^*\,a_{n,1},&&f_j(z) = \sum_{n=0}^{\infty}a_{n,j}\,z^n\,,~j=1,2\,.
\label{eq:1858}
\end{eqnarray}
Because of the properties \eqref{eq:1850} -- \eqref{eq:1852} the function $\Delta_k(z_2^*,z_1)$ is called the ``reproducing
kernel'' of the Hilbert space. It has a number of properties usually associated with the (more singular) ``delta-function''
$\delta(x_2-x_1)$ for other spaces of functions!

In the Hilbert space \eqref{eq:1847} a representation of the Lie algebra \eqref{eq:1714} is given by
\begin{equation}
  \label{eq:1853} \tl{K}_0 = z\frac{d}{dz} +k\,,~~
 \tl{K}_+ = z\,,~~\tl{K}_- = 2k\frac{d}{dz} +z\frac{d^2}{dz^2}\,. 
\end{equation}

\paragraph{Perelomov holomorphic functions} \hfill

The corresponding relations for the states $|k,\lambda \rangle$ are
\begin{eqnarray}
  \label{eq:1854}
 (f_2,f_1)_{k,\lambda} &\equiv& \int_{\mathbb{D}}d\tl{\mu}_k(\lambda)\,f_2^*(\lambda)f_1(\lambda)\,,\\
 d\tl{\mu}_k(\lambda)&=& \frac{2k-1}{\pi}
(1-|\lambda|^2)^{2k-2}\,|\lambda|d|\lambda|\,d\theta\,, \nonumber \\
(\tl{e}_{k,n_2},\tl{e}_{k,n_1})_{k, \lambda} = \delta_{n_2\,n_1}\,,~~\tl{e}_{k,n}(\lambda) &=& \sqrt{\frac{(2k)_n}{n!}}\,
\lambda^n\,,
\label{eq:1855}\end{eqnarray} \begin{eqnarray} \Delta_k(\lambda_2^*,\lambda_1)= \sum_{n=0}^{\infty} 
\tl{e}_{k,n}^*(z_2)\,\tl{e}_{k,n}(z_1) &=&
(1-\lambda_2^*\,\lambda_1)^{-2k}\,,\label{eq:1856} \\
\int_{\mathbb{D}}d\tl{\mu}_k(\lambda_2)\,\Delta_k(\lambda_2^*,\lambda_1)\,\tl{e}_{k,n}(\lambda_2) &=& 
\tl{e}_{k,n}(\lambda_1)\,, \label{eq:1857} \\
\int_{\mathbb{D}}d\tl{\mu}_k(\lambda_2)\,\Delta_k(\lambda_2^*,\lambda_1)\,f(\lambda_2) &=& f(\lambda_1)\,,~
f(\lambda) = \sum_{n=0}^{\infty} b_n\,\lambda^n\,, \label{eq:1859}\\
\int_{\mathbb{D}}d\tl{\mu}_k(\lambda)\,\Delta_k(\lambda_2^*,\lambda)\Delta_k(\lambda^*,\lambda_1)&=&
 \Delta_k(\lambda_2^*,\lambda_1)\,,
\label{eq:1860} \\
(f_2,f_1)_{k,\lambda} = \sum_{n=0}^{\infty} \frac{n!}{(2k)_n}\,b_{n,2}^*\,b_{n,1}\,,&&f_j(\lambda) = 
\sum_{n=0}^{\infty}b_{n,j}
\,\lambda^n\,,~j=1,2\,.\label{eq:1861}
\end{eqnarray}
As \cite{erdI1}
\begin{equation}
  \label{eq:1862}
  \int_0^1d|\lambda|^2\,(1-|\lambda|^2)^{2k-2}|\lambda|^{2n +1} = \frac{\Gamma(2k-1)\,n!}{\Gamma(2k+n)}\,,
\end{equation}
the factor $2k-1$ in the measure \eqref{eq:1854} is multiplied by $\Gamma(2k-1)$, yielding $\Gamma(2k)$, which
means that the integral and sums \eqref{eq:1854} -- \eqref{eq:1860} are well-defined for $k>0$. The right-hand side
of Eq.\ \eqref{eq:1861} may be used in oder to define the scalar product for all $k>0$. The properties \eqref{eq:1857} --
\eqref{eq:1860} can be interpreted as the completeness relation for the functions \eqref{eq:1855} where $k>0$.

In the Hilbert space \eqref{eq:1854} one has the following representation of the Lie algebra \eqref{eq:1714} by
self-adjoint operators
\begin{equation}
  \label{eq:1866} \tl{K}_0 =
\lambda\frac{d}{d\lambda}+k\,,~~ 
  \tl{K}_+ = 2k\,\lambda +
\lambda^2\frac{d}{d\lambda}\,,~~\tl{K}_- = \frac{d}{d\lambda}\,.
\end{equation}
\paragraph{Bargmann-Segal holomorphic functions} \hfill

The Hilbert space of holomorphic functions associated with the Schr\"odinger-Glauber coherent states \eqref{eq:1768}
was thoroughly discussed by Bargmann \cite{barg2}. About the same time such Hilbert spaces were also introduced by Segal
into quantum field theory \cite{segal2}  Such a Hilbert space has the following essential properties:
\begin{eqnarray}
  \label{eq:1867}
  (f_2,f_1)_{\alpha}&\equiv& \int_{\mathbb{C}}\,d\tl{\mu}(\alpha)\,f_2^*(\alpha)f_1(\alpha)\,, \\
d\tl{\mu}(\alpha)& =& 
\frac{d^2\alpha}{\pi}e^{-|\alpha|^2}\,, \nonumber \\
\tl{h}_n(\alpha)& =& \frac{\alpha^n}{\sqrt{n!}}\,,~~(\tl{h}_{n_2},\tl{h}_{n_1})_{\alpha} = \delta_{n_2,n_1}\,,
 \label{eq:1868}\\
\Delta(\alpha_2^*, \alpha_1) = \sum_{n=0}^{\infty}\tl{h}^*_n(\alpha_2)\,\tl{h}_n(\alpha_1) &=& e^{\alpha_2^*\,\alpha_1}\,,
\label{eq:1869} \\
\int_{\mathbb{C}}d\tl{\mu}(\alpha_2)\,\Delta(\alpha_2^*,\alpha_1)\,\tl{h}_{n}(\alpha_2) &=& 
\tl{h}_{n}(\alpha_1)\,, \label{eq:1870} \\
\int_{\mathbb{C}}d\tl{\mu}(\alpha_2)\,\Delta(\alpha_2^*,\alpha_1)\,f(\alpha_2) &=& f(\alpha_1)\,,~
f(\alpha) = \sum_{n=0}^{\infty} c_n\,\alpha^n\,, \label{eq:1871} \end{eqnarray} \begin{eqnarray}
\int_{\mathbb{C}}d\tl{\mu}(\alpha)\,\Delta(\alpha_2^*,\alpha)\Delta(\alpha^*,\alpha_1)&=&
 \Delta(\alpha_2^*,\alpha_1)\,,
\label{eq:1872} \\
(f_2,f_1)_{\alpha} = \sum_{n=0}^{\infty} n!\,c_{n,1}^*\,c_{n,2}\,,&&f_j(\alpha) = \sum_{n=0}^{\infty}c_{n,j}
\,\alpha^n\,,~j=1,2\,.\label{eq:1873}
\end{eqnarray}
Recall that $df^*(\alpha)/d\alpha = df(\alpha^*)/d\alpha = 0$ for a holomorphic function $f(\alpha)$.

The mutual adjoint annihilation and creation operators in the Hilbert space \eqref{eq:1867} are \cite{fock,barg2}
\begin{equation}
  \label{eq:1874}
  a= \frac{d}{d\alpha}\,,~~~~a^{\dagger} = \alpha\,,~~~~[a, a^{\dagger}] = 1\,.
\end{equation}
Inverting the relations \eqref{eq:1731} yields the following generators for the Lie algebra \eqref{eq:1714}
\begin{equation}
  \label{eq:1875}
  \tl{K}_0= N+k\,,~~~\tl{K}_+ = \alpha\,\sqrt{N+2k}\,,~~~\tl{K}_- = \sqrt{N+2k}\,\frac{d}{d\alpha}\,,~~~
N= \alpha\,\frac{d}{d\alpha}\,.
\end{equation}
\subsubsection{Probabilities for transitions to number states}
\paragraph{Barut-Girardello states} \hfill

From the expansions \eqref{eq:1765} -- \eqref{eq:1768} one immediately can read off the following transition probabilities
\begin{equation}
  \label{eq:1876}
  p_{k}(n \leftrightarrow z) = \frac{|z|^{2n}}{(2k)_n\,n!\,g_k(|z|^2)}\,,~~~
 p_{k}(n=0 \leftrightarrow z) = \frac{1}{g_k(|z|^2)}\,.
\end{equation}
In applications one would like to express $|z|$ in terms of the average number $\bar{n}_{k,z}\,$, here given by Eq.\
\eqref{eq:1790}. As the ratio $\rho_k(|z|)$ depends on $|z|\,$, too, the inversion $|z| = |z|(\bar{n}_{k,z})$ is not
immediate. But for large $|z|$ one has in leading order \cite{ka9}
\begin{equation}
  \label{eq:32}
  \rho_k(|z|) \to 1\,,~~g_k(|z|^2) \to \frac{\Gamma (2k)}{2\sqrt{\pi}}\,\frac{e^{2|z|}}{|z|^{2k-1/2}}~~\text{ for }~ |z| \to
\infty\,,
\end{equation}
so that asymptotically
\begin{equation}
  \label{eq:33}
   p_{k}(n \leftrightarrow z) \asymp \frac{2\sqrt{\pi}}{n!\,\Gamma (2k +n)}\,(\bar{n}_{k,z})^{2(n+k-1/4)}\,
e^{-2\,\bar{n}_{k,z}}\,.
\end{equation}
As the Barut-Girardello states have not yet been produced in a laboratory the distribution \eqref{eq:33} has not been 
tested experimentally (to the best of my knowledge)!
\paragraph{Perelomov states}\hfill

Here we get
\begin{equation}
  \label{eq:1863}
  p_{k}(n \leftrightarrow \lambda) = (1-|\lambda|^2)^{2k}\,\frac{(2k)_n}{n!}\,|\lambda|^{2n}\,,~~~
 p_{k}(n=0 \leftrightarrow \lambda) = (1-|\lambda|^2)^{2k}\,.
\end{equation}
Using the first of the relations \eqref{eq:1834} we can also write
\begin{equation}
  \label{eq:34}
   p_{k}(n \leftrightarrow \lambda) = \frac{2k\,(2k)_n}{(\bar{n}_{k,\lambda} +2k)\,n!}\left(\frac{\bar{n}_{k,\lambda}}{
\bar{n}_{k,\lambda} + 2k}\right)^n\,. 
\end{equation}
As the Perelomov states for $k=1/2$ can be  produced in the laboratory (see the next subsec.), the distribution
\eqref{eq:34} has been verified experimentally by counting photon numbers emanating from a Perelomov (squeezed) 
state \cite{kum}.
 \paragraph{Schr\"{o}dinger-Glauber states}\hfill

Here we have the usual Poisson distribution
\begin{equation}
  \label{eq:1864}
  p_k(n \leftrightarrow \alpha) = \frac{|\alpha|^{2n}}{n!}\,e^{-|\alpha|^2}\,,~~|\alpha|^2 = \bar{n}_{\alpha}\,.
\end{equation}
As to its experimental verification see subsec.\ 6.2.1 above.
\subsection{Physical dynamics described by the basic operators $\tl{K}_0,\,\tl{K}_+$  and $\tl{K}_-$}
The conventional annihilation and creation operators $a$ and $a^{\dagger}$ are a convenient and popular tool in order to
 build Hamiltonians which describe interactions between elementary excitations, particles and modes, be it scattering,
annihilation or creation of them. Completely similar one can construct physically useful model Hamiltonians from the three
basic operators $\tl{K}_0,\,\tl{K}_+$ and $\tl{K}_-$ (or $\tl{K}_1$ and $\tl{K}_2\,$). 

Actually there are already quite a
 number of such models in use,
especially in the field of quantum optics. They usually come in a form in which the $\tl{K}_j$ are expressed in terms of
 one or
 several pairs of $a$ and $a^{\dagger}$. I shall list several typical examples, without any claim of even partial
 completeness. I shall merely mention explicitly  some quite early and some very recent original papers, but otherwise
refer to the corresponding chapters in textbooks \cite{param} and their associated References. 

An early review on the
 dynamics of models expressed in terms of the generators $\tl{K}_j$ is Ref.\ \cite{datt}.
 Early papers using that
Lie algebra explicitly for the generation of squeezed states are Refs.\ \cite{wod1}. Usually all those applications are
discussed in the language of the group $SU(1,1)$. I have stressed in the Introduction and in sec.\ 3 why the language
 of the isomorphic symplectic
group $Sp(2,\mathbb{R})$ is more appropriate because of its potential for generalizations to higher dimensions. 

An essential model to start with is the one which we encountered in the context of the unitary transformation 
\eqref{eq:1800} which generates the self-adjoint interaction
\begin{equation}
  \label{eq:1797}
 (\sinh|w|)\,W(\theta,\,\vec{K})= (1/2)(\sinh|w|)(e^{-i\theta}\,\tl{K}_+ + e^{i\theta}\,\tl{K}_-) = (\sinh |w|)
 (\vec{n}\cdot \vec{K}_{\perp})
\end{equation}
of Eqs.\ \eqref{eq:2082} and \eqref{eq:2086}. The angle $\theta$ here plays the role of a mixing angle as to the operators
$\tl{K}_1$ and $\tl{K}_2$: For $\theta =0$ the term \eqref{eq:1797} is pure $\tl{K}_1$ and for $\theta = \pi/2$ pure
 $\tl{K}_2$. (As to properties of the classical mechanics counterpart of these interactions see subsec.\ 2.3)
\subsubsection{Generation of Perelomov coherent states}

As the operator $U_P$ from Eq.\ \eqref{eq:1800} generates the {\em Perelomov coherent states} $|k,\lambda\rangle$ 
from the ground state,
 the interaction
 \eqref{eq:1797} can be used to generate such states experimentally! 

In applications the operator $W$ from Eq.\ \eqref{eq:1797} is generally  multiplied by a ``classical''
 function \\ $G[g(\tl{t}),C(\tl{t},a)]$,
 containing coupling
constants $g(\tl{t})$ (possibly time-dependent) and (possibly) time-dependent external ``classical'' fields $C(\tl{t},a)$
 which themselves may depend
 on additional parameters $a$, e.g.\ second-order or third-order non-linear susceptibilities ($\chi^{(2)}$ or
 $\chi^{(3)}$) \cite{boyd}, spatial 
coordinates etc.

The potential
\begin{equation}
  \label{eq:1799}
  V = G[g(t),C(t,a)]\,W(\theta,\vec{\tl{K}})
\end{equation}
is then being dealt with in the interaction picture, where $V$ determines the time evolution of the {\em states} 
and the free
Hamiltonian $\hbar\,\om\,\tl{K}_0$ that of the {\em operators}. 

The interaction Hamiltonian \eqref{eq:1797} is linear in the operators $K_j$. Another possibility is to have interactions
which are bilinear in the operators $K_j$, e.g.\ proportional to $K_+K_-$ in the description of scattering processes (see
below). These can be diagonalized with the help of the Casimir relations \eqref{eq:1715}. 

\subsubsection{One-mode generated Lie algebra $\mathfrak{so}(1,2)$}

Already in section 3.5 we encountered the one-mode representations
\begin{equation}
  \label{eq:1798}
  \tl{K}_0 = \frac{1}{4}(2a^{\dagger}a +1)\,,~~\tl{K}_+ =\frac{1}{2}{a^{\dagger}}^2\,,~~\tl{K}_-= \frac{1}{2}a^2\,,~~\tl{K}_1=
\frac{1}{4}({a^{\dagger}}^2+a^2)\,,~~\tl{K}_2=\frac{1}{4i}({a^{\dagger}}^2-a^2)\,.
\end{equation}
Inserted into Eq.\ \eqref{eq:1797} the term $W$ describes the creation or annihilation of two identical modes (photons).

\paragraph{Degenerate parametric down-conversions and amplifications} \hfill 

 Such  processes  occur experimenally in so-called ``degenerate parametric down-conversions'' where a classical
 electromagnetic
(``pump'') wave of frequency $2\om$ generates two identical photons each with frequency $\om$ in a $\chi^{(2)}$ nonlinear
 medium and
amplifying one of the ``quadratures'' $(a + a^{\dagger})$ and $i\,(a^{\dagger}-a)$ and reducing the other. Thus,
in applications one often chooses $\theta = \pi/2$ in Eq.\ \eqref{eq:1799} in order to generate squeezed light 
(cf.\ Eq.\ \eqref{eq:1243}).

\paragraph{Squared hermitian amplitudes} \hfill 

The square of the hermitian field mode
\begin{equation}
  \label{eq:1812}
  E= \lambda\,(a\,e^{-i\om\,t} + a^{\dagger}\,e^{i\om\,t})\,,~\lambda \in \mathbb{R}\,,
\end{equation}
may be written in terms of the operators \eqref{eq:1798} as
\begin{equation}
  \label{eq:1813}
  E^2 = 4\lambda^2\,[\tl{K}_0 + \tl{K}_1\,\cos (2\om\,t) - \tl{K}_2\,\sin (2\om\,t)].
\end{equation}
This expression has also been used for the generation of squeezed light \cite{hille1}.

\subsubsection{Interactions bilinear in the $K_j$}

\paragraph{Optical Kerr effect} \hfill

In some materials a light beam has an additional term in its refractive index which is proportional to the intensity
of the light \cite{boyd2}, i.e.\ that extra part of the index is proportional to the square of the electric field. 
Phenomenologically this means that the polarization of the material is proportional to the 3rd power of the electric field,
with a nonlinear coefficient $\chi^{(3)}$. A very simple quantum mechanical model for the associated elementary process
 is given by the interaction term
\begin{equation}
  \label{eq:31}
  g\,\chi^{(3)}\,a^{\dagger}N\,a = g\,\chi^{(3)}a^{\dagger}a^{\dagger}a\,a \propto \chi^{(3)}\tl{K}_+\tl{K}_-\,,
\end{equation}
where, according to Eqs.\ \eqref{eq:1715} and \eqref{eq:259}, the product $\tl{K}_+\tl{K}_-$ can be replaced by
$\tl{K}_0(\tl{K}_0-1) +(3/16){\bf 1}\,$. Thus, the total Hamiltonian can be diagonalized in terms of the number states
$|k, n\rangle\,$, where $k=1/4$ and $=3/4$. 

\paragraph{Degenerate four-wave mixing} \hfill

The simple model interaction Hamiltonian \eqref{eq:31} may also be used in order to describe another optical process in
a non-linear medium with 3rd order susceptibility: Two high intensity classical optical light beams of the same frequency
$\om$ interact with a weak (quantum) beam with frequency $\om$, creating a fourth photon beam, again with the same
 frequency $\om$ and special properties of interest, e.g.\ squeezed light. The process, and the corresponding
 ``nondegenerate one'' mentioned below, is called ``four-wave mixing'' and
played a prominent role in the first stages of light squeezing \cite{scull2}.
The annihilation and creation of two photons are represented by the operators $a$ and $a^{\dagger}$. 

\subsubsection{Two-mode generated Lie algebra $\mathfrak{so}(1,2)$}

A much larger variety of unitary irreducible representations can be generated with two ``canonical'' annihilation and
 creation operators \cite{ka13}:
 \begin{equation}
  \label{eq:2092}
 \tl{K}_+=a_1^{\dagger}a_2^{\dagger}~,
~\tl{K}_-=a_1a_2~,~\tl{K}_0=\frac{1}{2}(a_1^{\dagger}a_1+a_2^{\dagger}a_2+1)\,,
\end{equation}
obey the
commutation relations \eqref{eq:1714}. 

 The tensor product $ \mbox{$\cal
H$}^{osc}_1\otimes \mbox{$\cal H$}^{osc}_2$ of the two harmonic
oscillator Hilbert spaces contains all the irreducible unitary
representations of the group 
 $SU(1,1)\cong SL(2,\mathbb{R})=Sp(2,\mathbb{R})$
 (for which
$k=1/2,1,3/2,\ldots$) in the following way: 

 Let $|n_j\rangle_j,~n_j =0,1,\ldots\,, \, j=1,2,$ be the eigenstates
of the number operators $N_j=a^{\dagger}_ja_j$, generated by $a^{\dagger}_j$ from the
oscillator ground states. 

 Then each of those two subspaces of
 $\mbox{$\cal H$}^{osc}_1\otimes \mbox{$\cal H$}^{osc}_2= \{|n_1
 \rangle_1\otimes
 |n_2\rangle_2\}$ with fixed $|n_1-n_2|\neq 0$  contains an irreducible
  representation
 with Bargmann index
\begin{equation}
  \label{eq:2093}
 k=1/2+|n_1-n_2|/2=1,3/2,2,\ldots,
\end{equation}
i.e.\ the operator $N_1-N_2$ commutes
  with all 3 operators in Eqs.\ \eqref{eq:2092}

 The number
  $n$ in the eigenvalue $n+k$ of $\tl{K}_0$ is given by
\begin{equation}
  \label{eq:2094}
  n=\min\{n_1,n_2\}~ (=0,1,2,\ldots )\,.
\end{equation}
For the ``diagonal'' case $n_2=n_1$ one gets the unitary
   representation with $k=1/2$. 

Inserting the operators \eqref{eq:2092} into the interaction \eqref{eq:1799} yields other examples of associated physical
processes:
\paragraph{Nondegenerate parametric down conversion and amplification}\hfill

In analogy to the degenerate case mentioned above here a classical light beam of frequency $2\om$ generates two
 photons of now {\em different} frequencies $\om_1$ and
 $\om_2$ with $2\om= \om_1 + \om_2$ in a nonlinear medium.
\paragraph{ Nondegenerate four-wave mixing}\hfill

Here the frequencies of the two pump beams and those of the photons are no longer equal. Now the operators $K_+$ and
$K_-$ in the effective Hamiltonian \eqref{eq:31} are replaced by those of Eq.\ \eqref{eq:2092}.

\paragraph{ Mach-Zehnder interferometer}\hfill

The group $SU(1.1) \cong Sp(2, \mathbb{R})$ has played a prominent role in the quantum optical descriptions of the
venerable Mach-Zehnder interferometer \cite{mach}.

\subsubsection{Generation of Barut-Girardello coherent states}

Contrary to the Perelomov coherent states the Barut-Girardello coherent states have not yet produced in the laboratory (to 
the best of my knowledge!). There exist, however,  a number of proposals how to generate them \cite{bar}. One problem is the
 the lack of a {\em unitary} operator analogously to Eq.\ \eqref{eq:1800} as already discussed in subsecs.\ 6.2.3 and 6.2.4.
\subsubsection{Holstein-Primakoff type generators}

The one-mode and the 2-mode versions of the generators $\tl{K}_0,\,\tl{K}_+$ and $\tl{K}_-$ from above can only produce representations
with $k=1/4,\,3/4$ and $k= 1/2,\,1,\,3/2,\,\ldots$. As we are especially interested in representations with small $k<
 1/4$ we have to use corresponding representations. Some of them will be discussed in the next Section. If one wants to
construct those with the help of the usual annihilation and creation operators one can try the nonlinear
 Holstein-Primakoff-type operators \cite{hol1}
 \begin{equation}
   \label{eq:2091}
   \tl{K}_0 = N+k\,,~~~\tl{K}_+ = a^{\dagger}\sqrt{N+2k}\,~~~\tl{K}_-=\sqrt{N+2k}\,a\,,~~~N=a^{\dagger}a\,.
 \end{equation}
Inserted into the interaction term \eqref{eq:1797} and \eqref{eq:1799} one has to find experimental ways in order to
generate a ground state with $k\neq 1/2$ (see also subsec.\ 9.1) and to implement the nonlinear  factor
 $\sqrt{N+2k}\,$ \cite{hol2}.
\subsubsection{Additional proposals for using symplectic groups in quantum optics}
There have been a number of papers with proposals to use symplectic groups $Sp(2n, \mathbb{R})\,,\, n > \,1\,,$ in quantum
optics which are merely quoted here \cite{prop}.

\section{Examples of explicit Hilbert spaces for the $\mathbf{(\vp,I)}$-model of the harmonic oscillator}
\subsection{The case $k=1/2$}
As a first step let us discuss the well-known quantum mechanics of the HO in the framework of concrete irreducible unitary 
representations of the group $Sp(2,\mathbb{R})$ with Bargmann index $k=1/2$ \cite{ka14}, before passing to the more
 general case with
$k \neq 1/2\,$:
\subsubsection{The Hardy space $H^2_+$ on the circle as the Hilbert space for the HO}
The simplest example is the ``Hardy (sub)space'' $H^2_+(S^1, d\vt)$ of the usual Hilbert space $L^2(S^1, d\vt)$ on the
unit circle $S^1$ with the scalar product
\begin{equation}
 \label{eq:1884}
  (f_2,f_1) = \frac{1}{2\pi}\,\int_{S^1}d\vt\,f_2^*(\vt)f_1(\vt)\,,
\end{equation}
and the orthonormal basis
\begin{equation}
  \label{eq:1880}
  e^{i\,n\,\vt}\,,~~ n \in \mathbb{Z}\,.
\end{equation}
The associated Hardy space $H^2_+(S^1,\,d\vt)$ is spanned by the basis consisting of the elements with non-negative $n$,
namely
\begin{equation}
  \label{eq:1881}
  e_n(\vt)= e^{i\,n\,\vt}\,,~~n=0,1,2,\cdots\,.
\end{equation}

 If we have  two Fourier series $\in H^2_+(S^1\,,d\vt)$,
\begin{equation}
  \label{eq:1882}
 f_1(\vt)=\sum_{n=0}^{\infty}a_n\,e^{i\,n\,\vt}\,,~
~f_2(\vt)= \sum_{n=0}^{\infty}b_n\,e^{ i\,n\,\vt}\,,
\end{equation}
they have the scalar product
\begin{equation}
 \label{eq:1883}
 (f_2,f_1)_+=
\frac{1}{2\pi}\int_{S^1} d\vt\,
f^*_2(\vt)f_1(\vt)=\sum_{n=0}^{\infty}b^*_n\,a_n~~.
\end{equation} 

The reproducing kernel 
 here has the form
\begin{equation}
  \label{eq:1898}
 \Delta(\varphi_2, \varphi_1) =
\sum_{n=0}^{\infty} e_n(\varphi_2)^*\,e_n(\varphi_1) =
(1-e^{i\,(\vp_1-\vp_2)})^{-1}\,,
\end{equation} with the usual property
\begin{equation}\label{eq:1899} \frac{1}{2\,\pi}
\int_0^{2\,\pi} d\vp_2 \, \Delta(\varphi_2, \varphi_1)\ e_n(\vp_2)
= e_n(\vp_1)\,.
\end{equation}

The kernel has a singularity (pole) for $\varphi_2 = \varphi_1$.
In calculations one has to replace
$\exp(i\,(\varphi_1-\varphi_2))$ by
$(1-\epsilon)\,\exp(i\,(\varphi_1-\varphi_2))$ and then take the
limit $\epsilon \to 0$ at the end. 

The $Sp(2, \mathbb{R})$ Lie algebra generators  for $k=1/2$ are 
\begin{eqnarray}
\tl{K}_0 &=& \frac{1}{i}\partial_{\vt} + \frac{1}{2}\,,\label{eq:1820} \\
\tl{K}_+ &=&
e^{i\,\vt}\,(\frac{1}{i}\partial_{\vt}+1)= e^{i\,\vt}(\tl{K}_0+\frac{1}{2})\,,\label{eq:1821} \\
\tl{K}_- & =& e^{-i\,\vt}\,\frac{1}{i}\partial_{\vt}=  e^{-i\,\vt}(\tl{K}_0 -\frac{1}{2}) = 
(\frac{1}{i}\partial_{\vt}+1)e^{-i\,\vt}\,. \label{eq:1824}
\end{eqnarray}

The right-hand side of the scalar product \eqref{eq:1883} coincides with the right-hand side of the scalar product
\eqref{eq:1861} for $k= 1/2$. Actually the functions \eqref{eq:1881} of the present Hilbert space $H^2_+(S^1,\,d\vt)$
 may be considered as
 limits of those from Eq.\ \eqref{eq:1855} with $k=1/2$
 for  $|\lambda| \to 1$: For $\lambda = |\lambda|\exp (i\vt)$ the operators \eqref{eq:1866} 
  become the operators \eqref{eq:1820} -- \eqref{eq:1824} in the limit  $|\lambda| \to 1$.

 For the  operators \eqref{eq:1820} -- \eqref{eq:1824} the relations \eqref{eq:1724} -- \eqref{eq:1726} take the form
\begin{eqnarray}
\tl{K}_0\,e_n(\vt) &=& (n+\frac{1}{2})\,e_n(\vt)\,, \label{eq:1822}\\
\tl{K}_+\,e_n(\vt)&=& (n+1)\,e_{n+1}(\vt)\,, \label{eq:1823} \\
\tl{K}_-\,e_n(\vt)&=& n\,e_{n-1}(\vt)\,. \label{eq:1825}\end{eqnarray}

The (dimensionless) Hamilton operator for the $(\vp,I)$-model of the HO now has the extremely simple explicit form
\begin{equation}
  \label{eq:1829}
  \tl{H}(\vec{K})= \tl{K}_0=\frac{1}{i}\partial_{\vt} + \frac{1}{2}\,,
\end{equation}
and the corresponding simple eigenfunctions \eqref{eq:1881}!

I would like to stress  again (like I did in Refs.\ \cite{ka1} and \cite{ka2}) that the mathematical variable $\vt$ used
 here is {\em not}
the canonically conjugate ``observable'' of the operator \eqref{eq:1829}: the angle $\vt$ is not a self-adjoint
 multiplication operator nor is $ \exp (i\,\vt)$ a unitary operator! The self-adjoint observables ``conjugate'' to
$\tl{K}_0$ are the operators $\tl{K}_1$ and $\tl{K}_2\,$!

The composite ladder operators
\begin{eqnarray}
  \label{eq:1826}
  A& =& (\tl{K}_0+1/2)^{-1/2}\tl{K}_- = \tl{K}_-\,(\tl{K}_0-1/2)^{-1/2}=e^{-i\,\vt}(\tl{K}_0-1/2)^{1/2}\,,\\
 A^{\dagger}&=&\tl{K}_+\,(\tl{K}_0+1/2)^{-1/2} = e^{i\,\vt}\,(\tl{K}_0+1/2)^{1/2}\,,\label{eq:1934}
\end{eqnarray}
have the desired properties
\begin{equation}
  \label{eq:1827}
  A\,e_{n}(\vt)= \sqrt{n}\,e_{n-1}(\vt)\,,~~A^{\dagger}\,e_n(\vt) = \sqrt{n+1}\,
e_{n+1}(\vt)\,,
\end{equation}
and, therefore, have the usual matrix elements \cite{mesa}. The same applies, of
course, to those of the composite operators $\tl{Q}$ and $\tl{P}$:
\begin{equation}
  \label{eq:1828}
  \tl{Q}=\frac{1}{\sqrt{2}}\,(A^{\dagger} +A)\,,~~~ \tl{P}=\frac{i}{\sqrt{2}}\,(A^{\dagger} -A)\,.
\end{equation}
Obviously we can reproduce all the quantum physical properties of the HO which - over decades - have been derived by 
means of the operators $\tl{Q}$ and $\tl{P}$ and the $(\tl{q},\tl{p})$-Hamiltonian \eqref{eq:1737}.

The (composite) number operator $N= A^{\dagger}A$ is as expected:
\begin{equation}
  \label{eq:1900}
  N= A^{\dagger}A = \tl{K}_+(\tl{K}_0+1/2)^{-1}\tl{K}_- = e^{i\,\vt}(\tl{K}_0+ 1/2)(\tl{K}_0 +1/2)^{-1}\,e^{-i\,\vt}\,(\tl{K}_0 - 1/2) = \frac{1}{i}\,
\partial_{\vt}\,.
\end{equation}

Remarks:
\begin{itemize}
\item The eigenfunctions \eqref{eq:1881} are periodic:
  \begin{equation}
    \label{eq:1933}
    e_n(\vt + 2\pi) = e_n(\vt)\,,
  \end{equation}
Further below we shall encounter unitarily equivalent quasi-periodic eigenfunctions.
\item The ground state of the Hamiltonian \eqref{eq:1829} is given by the number $1$:
  \begin{equation}
    \label{eq:1932} 
e_{n=0}(\vt) = 1 \,.
  \end{equation}
\item
The probability densities $p_n(\vt)$ associated with the ``number states'' \eqref{eq:1881} are completely flat:
\begin{equation}
  \label{eq:1931}
  p_n(\vt) = 1\,,~n=0,1,\cdots\,.
\end{equation}
\item The number state relations \eqref{eq:1745} -- \eqref{eq:1760} for general $k$ do, of course, hold in the present
special case $k=1/2\,$, too!
\end{itemize}

The time-dependent Schr\"odinger equation for a general state $\psi(\tl{t}, \vt)$ is given by
\begin{equation}
  \label{eq:1901}
  i\,\partial_{\tl{t}}\,\psi (\tl{t}, \vt) = \tl{K}_0\,\psi (\tl{t}, \vt)\,,
\end{equation}
which means that the eigenfunctions \eqref{eq:1881} have the time dependence
\begin{equation}
  \label{eq:1902}
  e_n(\tl{t}, \vt) = e^{-i\,\tl{E}_n\,\tl{t}}\,e_n(\vt) = e^{-i\tl{t}/2}\,e^{i\,n\,(\vt -\tl{t})}\,,~\tl{E}_n= n+1/2\,,
\end{equation}
 and $\psi (\tl{t}, \vt)$ may be expanded as
 \begin{equation}
   \label{eq:61}
   \psi (\tl{t}, \vt) = e^{-i\tl{t}/2}\, \sum_{n=0}^{\infty}c_n\,e^{i\,n\,(\vt -\tl{t})}\,,~~c_n =(e_n,\psi(\tl{t}=0))_+
 \end{equation}

The last two equations show again that the angle $\vt$ plays the role of a time variable (up to a sign) and that the circle
$S^1$ parametrized by $ \vt \in \mathbb{R} \bmod{2\pi}$ becomes ``unwrapped'' onto the time-axis, finitely or infinitely
 many
times, thus realizing an $m$-fold or a universal covering of the circle or of the group $U(1)$!

Introducing the usual quantities with physical dimensions, we get from Eqs.\ \eqref{eq:938} -- \eqref{eq:944}
\begin{eqnarray}
  \label{eq:1903}
  H &=& \hbar\,\omega \tl{K}_0\,,~H e_n(\vt) = E_n\,e_n(\vt)\,, \\E_n&=&\hbar\,\omega\,\tl{E}_n = \hbar\,\omega\,(n+1/2)\,, 
\label{eq:1904}\\
i\hbar\,\partial_t\,\psi(t,\vt)& =& H\,\psi(t,\vt)\,, \label{eq:1905} \\
e_n(t,\vt) &=& e^{-i(E_n/\hbar)\,t}\,e^{i\,n\,\vt} = e^{-i\,\omega\,t/2}\,e^{i\,n\,(\vt-\omega\,t)}\,.\label{eq:1906}
\end{eqnarray}
\subsubsection{Space reflections and time reversal}
According to Subsects.\ 4.5 and 5.1 we can implement the space reflections $\Pi$ and the time reversal $T$ as follows:
\begin{equation}
  \label{eq:1908}
  \Pi\,:~~~ \vt \to \vt \pm \pi\,,
\end{equation}
which implies
\begin{eqnarray}
  \label{eq:1909}
  \partial_{\vt} &\to& \partial_{\vt}\,, \\
\tl{K}_0 &\to& \tl{K}_0\,, \label{eq:1910} \\
K_{\pm} & \to & - K_{\pm}\,, \label{eq:1911} \\
\tl{Q} &\to& - \tl{Q}\,, \label{eq:1916} \\\tl{P} &\to& -\tl{P}\,, \label{eq:1912} \\
e_n(\vt) & \to & e_n(\vt \pm \pi) = (-1)^n\,e_n(\vt)\,. \label{eq:1913}
\end{eqnarray}
The last relation shows that the functions $e_n(\vt)$ have the same symmetry properties under reflections as the usual
Hermite functions \eqref{eq:998}.

Furthermore
\begin{equation}
  \label{eq:1914}
  T\,:~~~ \vt \to -\vt\,,~~i \to -i\,,
\end{equation}
yielding
\begin{eqnarray}
  \label{eq:1915}
  \frac{1}{i}\partial_{\vt} &\to &  \frac{1}{i}\partial_{\vt}\,, \\
\tl{K}_0 &\to& \tl{K}_0\,, \label{eq:1917} \\ K_{\pm} &\to& K_{\pm}\,, \label{eq:1918} \\
A\,,~A^{\dagger} &\to& A\,,~A^{\dagger}\,, \label{eq:1919} \\
\tl{Q} & \to & \tl{Q}\,, \label{eq:1920} \\
 \tl{P} & \to & - \tl{P}\,,\label{eq:1921} \\
e_n(\vt) &\to& [e_n(-\vt)]^* = e_n(\vt)\,.
\end{eqnarray}
\subsubsection{Perturbations}
Like in the classical case (Eqs.\ \eqref{eq:953} -- \eqref{eq:1907}) external time-dependent perturbations of the Hamilton
operator \eqref{eq:1829} can be integrated immediately: Take
\begin{equation}
  \label{eq:1922}
  \tl{H} = \tl{K}_0 + f(\tl{t})\,,
\end{equation}
where $f(\tl{t})$ is a given real function of time. Then the usual product separation of variables gives the following
solution of the time-dependent Schr\"odinger Eq.\
\begin{equation}
  \label{eq:1923}
  i\partial_{\tl{t}}\,\psi(\tl{t}, \vt) = [\tl{K}_0 + f(\tl{t})]\,\psi(\tl{t}, \vt)\,.
\end{equation}
The ansatz
\begin{equation}
  \label{eq:1924}
  \psi(\tl{t}, \vt) = v(\tl{t})\,u(\vt)
\end{equation}
yields
\begin{equation}
  \label{eq:1925}
   \{i\,[\partial_{\tl{t}}\,v(\tl{t})]/v(\tl{t})\} - f(\tl{t}) = \{\frac{1}{i}[\partial_{\vt}\,u(\vt)]/u(\vt)\} +\frac{1}{2}
 = \tl{E}
 = \text{ const. },
\end{equation}
with the (normalized) solution
\begin{equation}
  \label{eq:1926}
  v(\tl{t}) = e^{ -i[\tl{E}\,\tl{t} + \int_0^{\tl{t}}d\tau f(\tau)\,]}\,.
\end{equation}
For $u(\vt)$ we can take
\begin{equation}
  \label{eq:1927}
  u(\vt) = e_n(\vt)\,,~\text{ with } \tl{E} =  n + 1/2\,,
\end{equation}
or appropriate superpositions.

Thus, the perturbation $f(\tl{t})$ causes a time-dependent modification of the phase $\tl{E}\,\tl{t}$.  \\
If 
\begin{equation}
  \label{eq:1936}
 f(\tl{t}) = a = \text{ const.\,, } 
\end{equation}
 then we have 
\begin{equation}
  \label{eq:1928}
 v(\tl{t}) =e^{ -i\,( \tl{E}  +a)\,\tl{t}}\,,
\end{equation}
i.e.\ we have introduced an effective (dynamical) $k \neq 1/2$\,! For an explicit example see subsec.\ 9.1\,.

For the periodic perturbation
\begin{equation}
  \label{eq:1935}
  f(\tl{t}) = \epsilon \cos (\tl{\sigma}\, \tl{t})\,,~ 
\end{equation}
we get for $v(\tl{t})$ the time - dependent phase factor
\begin{equation}
  \label{eq:1937}
  v(\tl{t}) = e^{ -i\,[\tl{E}\,\tl{t} +(\epsilon/\tl{\sigma})\,\sin (\tl{\sigma}\,\tl{t})]}\,.
\end{equation}
Similarly we have for the slightly different  perturbation
\begin{equation}
  \label{eq:1929}
  \tl{K}_0 \to [1 + g(\tl{t})]\,\tl{K}_0
\end{equation}
a corresponding phase factor
\begin{equation}
  \label{eq:1930}
  e^{ -i\tl{E}\,\tl{t}} \to e^{ -i\tl{E}[\tl{t} + \int_0^{\tl{t}}d\tau g(\tau)\,]}\,.
\end{equation}
If one inserts for $g(\tl{t})$ the same expressions as for $f(\tl{t})$ in Eqs.\ \eqref{eq:1936} and \eqref{eq:1935} one
gets the corresponding similar expressions for the phase factor \eqref{eq:1930}.

\subsubsection{A unitary transformation}
The following unitary transformation is of interest, especially later for the more general case $k\neq 1/2$:

In the above description of the states \eqref{eq:1819} and the operators \eqref{eq:1820} -- \eqref{eq:1824} the dependence
on the index $k = 1/2$ is contained in the operators. We shall see below that in the general case we have 
\begin{equation}
  \label{eq:1893}
 \tl{K}_0 = \frac{1}{i}\partial_{\vt} + k\,,~ \tl{K}_+ = e^{i\,\vt}\,(\frac{1}{i}\partial_{\vt}+2k)\,,~
\tl{K}_-  = e^{-i\,\vt}\,\frac{1}{i}\partial_{\vt}\,. 
\end{equation}
The unitary transformation in question is defined by the replacement
\begin{equation}
  \label{eq:1894}
  e_n(\vt) = e^{i\,n\,\vt} \to e_{1/2,\,n}(\vt) = e^{i(n+1/2)\vt}\,,~n=0,1,\cdots\,.
\end{equation}
It shifts the ground state energy characterized by $k= 1/2$ from the Hamiltonian \eqref{eq:1829} to the eigenfunctions
\eqref{eq:1881}.

The operators \eqref{eq:1820} -- \eqref{eq:1824} now take the form
\begin{eqnarray}
  \label{eq:1895}
  \tl{K}_0 &=& \frac{1}{i}\partial_{\vt}\,, \\
\tl{K}_+ &=&
e^{i\,\vt}\,(\frac{1}{i}\partial_{\vt}+1/2)\,,\label{eq:1896} \\
\tl{K}_- & =& e^{-i\,\vt}\,(\frac{1}{i}\partial_{\vt}-1/2)\,.\label{eq:1897}
\end{eqnarray}
The eigenfunctions \eqref{eq:1894} are  only quasi-periodic:
\begin{equation}
  \label{eq:1938}
  e_{1/2,\,n}(\vt+2\pi) = e^{i\,\pi/2}\, e_{1/2,\,n}(\vt)
\end{equation}

The relations \eqref{eq:1822} -- \eqref{eq:1828} remain unchanged.

\subsubsection{Coherent state wave functions  and  their probability densities}

Passing to the $(\vp,I)$-model of the HO and its associated $Sp(2,\mathbb{R})$-structure yields additional
 information, even for $k=1/2$:
\paragraph{Wave functions on $S^1$} \hfill

  We have two additional coherent states: Setting $k=1/2$ and $|k=1/2,n \rangle = e_n(\vt)$ in Eqs.\
\eqref{eq:1765} -- \eqref{eq:1767} the series can be summed immediately, yielding
\begin{equation}
  \label{eq:1830}
  |k=1/2,z \rangle (\vt) \equiv f_z(\vt) = \frac{e^{ z\,e^{i\,\vt}}}{\sqrt{I_0(2|z|)}}
\end{equation}
and
\begin{equation}
  \label{eq:1831}
  |k=1/2,\lambda \rangle (\vt) \equiv f_{\lambda}(\vt) = \frac{(1-|\lambda|^2)^{1/2}}{1-\lambda\,e^{ i\,\vt}}\,.
\end{equation}
These new coherent state functions have all the properties listed in sec.\ 6 for general $k\,$.

The series \eqref{eq:1768} cannot be summed in an elementary way but yields
\begin{eqnarray}
  \label{eq:1832}
  |k=1/2, \alpha \rangle (\vt)&\equiv& f_{\alpha}(\vt) = e^{ -|\alpha|^2 /2}\,\hat{f}_{\alpha}(\vt)\,, 
\\ \hat{f}_{\alpha}(\vt) &=&
\sum_{n=0}^{\infty}\,
\frac{(\alpha\,e^{ i\,\vt})^n}{\sqrt{n!}}=\sum_{n=0}^{\infty}\,
\frac{(|\alpha|\,e^{ i\,(\vt-\beta)})^n}{\sqrt{n!}}\,.\label{eq:2038}
\end{eqnarray}
The function $\hat{f}_{\alpha}$ in Eq.\ \eqref{eq:2038} is an entire function \cite{ent} of its complex argument
\begin{equation}
  \label{eq:2039}
  \zeta = |\alpha|\,e^{i\,(\vt-\beta)}\,.
\end{equation}
The growth of such functions for large $|\alpha|$ has been investigated  for more than a century \cite{ent}.
Application of standard  saddle point methods \cite{sadd} yields for functions like
\begin{equation}
  \label{eq:2040}
  f^{(\rho)}(\zeta) = \sum_{n=0}^{\infty}\frac{\zeta^n}{(n!)^{1/\rho}}
\end{equation}
the following asymptotic expansion \cite{sadd2}
\begin{equation}
  \label{eq:2041}
   f^{(\rho)}(\zeta)  \asymp  \sqrt{\rho}\,(2\pi)^{(1-1/\rho)/2}\,\zeta^{ (\rho-1)/2}e^{ \zeta^{\rho}/\rho}
 \text{ for } |\zeta| \to \infty\,,~ |\arg (\zeta)| \le \frac{\pi}{2\rho} -\epsilon, \epsilon > 0\,. \end{equation}
For that part of the complex plane where the function \eqref{eq:2040} decreases with increasing $|\zeta|$,
 Ref.\ \cite{ev} gives the estimate
\begin{equation}
f^{(\rho)}(\zeta)  \asymp  [1-\sin (\pi/\rho)/\pi]\,\frac{1}{\zeta\,(\ln \zeta)^{ 1/\rho}} \text{ for }
|\zeta| \to \infty\,,~ \frac{\pi}{2\rho} + \epsilon \leq |\arg (\zeta)| \leq \pi\,. \label{eq:2042}
\end{equation}
As the assumptions made in Ref.\ \cite{ev} include the exactly known case $\rho =1$ the estimate \eqref{eq:2042} does not
appear to be a good one!

The limits \eqref{eq:2041} for $\arg (\zeta)$ come from  the requirement $\Re (\zeta^{\rho}) > 0\,$. They also imply
$ \rho \geq 1/2$.
The result for the exponential growth in the sector $|\arg (\zeta)| \le \frac{\pi}{2\rho} -\epsilon$ shows 
 $f^{(\rho)}(\zeta)$ to be of ``order'' $\rho$ and of ``type'' $1/\rho$ there.

The function $\hat{f}_{\alpha}$ from Eq.\ \eqref{eq:2038} has $\rho = 2$ and therefore we get for the wave function
\eqref{eq:1832}
\begin{equation}
  \label{eq:2043}
  f_{\alpha}(\vt) \asymp (2\pi)^{1/4}\,\sqrt{2\,|\alpha|}e^{i\,(\vt-\beta)/2}\,e^{-|\alpha|^2[1-e^{2i(\vt-\beta)}]/2}\,,~
\text{ for } |\alpha| \to \infty\,,~ |\vt-\beta| \le \frac{\pi}{4} -\epsilon\,.
\end{equation} 
\paragraph{Probability densities} \hfill

The probability density of the wave function \eqref{eq:1830} is given by
\begin{equation}
  \label{eq:1885}
  p_z(\vt) = |f_z(\vt)|^2 = \frac{e^{ 2|z|\cos (\vt-\phi)}}{I_0(2|z|)}\,.
\end{equation}
For large $|z|$ we have \cite{knu1}
\begin{equation}
  \label{eq:1886}
  I_0(2|z|) \asymp \frac{e^{ 2|z|}}{2\sqrt{\pi\,|z|}}[1 + O(1/|z|)]\,,
\end{equation}
so that
\begin{equation}
  \label{eq:1887}
  p_z(\vt) \asymp 2\sqrt{\pi\,|z|}\,e^{ -2|z|[1-\cos(\vt-\phi)]}[1+O(1/|z|)]\,.
\end{equation}
For $|\vt -\phi| \ll 1$ the density \eqref{eq:1887} takes (locally) an approximate Gaussian form:
\begin{equation}
  \label{eq:1888}
  p_z(\vt) \asymp 2\sqrt{\pi\,|z|}\,e^{ -|z|(\vt-\phi)^2}[1+O(1/|z|)]\, \text{ for large } |z|\,.
\end{equation}
The last relation shows that for large $|z|$ (the classical limit) the density $p_z(\vt)$ has a sharp peak at
$\vt = \phi = - \arg (z)\,$, so that in the correspondence limit $|z| \to \infty$ the variable $\vt$ approaches the
``classical'' angle $\phi\,$.

As $p_z(\vt+2\pi)=p_z(\vt)$ and $p_z(\vt)$  an even function of $\vt-\phi$ it may be expanded into a Fourier series
 with respect
 to $\cos (n\,\vt)$:
Using the relation \cite{wa}
\begin{equation}
  \label{eq:1889}
  \frac{1}{2\pi}\int_{0}^{2\pi}d\vt\,e^{ 2|z|\,\cos \vt}\cos (n\,\vt) = I_n(2|z|)\,,
\end{equation}
we get
\begin{equation}
  \label{eq:1890}
  p_z(\vt)= \frac{1}{I_0(2|z|)}\,\{I_0(2|z|) + 2\sum_{n=1}^{\infty}I_n(2|z|)\,\cos [n\,(\vt-\phi)]\}\,.
\end{equation}

For the probability density of the wave function \eqref{eq:1831} we have
\begin{equation}
  \label{eq:1891}
  p_{\lambda}(\vt) = \frac{1-|\lambda|^2}{1-2|\lambda|\,\cos (\vt-\theta) +|\lambda|^2}\,.
\end{equation}
It has the properties
\begin{eqnarray}
  \label{eq:1817}
  p_{\lambda}(\vt) &\to& 1 \text{ for } |\lambda| \to 0\,, \\
p_{\lambda}(\vt) &\approx& \frac{\epsilon}{(1-\epsilon)[1-\cos(\vt-\theta)]}~ 
 \text{ for } |\lambda| = 1-\epsilon\,,~ 0 < \epsilon \ll 1\,,~
\cos (\vt - \theta) \neq 1\,,
\label{eq:1818} \\ p_{\lambda}(\vt) &=& 1+ 2\sum_{n=1}^{\infty}\cos n(\vt-\theta)\,|\lambda|^n\,.\label{eq:1819}
\end{eqnarray}
Eq.\ \eqref{eq:1818} shows that for $|\lambda| \to 1^-$ (the classical limit) $p_{\lambda}(\vt)$ is strongly peaked at
$\vt = \theta$.

For calculating the coefficients of the Fourier series \eqref{eq:1819} the relation \cite{grary2a}
\begin{equation}
  \label{eq:1892}
  \int_0^{2\pi} d\vt\,\frac{\cos n\,\vt}{1-2|\lambda|\cos \vt +|\lambda|^2} = \frac{2\pi\,|\lambda|^n}{1-|\lambda|^2}\,,~
|\lambda| < 1\,,
\end{equation}
has been used.

The function \eqref{eq:1891} is well-known in the mathematical literature as the ``Poisson kernel'' $P_{|\lambda|}(\vt-
\phi)$ for the representation of harmonic functions inside the unit disc \cite{con} by functions on the boundary
$\partial\mathbb{D} = S^1$.

The exact probability density $p_{\alpha}(\vt)$ for the wave function \eqref{eq:1832} appears somewhat ``unruly'':
\begin{equation}
  \label{eq:1942}
  p_{\alpha}(\vt) = |f_{\alpha}(\vt)|^2 = e^{ -|\alpha|^2}\,\sum_{n_1,n_2 =0}^{\infty}
\frac{|\alpha|^{ n_1+n_2}}{\sqrt{n_1!\,n_2!}}\,e^{ i\,(\vt-\beta)(n_1-n_2)}\,.
\end{equation}
More instructive is the density for the asymptotic expansion \eqref{eq:2043}:
\begin{equation}
  \label{eq:2044}
  p_{\alpha}(\vt) \asymp 2\sqrt{2\pi}\,|\alpha|\,e^{-|\alpha|^2[1-\cos 2(\vt-\beta)]} \text{ for large } |\alpha|\,,
\end{equation}
which for $|\vt-\beta| \ll 1\,,\,|\alpha|\,|\vt-\beta|$ finite, becomes a Gaussian distribution, too:
\begin{equation}
  \label{eq:2045}
  p_{\alpha}(\vt)  \approx  2\sqrt{2\pi}\,|\alpha|\,e^{-2|\alpha|^2(\vt-\beta)^2} \text{ for } |\vt-\beta| \ll 1\,,~
 |\alpha| \to
\infty\,,~|\alpha|\,|\vt-\beta| \text{ finite }\,.
\end{equation}
As $p_{\alpha}(\vt)$ is a periodical and even function of  $\vt-\beta$ it may be Fourier expanded,
but the result does not appear to be very instructive.  

\subsubsection{Expectation values and transition probabilities}
All the properties of the 3 types of coherent states listed in sec.\ 6 for general $k$ do hold, of course, for the
special value $k = 1/2$, too. I, therefore, mention here just a few  special features: 

We have (cf.\ Eqs.\ \eqref{eq:1766} and \eqref{eq:1771})
\begin{equation}
  \label{eq:1949}
  g_{1/2}(|z|^2) = I_0(2|z|)\,,~\rho_{1/2}(|z|) = \frac{I_1(2|z|)}{I_0(2|z|)}\,.
\end{equation}
Remarkable is that now (cf.\ Eq.\ \eqref{eq:1833})
\begin{equation}
  \label{eq:1950}
\langle N^2 \rangle_{1/2,z} = |z|^2\,,  
\end{equation}
which provides a direct ``measurement'' of the modulus $|z|$.
For the transition probabilities \eqref{eq:1876} and \eqref{eq:1863} we get
\begin{equation}
  \label{eq:1951}
  p (1/2,\,n \leftrightarrow z)  = \frac{|z|^{2n}}{(n!)^2\,I_0(2|z|)} \asymp \frac{2\sqrt{\pi}|z|^{2n +1/2}}{
(n!)^2}\,e^{-2|z|} \text{ for } |z| \to \infty\,~.
\end{equation}
\begin{equation}
  \label{eq:1952}
  p (1/2,\,n \leftrightarrow \lambda) = (1-|\lambda|^2)\,|\lambda|^{2n}= \frac{1}{\bar{n}_{\lambda}+1}\left(\frac{\bar{n}_{
\lambda}}{\bar{n}_{\lambda}+1}\right)^n\,,~\bar{n}_{\lambda} \equiv \bar{n}_{1/2,\,\lambda}\,.
\end{equation}
The last probability may (formally) be interpreted in the context of Bose-Einstein statistics \cite{be}: Assume that
 a system of free
 Bose-Einstein quanta has distinct energy levels $E_{\nu}\,,\,\nu = 0,1,\ldots$ and is in a heat bath with inverse
temperature $\beta =1/(k_BT)$ and chemical potential $\mu$. Then
\begin{equation}
  \label{eq:1953}
  (1-|\lambda|^2)\,|\lambda|^{2n}\,,~|\lambda|^2 = e^{-\beta(E_{\nu}-\mu)}
\end{equation}
is the probability to find $n$ quanta in a state with energy $E_{\nu}$.

As already mentioned previously the distribution \eqref{eq:1952} has been verified experimentally \cite{kum}.

From Eq.\ \eqref{eq:1801} we get
\begin{equation}
  \label{eq:1954}
  (f_{\lambda},f_{z})_+ = \frac{(1-|\lambda|^2)^{1/2}}{\sqrt{I_0(2|z|)}}\,e^{\lambda^*\,z}\,,~|(f_{\lambda},f_{z})_+|^2 =
\frac{1-|\lambda|^2}{I_0(2|z|)}\,e^{2|\lambda|\,|z|\,\cos(\phi-\theta) }\,.
\end{equation}
Furthermore \cite{ka9}
\begin{eqnarray}
  \label{eq:1955}
 (f_{\alpha},f_{z})_+ &=& \frac{e^{\ds -|\alpha|^2/2}}{\sqrt{I_0(2|z|)}}\,\sum_{n=0}^{\infty}\frac{(\alpha^*\,z)^n}{
(n!)^{3/2}}\,,\\
(f_{\alpha},f_{\lambda})_+&=& e^{\ds -|\alpha|^2/2}\,(1-|\lambda|^2)^{1/2}\,\sum_{n=0}^{\infty}\frac{(\alpha^*\,\lambda)^n}{
\sqrt{n!}}\,. \label{eq:1956}
\end{eqnarray}
In evaluating the series \eqref{eq:1955} and \eqref{eq:1956} we encounter the same problems as for the series
\eqref{eq:2038}. The asymptotic expansion \eqref{eq:2041} yields for the transition probabilities
\begin{eqnarray}
  \label{eq:1939}
 |(f_{\alpha},f_{z})_+|^2 &\asymp & \frac{2}{3\sqrt{2\pi}\,|\alpha\,z|^{1/3}}\,\frac{e^{-|\alpha|^2}}{I_0(2|z|)}\,
e^{3|\alpha\,z|^{2/3}\,\cos[2(\beta-\phi)/3)]} \\ && \text{ for large } |\alpha\,z|\,,~|\beta-\phi| \leq 3\pi/4
 -\epsilon\,, \nonumber \\
|(f_{\alpha},f_{\lambda})_+|^2 & \asymp & 2\sqrt{2\pi}|\lambda\,\alpha|\,e^{-|\alpha|^2}\,(1-|\lambda|^2)\,
e^{|\lambda\,\alpha|^2\,
\cos 2(\beta-\theta)}\\ && \text{ for large } |\alpha|\,, |\beta-\theta| \leq \pi/4 -\epsilon\,.\label{eq:1940} \nonumber
\end{eqnarray}
\subsubsection{Eigenfunctions of $\tl{K}_1$ and $\tl{K}_2$}
Like the operators $\tl{Q}$ and $\tl{P}$, which as generators of non-compact groups in general have a continuous spectrum,
the self-adjoint operators $\tl{K}_1$ and $\tl{K}_2$ as generators of non-compact groups have a real continuous spectrum.
 Their
 ``eigenfunctions'' may be determined as solutions of differential eqs.:

It follows from Eqs.\ \eqref{eq:1821} and \eqref{eq:1824} that
\begin{equation}
  \label{eq:1957}
  \tl{K}_1 = \frac{1}{2}\,(\tl{K}_+ +\tl{K}_-) = \cos \vt \frac{1}{i}\partial_{\vt} +\frac{1}{2}e^{i\vt}
\,,~ \tl{K}_2 = \frac{1}{2i}\,(\tl{K}_+ - \tl{K}_-) = \sin \vt \frac{1}{i}\partial_{\vt} +\frac{1}{2i}e^{i\vt}\,.
\end{equation}
It is helpful to observe that $\tl{K}_1$ is obtained from $\tl{K}_2$ by the substituation $\vt \to \vt + \pi/2$.
The eigenvalue equation
\begin{equation}
  \label{eq:1958}
  \tl{K}_2\,f_{h_2}(\vt) = h_2\,f_{h_2}(\vt)\,,~h_2 \in \mathbb{R}\,,
\end{equation}
leads to
\begin{equation}
  \label{eq:1959}
  (\partial_{\vt}f_{h_2})/f_{h_2} = \frac{i\,h_2}{\sin \vt}-\frac{1}{2}\,(\cot \vt +i)\,.
\end{equation}
As $\sin \vt$ and $\tan(\vt/2)$ are positive in the (open) interval $(0, \pi)$ and negative in $(\pi, 2\pi)$ one has
 to treat the
 two intervals
slightly differently. For the first interval we get
\begin{eqnarray}
  \label{eq:1960}
  f_{h_2}(\vt) &=& C_1 e^{-i\vt/2}(\sin\vt)^{-1/2}\,[\tan(\vt/2)]^{i\,h_2}\\ &= & \frac{C_1}{\sqrt{2}}\,e^{-i\vt/2}
\,[\sin (\vt/2)]^{i\,h_2 -1/2}\,[\cos (\vt/2)]^{-i\,h_2-1/2}\,, \nonumber \\ 
&& C_1 = \text{const.}\,,~\vt \in (0, \pi)\,. \nonumber
\end{eqnarray}
For the second we get, with $ (\sin \vt)^{-1/2} = e^{i\pi/2}(|\sin \vt|)^{-1/2}$ and $\ln \tan (\vt/2) = \ln |\tan (\vt/2)| +
i\pi$,
\begin{eqnarray}
  \label{eq:1961}
  f_{h_2}(\vt) &=& C_2 \,e^{-i\pi/2}\,e^{-\pi\,h_2}\,e^{-i\vt/2}(|\sin\vt|)^{-1/2}\,[|\tan(\vt/2)|]^{i\,h_2}\\
&=&\frac{C_2}{\sqrt{2}} \,e^{-i\pi/2}\,e^{-\pi\,h_2}\,e^{-i\vt/2}\,[\sin (\vt/2)]^{i\,h_2 -1/2}\,[|\cos (\vt/2)|]^{-i\,h_2-1/2}\,, \nonumber \\
&& C_2 = \text{const.}\,,~\vt \in (\pi, 2\pi)\,. \nonumber
\end{eqnarray}
The three constant factors in the last expression may be combined to $C_1 =C_2 \,e^{-i\pi/2}\,e^{-\pi\,h_2}$. 
For $\vt = 0,\pi$ the functions \eqref{eq:1960} become singular, so do the functions \eqref{eq:1961} for $\vt = \pi, 2\pi$.
The constant
$C_1$ can be determined like in the case of plane waves:
Substituting
\begin{equation}
  \label{eq:1962}
  u(\vt) = \ln [\tan (\vt/2)]\,,~du = \frac{d\vt}{\sin \vt}\,,~ u(\vt \to 0^+) \to -\infty\,,~u(\vt \to \pi^-)
 \to +\infty\,,
\end{equation}
into
\begin{equation}
  \label{eq:1963}
  \frac{1}{2\pi}\int_{0}^{\pi}d\vt\,f_{h_2^{\prime}}^*(\vt)\,f_{h_2}(\vt) = \frac{|C|^2}{2\pi}\int_0^{\pi}
\frac{d\vt}{\sin \vt}[\tan (\vt/2)]^{i(h_2-h_2^{\prime})}
\end{equation}
yields
\begin{equation}
  \label{eq:1964}
   \frac{1}{2\pi}\int_{0}^{\pi}d\vt\,f_{h_2^{\prime}}^*(\vt)\,f_{h_2}(\vt) = \frac{|C|^2}{2\pi}\int_{-\infty}^{\infty}du\,
e^{iu\,(h_2 -h_2^{\prime})} =
|C_1|^2 \delta (h_2 -h_2^{\prime})\,.
\end{equation}
The interval $(\pi, 2\pi)$ gives the same contribution, 
so that the ``normalized'' eigenfunctions of $\tl{K}_2$ are 
\begin{eqnarray}
  \label{eq:1965}
  f_{h_2}(\vt) &=& e^{-i\vt/2}(2|\sin\vt|)^{-1/2}\,[|\tan(\vt/2)|]^{i\,h_2}\\
&=& \frac{1}{2}\,e^{-i\vt/2}
\,[\sin (\vt/2)]^{i\,h_2 -1/2}\,[\cos (\vt/2)]^{-i\,h_2-1/2}\nonumber
\,, \\ && \vt \in (0,\pi)\,,(\pi, 2\pi)\,,~h_2 \nonumber
\in \mathbb{R}\,.
\end{eqnarray} 
Implementing the substitution $\vt + \pi/2$ we get - up to an irrelevant phase factor - the eigenfuctions of $\tl{K}_1$:
\begin{eqnarray}
  \label{eq:1966}
   f_{h_1}(\vt)& =& e^{-i\vt/2}(2|\cos\vt|)^{-1/2}\,[|\tan(\vt/2 +\pi/4)|]^{i\,h_1}\,,\\
&&\vt \in (-\pi/2 ,\pi/2)\,,
(\pi/2, 3\pi/2)\,,~h_1\in \mathbb{R}\,,\nonumber \\
&& \tan(\vt/2+\pi/4)=(\sin \vt +1)/\cos \vt\,.~~ \nonumber
\end{eqnarray}

For the coefficients $c_n$ in the expansion
\begin{equation}
  \label{eq:1967}
  f_{h_2}(\vt) = \sum_{n=0}^{\infty}c_n\,e^{i\,n\,\vt}
\end{equation}
one gets \cite{erdI2}
\begin{eqnarray}
  \label{eq:1968}
  c_n&=& \frac{1}{2\pi}\,\int_0^{2\pi}d\vt\,f_{h_2}(\vt)\,e^{-i\,n\,\vt} \\ & =&  \frac{1}{2\pi}\,\int_0^{\pi}
d\vp\,(\cos \vp)^{-i\,h_2 -1/2}\,(\sin \vp)^{i\,h_2 -1/2}\,e^{-2i(n+1/2)\,\vp}\nonumber  \\ &=&
e^{-i\,\pi(n+1/4 -i\,h_2/2)}\,\frac{\Gamma (1/2 +i\,h_2)}{n!\,\Gamma (1/2 -n +i\,h_2)}\,F(1/2+i\,h_2, -n; 1/2-n +i\,h_2;
z=-1) \nonumber \\
&=& (-1)^n\,e^{-i\pi/4}\,e^{-h_2/2}\,\sum_{m=0}^n(-1)^m\,\frac{\Gamma (1/2+i\,h_2+m)\,(-n)_m}{\Gamma (1/2+i\,h_2 -n
 +m)\,m!}\,,
 \nonumber \end{eqnarray}
where
\begin{equation} \label{eq:1969}
F(a, b; c;z)= \sum_{m=0}^{\infty} \frac{(a)_m\,(b)_m}{(c)_m\,m!}\,z^m
\end{equation}
is the standard series for the hypergeometric function.

The relation \eqref{eq:1968} holds for $h_2 > 0$. For $h_2 < 0$ one has to replace $h_2$ in Eq.\ \eqref{eq:1968} by
$|h_2|$. 

Examples:
\begin{equation}
  \label{eq:1970}
  c_0 = e^{-i\pi/4}\,e^{-\pi\,h_2/2}\,,~|c_0|^2 = e^{-\pi\,h_2}\,;~~ c_1 =-2i\, e^{-i\pi/4}\,h_2\,
e^{-\pi\,h_2/2}\,,~|c_1|^2 = 4\,h_2^2\,e^{-\pi\,h_2}\,.
\end{equation}
\subsubsection{Relationship to the conventional description of the HO on $L^2(\mathbb{R},\,dx)$}
The relationship between the quantum mechanical description of the HO in the above Hilbert space $H_+^2(S^1,\,d\vt)$
and the usual one on $L^2(\mathbb{R},\,d\xi)$ has been discussed in some detail in chap.\ 4 of Ref.\ \cite{ka1}. I here
merely summarize the main steps:
\begin{enumerate} \item  The space $H_+^2(S^1,\,d\vt)$ is mapped unitarily onto the Hardy space
 $H_+^2(\mathbb{R},\,d\xi)$ of the real
line, the elements of which are boundary values $\lim_{\eta \to 0^+}g(z=\xi +i\,\eta)$  of functions which are holomorphic
 in the upper half ($\eta >0$) of the complex plane.  
\item The space $L^2(\mathbb{R},\,d\xi)$ is projected on  $H_+^2(\mathbb{R},\,d\xi)$ by the following Fourier tranformations
  \begin{eqnarray}
    \label{eq:2089}
    \hat{g}(p) &=& \frac{1}{\sqrt{2\pi}}\int_{-\infty}^{\infty}d\xi\,g(\xi)\,e^{-i\xi p}\,,~~~g(\xi) \in L^2(\mathbb{R},
\,d\xi)\,, \\
g^{(+)}(\xi)&=& \frac{1}{\sqrt{2\pi}}\int_{0}^{\infty}dp\,\hat{g}(p)\,e^{ip\xi}\,,~~~g^{(+)}(\xi) \in 
H_+^2(\mathbb{R},\,d\xi)\,.\label{eq:2169}
  \end{eqnarray}
\end{enumerate}
\subsection{The general case $k >0$}
In case nature ``allows'' for quantized harmonic oscillators with ground state energies for which $k \neq 1/2\,$, 
especially $k \in (0,1/2)\,$,  then one
needs corresponding Hilbert spaces for the description of such systems. I shall briefly mention three examples which may be
 useful and which are all unitarily equivalent: The Hilbert space of holomorphic functions on the unit circle as
 described by the Eqs.\ \eqref{eq:1854} -- \eqref{eq:1866} in the subsection 6.4.3 above, Hilbert spaces associated with the
Hardy space on the circle given by Eqs.\ \eqref{eq:1881} -- \eqref{eq:1883} and the Hilbert space $L^2([0, \infty), du)$
on the positive real line with Laguerre's functions as basis.

One can use the Hardy space \eqref{eq:1854} -- \eqref{eq:1866} itself by using a Holstein-Primakoff variant \cite{hol1}
for the Lie algebra generators
\begin{eqnarray}
  \label{eq:1971}
  \tl{K}_0 &=& \frac{1}{i}\,\partial_{\vt}+k\,, \\ \tl{K}_+ &=& e^{i\,\vt}\,[(N+2k)(N+1)]^{1/2}\,,~N=\frac{1}{i}\,
\partial_{\vt} \,.\label{eq:1972} \\
\tl{K}_- &=& [(N+2k)(N+1)]^{1/2}\,e^{-i\,\vt}\,. \label{eq:1973}
\end{eqnarray}
These operators have the properties \eqref{eq:1724} -- \eqref{eq:1726} when applied to the basis \eqref{eq:1819} and one
has $(f_2,\tl{K}_+f_1)_+ = (\tl{K}_-f_2,f_1)$ for functions \eqref{eq:1882}. For $k=1/2$ the operators
 \eqref{eq:1971}-\eqref{eq:1973}
reduce to the ones in Eqs.\  \eqref{eq:1820} -- \eqref{eq:1824}. For $k\neq 1/2$ the roots in the expressions
 \eqref{eq:1972} and \eqref{eq:1973}
become cumbersome and unpleasent to deal with. They will not be discussed here further. They might, however, be quite
useful under certain circumstances.
\subsubsection{Hilbert space of holomorphic functions on the unit disc}
In subsec.\ 6.4.3 above I have indicated  in connection with Eqs.\ \eqref{eq:1862} and \eqref{eq:1866} that the Hilbert
space of holomorphic functions on the unit disc $\mathbb{D} = \{\lambda \in \mathbb{C}\,,~|\lambda| <1\}$ with the
scalar product \eqref{eq:1854} can provide irreducible unitary representations of the group $SO^{\uparrow}(1,2)$ and all
its covering groups with $k>0$, the self-adjoint generators given by Eq.\ \eqref{eq:1899} (see also Appendix B).

The complex numbers $\lambda \in \mathbb{D}$ were introduced in Eq.\ \eqref{eq:1762} as eigenvalues of the operator
 \eqref{eq:1763}. It appears helpful to introduce a new complex variable $\om \in \mathbb{D}$ (not to be confused with the
circular frequency) in order to distinguish
the Hilbert space variable in Eqs.\ \eqref{eq:1854} -- \eqref{eq:1866} from the eigenvalue $\lambda$. So we have
\begin{eqnarray}
  \label{eq:1974}
  (f_2,f_1)_{k,\om} &\equiv& \int_{\mathbb{D}}d\tl{\mu}_k(\om)\,f_2^*(\om)f_1(\om)\,,\\
 d\tl{\mu}_k(\om)&=& \frac{2k-1}{\pi}
(1-|\om|^2)^{2k-2}\,|\om|d|\om|\,d\theta\,, \nonumber \\
\tl{e}_{k,n}(\om) &=& \sqrt{\frac{(2k)_n}{n!}}\,\om^n\,,~(\tl{e}_{k,n_2},\tl{e}_{k,n_1})_{k, \om} = \delta_{n_2\,n_1}\,,
\label{eq:1975}\\ 
(f_2,f_1)_{k,\om} = \sum_{n=0}^{\infty} \frac{n!}{(2k)_n}\,b_{n,2}^*\,b_{n,1}\,,&&f_j(\om) = \sum_{n=0}^{\infty}b_{n,j}
\,\om^n\,,~j=1,2\,,\label{eq:1976} 
\end{eqnarray}
 and
 \begin{equation}
   \label{eq:1977}
   \tl{K}_0 =\om\frac{d}{d\om}+k\,,~~ 
  \tl{K}_+ = \om\,(2k +
\om\frac{d}{d\om})\,,~~\tl{K}_- = \frac{d}{d\om}\,,
 \end{equation}
with the usual properties
\begin{eqnarray}
  \label{eq:1978}
  \tl{K}_0\,\tl{e}_{k,n}&=&(n+k)\tl{e}_{k,n}\,, \\\tl{K}_+\,\tl{e}_{k,n}&=& \sqrt{(2k+n)(n+1)}\,\tl{e}_{k,n+1}\,,\label{eq:1980} \\
\tl{K}_-\,\tl{e}_{k,n}&=& \sqrt{(2k+n-1)n}\,\tl{e}_{k,n-1}\,.\label{eq:1979}
\end{eqnarray}
The associated ladder operators
\begin{equation}
  \label{eq:1984}
  A = (\tl{K}_0+k)^{-1/2}\,\tl{K}_-\,,~~A^{\dagger} = \tl{K}_+\,(\tl{K}_0+k)^{-1/2}
\end{equation}
have the conventional {\em k-independent} Fock space properties
\begin{equation}
  \label{eq:1985}
  A\,\tl{e}_{k,n} = \sqrt{n}\,\tl{e}_{k,n-1}\,,~~ A^{\dagger}\,\tl{e}_{k,n} = \sqrt{n+1}\,\tl{e}_{k,n+1}\,.
\end{equation}
Inserting the number state basis functions \eqref{eq:1975} into the right-hand sides of the Eqs.\ \eqref{eq:1765},
\eqref{eq:1767} and \eqref{eq:1768} yields the coherent state functions of $\om$:
\begin{eqnarray}
  \label{eq:1981}
  |k,z\rangle (\om) \equiv f_{k,z}(\om)&=& \frac{e^{ z\,\om}}{\sqrt{g_k(|z|^2)}}\,, \\
 |k,\lambda \rangle (\om) \equiv f_{k,\lambda}(\om)&=& \frac{(1-|\lambda|^2)^k}{(1-\lambda\,\om)^{2k}} \,,\label{eq:1982}\\
 |k,\alpha\rangle (\om) \equiv f_{k,\alpha}(\om)&=&e^{ -|\alpha|^2/2}\,\sum_{n=0}^{\infty}\frac{\sqrt{(2k)_n}}{n!}\,
(\alpha\,\om)^n\,.\label{eq:1983}
\end{eqnarray}
The general properties of the three types of  coherent states as discussed in subsecs.\ 6.1 and 6.2 are, of course, 
here valid, too, and will not be repeated.
\subsubsection{Hilbert spaces related to the Hardy space on the circle}
The scalar product \eqref{eq:1976} as a series can be implemented on the Hardy space $H_+^2(S^1,d\vt)$
  in the following way: 

Let us introduce \cite{ka15} the following positive definite (self-adjoint) operator $A_k$ by
\begin{equation}
  \label{eq:1986}
  A_k\,e_n(\vt)= \frac{n!}{(2k)_n}\,e_n(\vt)\,,~~e_n(\vt) =e^{i\,n\,\vt}\,,~n=0,1,\ldots\,.
\end{equation}
Then we can define an additional scalar product for functions
\begin{equation}
  \label{eq:1987}
  f_j(\vt)=\sum_{n=0}^{\infty}c_{n,j}e_n(\vt)\,,~j=1,2,
\end{equation}
by
\begin{equation}
  \label{eq:1988}
  (f_2,f_1)_{k,+} \equiv (f_2,A_k\,f_1)_+ = \sum_{n=0}^{\infty}\frac{n!}{(2k)_n}\,c^*_{n,2}\,c_{n,1}\,.
\end{equation}
The series here is obviously of the same type as the one in Eq.\ \eqref{eq:1976}. As
\begin{equation}
  \label{eq:1989}
  \frac{n!}{(2k)_n}  = \left\{\begin{array}{rcl} <1 & \text{ for }& k > 1/2\,,~n>0 \\
=1 & \text{ for } & k= 1/2\,, \\ > 1 & \text{ for }& 0 < k < 1/2\,, n>0\,, \end{array} \right.
\end{equation}
one might suspect that these coefficients affect the convergence properties of the series \eqref{eq:1988}.
However, as
\begin{equation}
  \label{eq:1990}
  \lim_{n \to \infty}\left(\frac{n!}{(2k)_n}\right)^{1/n} = 1~ \text{ for }~ k >0,
\end{equation}
the radius of convergence of that series is the same with or without the factor \eqref{eq:1989} (according to the
Cauchy criterium \cite{whit})!

Let us denote the (Hardy space associated) Hilbert space with the scalar product \eqref{eq:1988} by $H^2_{k,+}(S^1,d\vt)$.
An orthonormal basis in this Hilbert space is given by
\begin{equation}
  \label{eq:1991}
  \hat{e}_{k,n}(\vt)= \sqrt{\frac{(2k)_n}{n!}}\,e_n(\vt)\,,~( \hat{e}_{k,n_2}, \hat{e}_{k,n_1})_{k,+} = \delta_{n_2n_1}\,.
\end{equation}
From the expressions \eqref{eq:1977} one can infer (taking the limit $\om \to \exp (i\,\vt)$ that
\begin{equation}
  \label{eq:1992}
  \tl{K}_0 = \frac{1}{i}\,\partial_{\vt}+ k\,,~\tl{K}_+=e^{i\,\vt}(\frac{1}{i}\,\partial_{\vt}+ 2k)\,,~\tl{K}_-= e^{-i\,\vt}\,
\frac{1}{i}\,\partial_{\vt}\,,
\end{equation}
with the right properties for the basis \eqref{eq:1991}:
\begin{eqnarray}
  \label{eq:1993}
  \tl{K}_0\,\hat{e}_{k,n}&=& (n+k)\,\hat{e}_{k,n}\,,\\ \tl{K}_+\,\hat{e}_{k,n}&=& \sqrt{(2k+n)(n+1)}\,\hat{e}_{k,n+1}\,,
\label{eq:1994}
\\ \tl{K}_-\,\hat{e}_{k,n} &=& \sqrt{(2k+n-1)n}\,\hat{e}_{k,n-1}\,. \label{eq:1995}
\end{eqnarray}
The operators \eqref{eq:1992} do not have these properties with respect to the basis $e_n(\vt)$!
 Correspondingly the operators $\tl{K}_+$ and $\tl{K}_-$
are adjoint to each other only with respect to the scalar product \eqref{eq:1988}, not with respect to \eqref{eq:1883}.
Their adjointness as  to \eqref{eq:1988} can be verified by taking two series
\begin{equation}
  \label{eq:1996}
  f_j(\vt)= \sum_{n=0}^{\infty}a_{n,j}\,\hat{e}_{k,n}(\vt)\,,~j=1,2,
\end{equation}
and showing that $(\tl{K}_-f_2,f_1)_{k,+} = (f_2,\tl{K}_+f_1)_{k,+}$!

Note that
\begin{eqnarray}
  \label{eq:2005}
  (e_{n_2},\hat{e}_{k,n_1})_+ &=& (\hat{e}_{k,n_1},e_{n_2})_+ = \sqrt{\frac{(2k)_{n_1}}{n_1!}}\,\delta_{n_2\,n_1}\,; \\
(e_{n_2},\hat{e}_{k,n_1})_{k,+}& =& (\hat{e}_{k,n_1},e_{n_2})_{k,+} = \sqrt{\frac{n_1!}{(2k)_{n_1}}}\,\delta_{n_2\,n_1}\,;
\label{eq:2006} \\(\hat{e}_{k,n_2},\hat{e}_{k,n_1})_+ &=& \frac{(2k)_{n_1}}{n_1!}\,\delta_{n_2\,n_1}\,, \;\;\;\;
 (e_{n_1},e_{n_2})_{k,+} = \frac{n_1!}{(2k)_{n_1}}\,\delta_{n_2\,n_1}\,.\label{eq:2007}
\end{eqnarray}

The Fock space ladder operators $A$ and $A^{\dagger}$ associated with the Lie algebra generators \eqref{eq:1992} are
given in the same way as in Eq.\ \eqref{eq:1984}. 
\paragraph{Coherent state wave functions} \hfill

Analogously to the relations \eqref{eq:1981} -- \eqref{eq:1983} we obtain on $H^2_{k,+}$ the following coherent state
wave functions by using the basis \eqref{eq:1991}:
\begin{eqnarray}
  \label{eq:1997}
  |k,z\rangle (\vt) \equiv f_{k,z}(\vt)&=& \frac{e^{ z\,e^{i\,\vt}}}{\sqrt{g_k(|z|^2)}}\,, \\
 |k,\lambda \rangle (\vt) \equiv f_{k,\lambda}(\vt)&=& \frac{(1-|\lambda|^2)}{(1-\lambda\,e^{ i\,\vt})^{2k}}
 \,, \label{eq:1998}\\
 |k,\alpha\rangle (\vt) \equiv f_{k,\alpha}(\vt)&=&e^{ -|\alpha|^2/2}\,\sum_{n=0}^{\infty}\frac{\sqrt{(2k)_n}}{n!}\,
(\alpha\,e^{ i\,\vt})^n\,.\label{eq:1999}
\end{eqnarray}

The reproducing kernel on $H^2_{k,+}$ is given by
\begin{equation}
  \label{eq:2000}
  \hat{A}_k(\vt_2-\vt_1) = \sum_{n=0}^{\infty}\hat{e}^*_{k,n}(\vt_2)\,\hat{e}_{k,n}(\vt_1) =
 [1-e^{ i\,(\vt_1-\vt_2)}]^{-2k}= \hat{A}_k^*(\vt_1-\vt_2)\,.
\end{equation}
According to the relations \eqref{eq:2005} -- \eqref{eq:2007} it has the properties 
\begin{eqnarray}
  \label{eq:2008}
  (\hat{A}_k(1,2),\,\hat{e}_{k,m}(2))_{k,+}&=& \hat{e}_{k,m}(\vt_1)\,,\\
 (\hat{A}_k(1,2),\,\hat{e}_{k,m}(2))_{+}&=& \frac{(2k)_m}{m!}\, \hat{e}_{k,m}(\vt_1)\,, \label{eq:2009} \\
(\hat{A}_k(1,2),\,e_{m}(2))_{k,+}&=& \sqrt{\frac{m!}{(2k)_m}}\,\hat{e}_{k,m}(\vt_1)= e_m(\vt_1)\,,\label{eq:2010} \\
(\hat{A}_k(1,2),\,e_{m}(2))_{+}&=& \sqrt{\frac{(2k)_m}{m!}}\,\hat{e}_{k,m}(\vt_1)= \frac{(2k)_m}{m!}\,e_m(\vt_1)\,.
\label{eq:2011}
\end{eqnarray}
The numbers $1$ and $2$ mean the variables $\vt_1$ and $\vt_2$, the latter being integration variable.

\paragraph{A unitary transformation}\hfill

In the above discussion the $k$-dependence of the representation is contained in the operators \eqref{eq:1992}, not
in the basis $e_n(\vt)$ of $H^2_+$ we started from. Like in subsection 7.1.4 one can shift the $k$-dependence partially
 from the operators to the basis by a unitary transformation:
\begin{equation}
  \label{eq:2012}
  e_n(\vt) = e^{i\,n\,\vt} \to e_{k,n}(\vt) = e^{i\,(n+k)\,\vt}\,,
\end{equation}
the  generators \eqref{eq:1992} now taking the form
\begin{equation}
  \label{eq:2013}
   \tl{K}_0 = \frac{1}{i}\,\partial_{\vt}\,,~\tl{K}_+=e^{i\,\vt}(\frac{1}{i}\,\partial_{\vt}+ k)\,,~\tl{K}_-= e^{-i\,\vt}\,
(\frac{1}{i}\,\partial_{\vt}-k)\,.
\end{equation}
The operators \eqref{eq:2013} act in a Hilbert space $H^2_{k,+}$, now with the orthonormal basis
\begin{equation}
  \label{eq:2014}
  \hat{e}_{k,n}(\vt) =\sqrt{\frac{(2k)_n}{n!}}\,e_{k,n}(\vt)\,.
\end{equation}
The basis functions \eqref{eq:2012} are no longer periodic but quasi-periodic:
\begin{equation}
  \label{eq:2015}
  e_{k,n}(\vt +2\pi) = e^{2i\,k\,\pi}\,e_{k,n}(\vt)\,.
\end{equation}
These functions are special Bloch-type wave functions on the circle \cite{ka2}\,.
\subsubsection{Hilbert space on the positive real line}

There exists a unitary mapping \cite{barg3} from the Hilbert space of holomorphic functions on the unit
 disc as characterized by
the Eqs.\ \eqref{eq:1974} and \eqref{eq:1975} to the Hilbert space $L^2(\mathbb{R}_+,du)$, where $\mathbb{R}_+ =
 [0, \infty)$, i.e.\ we have the scalar product
 \begin{equation}
   \label{eq:2016}
   (f_2,f_1) = \int_0^{\infty}du\,f_2^*(u)\,f_1(u)
 \end{equation}
for functions $f(u)$ on $\mathbb{R}_+$. The standard orthonormal basis on this space are Laguerre's functions \cite{erdII1},
slightly adapted for our purposes,
\begin{equation}
  \label{eq:2001}
  \br{e}_{k\,,n}(u) = \sqrt{\frac{n!}{\Gamma (2k)\,(2k)_n}}\,u^{ k-1/2}\,e^{-u/2}\,L^{2k-1}_n(u)\,,~k > 0\,, 
\end{equation}
where the functions $L^{\alpha}_n(u)$ are Laguerre's polynomials
\begin{equation}
  \label{eq:2002}
  L^{\alpha}_n(u) = \sum_{m=0}^{m=n} \binom{n+\alpha}{n-m}\,\frac{(-u)^m}{m!}\,,~~L^{\alpha}_n(0) = 
\frac{(\alpha+1)_n}{n!}\,.
\end{equation}
These have the generating function \cite{erdII1}
\begin{equation}
  \label{eq:2003}
  \sum_{n=0}^{\infty} L^{2k-1}_n(u)\,\om^n = (1-\om)^{-2k}\,e^{- u\,\om/(1-\om )}\,,~ \om \in \mathbb{D}\,.
\end{equation}
This implies that
\begin{equation}
  \label{eq:2004}
  B_k(\om,u) = \sum_{n=0}^{\infty} \tl{e}_{k,n}(\om)\,\br{e}_{k,n}(u) = \frac{1}{\sqrt{\Gamma (2k)}}\,(1-\om)^{-2k}\,
u^{k-1/2}\,e^{ -(u/2)\,(1+\om)/(1-\om)}\,,
\end{equation}
where $\tl{e}_{k,n}(\om)$ denotes the basis \eqref{eq:1975}. The function
$B_k(\om,u)$ is by construction  the kernel of a unitary transformation from the basis $\br{e}_{k,n}$ to the basis
$\tl{e}_{k,n}$, $B_{k}^*(\om,u)$ being the kernel for the inverse transformation:
\begin{equation}
  \label{eq:2017}
  \int_0^{\infty}du\,B_k(\om,u)\,\br{e}_{k,n} = \tl{e}_{k,n}(\om)\,,~~~~\int_{\mathbb{D}}d\tl{\mu}_k(\om)\,B^*_k(\om,u)\,
\tl{e}_{k,n}(\om) = \br{e}_{k,n}(u)\,.
\end{equation}
One can show \cite{bo4} that the operators $\tl{K}_0$,$\,\tl{K}_1$ and $\tl{K}_2$ now have the form
\begin{eqnarray}
  \label{eq:2018}
  \tl{K}_0 &=& -u\,\frac{d^2}{du^2} -\frac{d}{du} + \frac{(2k-1)^2}{4u} +\frac{u}{4}\,,\;\;\;\; \tl{K}_0\,
 \br{e}_{k\,,n}(u) =
(n+k)\, \br{e}_{k\,,n}(u)\,, \\
 \tl{K}_1 &=& -u\,\frac{d^2}{du^2} -\frac{d}{du} + \frac{(2k-1)^2}{4u} -\frac{u}{4}\,, \label{eq:2019} \\
\tl{K}_2 &=& \frac{1}{i}(u\,\frac{d}{du} +1/2)\,.\label{eq:2020}
\end{eqnarray}
As
\begin{equation}
  \label{eq:2021}
  \tl{K}_0-\tl{K}_1 = \frac{u}{2}\,,
\end{equation}
the integration variable $u$ may be associated with the classical quantity 
\begin{equation}
  \label{eq:2022}
  \tl{h}_0-\tl{h}_1 = \tl{I}(1-\cos \vp) = 2\tl{I}\sin^2(\vp/2)\,,
\end{equation}
 that is to say we have the correspondence
\begin{equation}
  \label{eq:2023}
  u \leftrightarrow 4\tl{I}\sin^2(\vp/2) \geq 0\,.
\end{equation}

Inserting $\br{e}_{k,n}(u)$ for the general number state $|k,n\rangle$ into the series \eqref{eq:1765}, \eqref{eq:1767}
and \eqref{eq:1768} yields the following coherent state wave functions
\begin{eqnarray}
  \label{eq:2024}
  |k,z \rangle (u)\equiv f_z(u) &=& \frac{u^{k-1/2}\,e^{ -u/2}}{\sqrt{g_k(|z|^2)}}\,\sum_{n=0}^{\infty}\frac{z^n}{
\Gamma (2k+n)}\,
L^{2k-1}_n(u) \\ &=&\frac{u^{k-1/2}\,e^{(z -u/2)}}{\sqrt{g_k(|z|^2)}}\,(u\,z)^{-k+1/2}\,J_{2k-1}(2\sqrt{u\,z})
\nonumber \\ &=& \frac{u^{k-1/2}\,e^{ (z -u/2)}}{\sqrt{g_k(|z|^2)}}\,\sum_{n=0}^{\infty}\frac{(-u\,z)^n}{n!\,\Gamma 
(2k+n)} \nonumber \\ &=& u^{k-1/2}\,e^{(z-u/2)}\,\frac{g_k(-u\,z)}{\sqrt{g_k(|z|^2)}}\,, \nonumber
\end{eqnarray}
where the relations \cite{erdII1}
\begin{eqnarray}
  \label{eq:2025}
  \sum_{n=0}^{\infty}\frac{z^n}{\Gamma (2k+n)}\,L^{2k-1}_n(u)& =& e^{ z}\,(u\,z)^{-k+1/2}\,J_{2k-1}(2\sqrt{u\,z})\,,\\
J_{\nu}(\zeta)& =& (\zeta/2)^{\nu}\sum_{n=0}^{\infty}\frac{(-1)^n}{n!\,\Gamma (\nu +n +1)}(\zeta/2)^{2n}\,, \label{eq:2026}
\end{eqnarray}
and \eqref{eq:1766} have been used.

The relation \eqref{eq:2003} implies
\begin{equation}
  \label{eq:2027}
 |k,\lambda \rangle (u) \equiv f_{\lambda}(u) = \frac{(1-|\lambda|^2)^k\,(1-\lambda)^{-2k}}{\sqrt{\Gamma (2k)}}\,
u^{k-1/2}\,e^{ -(u/2)(1+\lambda)/(1-\lambda)}\,.
\end{equation}
Finally
\begin{equation}
  \label{eq:2028}
  |k,\alpha \rangle (u) \equiv f_{\alpha}(u) = \frac{1}{\sqrt{\Gamma (2k)}}\,e^{ -(|\alpha|^2+u)/2}\,u^{k-1/2}\,
\sum_{n=0}^{\infty}\frac{\alpha^n}{\sqrt{(2k)_n}}\,L^{2k-1}_n(u)\,.
\end{equation}
Let us have a brief look at the behaviour of the probability densities
\begin{equation}
  \label{eq:2029}
  p_{k,n}(u) = |\br{e}_{k,n}(u)|^2 = \frac{n!}{\Gamma (2k)\,(2k)_n}\,u^{2k-1}\,e^{ -u}|L^{2k-1}_n(u)|^2
\end{equation}
for small $u$ as a function of $k$. Because of the second of the relations \eqref{eq:2002} we have
\begin{equation}
  \label{eq:2030}
  p_{k,n}(u \to 0^+) \approx \frac{1}{\Gamma (2k)}\,u^{2k-1}\,.
\end{equation}
Thus $p_{k,n}(u)$ vanish in the limit $u \to 0$ for $k >1/2$, has the finite value $1$ for $k=1/2$ and diverges
for $ 0 <k < 1/2\,$ (but is still integrable). Notice that the behaviour \eqref{eq:2030} is independent of $n\,.$ \\
As $\Gamma (2k)$ behaves like $1/(2k)$  near $k = 0$ we have
\begin{equation}
  \label{eq:2031}
  p_{k,n}(u \to 0^+; k \to 0^+) \approx 2k\,u^{2k-1}\,. 
\end{equation}
The ground state probability density is
\begin{equation}
  \label{eq:2032}
  p_{k,n=0}(u) = \frac{1}{\Gamma (2k)}\,u^{2k-1}\,e^{ -u}\,.
\end{equation}
For $k > 1/2$ it has a maximum at $u = u_0= 2k-1$.

The eigenfunctions $f_{h_2}(u)$ of the operator  $\tl{K}_2$ in Eq.\ \eqref{eq:2020} can easily be found as
\begin{equation}
  \label{eq:2033}
  f_{h_2}(u) = \frac{1}{\sqrt{2\pi}}\,u^{i\,h_2 -1/2}\,,~~\int_0^{\infty}du\,f_{h_2^{\prime}}(u)\,f_{h_2}(u) = \delta
 (h_2^{\prime}-h_2)\,.
\end{equation}
The last relation can be verified by the substitution $u=e^v\,,\,du = u\,dv$. The eigenfunctions, which are independent 
of $k$, can be used for
Mellin transformations \cite{mell}
\begin{equation}
  \label{eq:2034}
  \hat{g}(s) = \int_0^{\infty}du\,g(u)\,u^{s-1}\,,~~ s=ih_2 + 1/2\,,
\end{equation}
with the inversion
\begin{equation}
  \label{eq:2035}
  g(u) = \frac{1}{2\pi i}\,\int_{1/2-i\infty}^{1/2 + i\infty}ds\,u^{-s}\,\hat{g}(s)\,.
\end{equation}
The substitution $u=e^v$ shows the close relationship of the Mellin transform to the Fourier transform.

The eigenfunctions of $\tl{K}_1$ are more complicated \cite{erdI2a}:
\begin{equation}
  \label{eq:2036}
  f_{h_1}(u) = C\,u^{k-1/2}\,e^{ -iu/2}\,\Phi(k-ih_1;2k;iu)\,,~ C= \text{ const.}\,,
\end{equation}
where $\Phi(a;c;z)$ is the confluent hypergeometric  series
\begin{equation}
  \label{eq:2037}
 \Phi(a;c;z) = \sum_{n=0}^{\infty}\frac{(a)_n}{(c)_n\,n!}\,z^n\,. 
\end{equation}
\section{On the ground state of the quantized free electromagnetic field in a cavity}
\subsection{The electromagnetic field in a cavity as a set of harmonic oscillators}
 The standing free electromagnetic waves in a cavity can be interpreted as a denumerable set of harmonic oscillators
each of them having the ground state energy \eqref{eq:827}, the sum of which is infinite! This ``nuisance''
led to the concept of ``normal-ordering'', which just means to ignore the infinite ground state energies. On the other
hand, subtracting two such infinities leads to the Casimir effect \cite{casi,mil1,milt}, a quantum (``vacuum'') 
force between two ideally
 conducting plates, now experimentally verified  \cite{milt2}. The effect can, however, also be
derived without refering to vacuum energies and their fluctuations, by subtracting appropriate Green's functions 
associated with certain boundary conditions \cite{schw}.

 The issue of quantum vacuum energies assumes ``cosmic'' dimensions
in the context of the cosmological constant in Einstein's theory of gravity. The usual estimates for that
 constant are essentially based on the value \eqref{eq:827}. Those estimates turn out to be up to more than 100 orders
 of magnitudes larger than
 the experimentally determined value, the estimate depending on the cutoff chosen.  This discrepancy obviously constitutes
 the most urgent
 and provocative challenge as to the quantitative powers of physical theories. 
The issue has become very acute recently by the observation of an appreciable ``dark energy'' in the universe (about 
75\% of all matter),
 very likely related to the gravitational cosmological constant and the associated ``vacuum energies'' 
\cite{wein,car,peeb,volo,cope,pad,strau,nobb}. 

It is obvious that the much richer spectrum of possible ground states for the HO Hamiltonian \eqref{eq:923} can shed new
light on the subject. I here shall  only point out the crucial part of the issue without going into further details.

I first recall the main elements as to the formulation of standing waves in a cubic cavity with side lengths $L$
in terms of harmonic oscillators \cite{elemag}, with the slight generalization (compared to most textbooks)
 to allow for relative
dielectric constants $\epsilon$ and relative magnetic permeabilities $\mu$ different from the vacuum values $\epsilon =1,\,
\mu =1$:

In the Coulomb gauge Maxwell's equations without sources are given by
\begin{equation}
  \label{eq:1941}
  \frac{1}{v^2}\,\partial_t^2\,\vec{A} - \Delta \vec{A} =0\,;~~\text{div}\vec{A} =0\,,~~\vec{E}=-\partial_t\vec{A}\,,
~~\vec{B}
= \text{curl}\vec{A}\,,~~v = c/n\,,~~n= +\sqrt{\epsilon\,\mu}\,.
\end{equation}
Postulating periodic boundary conditions for the vector potential
leads to the solution type
\begin{equation}
  \label{eq:1944}
  \vec{A}(t,\vec{x}) = \frac{1}{\sqrt{\epsilon_0\,L^3}}\, \sum_{\vec{m}\in \mathbb{Z}^3}\vec{A}_{\vec{l}\,}(t)
\,e^{i\vec{l}\cdot\vec{x}}+\vec{A}_{\vec{l}\,}^*(t)\,
e^{-i\vec{l}\cdot\vec{x}}\,,
\end{equation}
where
\begin{equation}
  \label{eq:1945}
  \vec{l} = \frac{2\pi}{L}\,\vec{m}\,,~\vec{m} = (m_1,m_2,m_3)\,,~m_j \in \mathbb{Z}\,,~j=1,2,3\,;~~\vec{l}\cdot 
\vec{A}_{\vec{l}\,}(t) = 0\,.
\end{equation} 
 The time-dependent factors $\vec{A}_{\vec{l}\,}(t)$ obey the HO equations
\begin{equation}
  \label{eq:1946}
  \partial_t^2\,\vec{A}_{\vec{l}}\,(t)+\omega^2(\vec{l}\,)\,\vec{A}_{\vec{l}}\,(t)=0\,,~\omega^2(\vec{l}\,) = v^2\,
\vec{l}^{\,\,2}= \frac{c^2}{n^2}\,\vec{l}^{\,\,2}\,,~
\omega\,(\vec{l}\,)  \, \ge 0\,,
\end{equation}
with the solutions
\begin{equation}
  \label{eq:1947}
  \vec{A}_{\vec{l}}(t) = \sum_{\lambda = 1}^2 \tl{c}_{\vec{l}\,\lambda}(t)\,\vec{\epsilon}_{\vec{l}\,\lambda}
+ \tl{c}_{\vec{l}\,\lambda}^*(t)\,\vec{\epsilon}_{\vec{l}\,\lambda}\,,~~\tl{c}_{\vec{l}\,\lambda}(t) =c_{\vec{l}\,\lambda}\,
e^{-i\omega\,(\vec{l}\,)\,t}\,,
\end{equation}
where the $\vec{\epsilon}_{\,\vec{l}\,\lambda}$ are two  polarization vectors.  

Inserting the associated electric and magnetic fields into the integral for the electromagnetic field energy in the cavity,
\begin{equation}
  \label{eq:2052}
  E_{em}(\text{cavity}) = \frac{1}{2}\,\int_{cavity}d^3x\,[\epsilon\,\epsilon_0\,\vec{E}^{\,2}(t,\vec{x}) +
\frac{1}{\mu\,\mu_0}\,\vec{B}^{\,2}(t,\vec{x})]
\,,~~ \epsilon_0\,\mu_0 = 1/c^2\,,
\end{equation}
and observing that for plane wave solutions \cite{jack}
\begin{equation}
  \label{eq:2170}
  \vec{B}^2 = \epsilon\,\epsilon_0\,\mu\,\mu_0\,\vec{E}^2 =\frac{n^2}{c^2}\,\vec{E}^2\,,
\end{equation}
yields
\begin{equation}
  \label{eq:2053}
  E_{em}(\text{cavity}) = 2\epsilon\, \sum_{\vec{m}}\sum_{\lambda}[\omega(\vec{l}\,)]^2\,
|\tl{c}_{\vec{l}\,\lambda}|^2\,.
\end{equation}
Notice that $|\tl{c}_{\,\vec{l},\,\lambda}(t)|^2 =|c_{\,\vec{l}\,\lambda}|^2\,$.
Defining
\begin{equation}
  \label{eq:2056}
  \tl{c}_{\vec{l}\,\lambda} = \frac{1}{2}\,[q_{\,\vec{l}\,\lambda}+\frac{i}{\omega(\vec{l}\,)}\,p_{\,\vec{l}\,\lambda}]\,,~~ 
p_{\,\vec{l}\,\lambda} =\dot{q}_{\,\vec{l}\,\lambda}\,,
\end{equation}
the expression \eqref{eq:2053} finally becomes
\begin{equation}
  \label{eq:2057}
   E_{em}(\text{cavity}) = H_{em}(q,p) =\frac{\epsilon}{2}\sum_{\vec{m}}\sum_{\lambda}[p_{\vec{l}\,\lambda}^2+
\omega^2(\vec{l}\,)\,
q_{\,\vec{l}\,\lambda}^2]\,,
\end{equation}
where the individual terms
\begin{equation}
  \label{eq:2059}
  H_{\vec{l}\,\lambda}(q,p) = \frac{1}{2}\, [p_{\,\vec{l}\,\lambda}^2+\omega^2(\vec{l}\,)\,
q_{\,\vec{l}\,\lambda}^2]
\end{equation}
are independent of time!

As can be seen from Maxwell's eqs.\ \eqref{eq:1941}, the  electric field provides the canonical momenta,
 the magnetic field (via its vector potential) the canonical coordinates \cite{loui}.

The standard quantization procedure is now obvious: The classical quantities $q_{\,\vec{l}\,\lambda}$ and
 $p_{\,\vec{l}\,\lambda}$ are 
promoted to operators $Q_{\,\vec{l}\,\lambda}$ and $P_{\,\vec{l}\,\lambda}$, having the commutation relations
\begin{equation}
  \label{eq:2058}
  [Q_{\,\vec{l}\,\lambda}\,,\, P_{\,\vec{l}^{\prime}\,\lambda^{\prime}}]= i\,\hbar\, \delta_{\vec{l}\;\vec{l}^{\,\prime}\,}\,
\delta_{\lambda\,\lambda^{\prime}}\,.
\end{equation}
A very minor point may be worth mentioning here: The right-hand side of the energy \eqref{eq:2057} does not contain a
mass term. As the dimension of the energy is given, $[L^2\,T^{-2}\,M]$, the quantities $q_{\,\vec{l}\,\lambda}$ and
 $p_{\,\vec{l}\,\lambda}$ here have dimensions $[M^{1/2}\,L]$ and $[M^{1/2}\,L\,T^{-1}]$, respectively. But their product still
has the dimension of an action, $[M\,L^2\,T^{-1}]$!

We now come to the point of departure: Assuming (for a moment) $\epsilon =1,\,\mu =1$ and introducing angle and action
 variable for each mode,
\begin{equation}
  \label{eq:2060}
  q_{\,\vec{l}\;\lambda} = \sqrt{2\,I_{\,\vec{l}\;\lambda}/\om (\vec{l}\,)}\,\cos \vp_{\,\vec{l}\;\lambda}\,,~~~
 p_{\,\vec{l}\;\lambda} = -\sqrt{2\, \om (\vec{l}\,)\,I_{\,\vec{l}\;\lambda}}\,\sin \vp_{\,\vec{l}\;\lambda}\,,
\end{equation}
yields
\begin{equation}
  \label{eq:2061}
   H_{\,\vec{l}\,\lambda}(\vp,I) = \om (\vec{l}\,)\,I_{\,\vec{l}\;\lambda}\,,~~H_{em}(\vp,\,I) =\sum_{\vec{m}}\sum_{\lambda}
H_{\vec{l}\,\lambda}\,.
\end{equation}

Quantization proceeds now as discussed above for the angle-action model of the HO:
Each of the $ H_{\,\vec{l}\,\lambda}$ is replaced by an operator
\begin{equation}
  \label{eq:2062}
 H_{\,\vec{l}\,\lambda}(\vec{K}) = \hbar\,\om (\vec{l}\,)\,\tl{K}_0(\vec{l}\,,\lambda)\,,~~ H_{em}(\vec{K}) =
\sum_{\vec{m}}\sum_{\lambda}
\oplus H_{\,\vec{l}\,\lambda}(\vec{K})\,.
\end{equation}
Each $\tl{K}_0(\vec{l}\,,\lambda)$ acts irreducibly in a Hilbert space that carries a unitary representation with Bargmann
index $k$, together with the operators $\tl{K}_1(\vec{l}\,,\lambda)$ and $\tl{K}_2(\vec{l}\,,\lambda)$ or the ladder
 operators
$\tl{K}_+(\vec{l}\,,\lambda)$ and $\tl{K}_-(\vec{l}\,,\lambda)$. Because of the required bosonic exchange symmetries I
here assume the
same Bargmann index $k$ for all representations. I here do not enter the important subject of constructing and
 analyzing
the quantized free or even interacting electromagnetic fields themselves in terms of  the operators
 $\tl{K}_j(\vec{l}\,,\lambda)$ etc.
The usual $k$-independent annihilation and creation operators associated with the fields themselves are given by
\begin{equation}
  \label{eq:2066}
  A_{\,\vec{l},\,\lambda} = [\tl{K}_0(\vec{l}\,,\lambda)+k]^{-1/2}\,\tl{K}_-(\vec{l}\,,\lambda))\,,~~~
 A_{\,\vec{l},\,\lambda}^{\dagger} = \tl{K}_+(\vec{l}\,,\lambda)\,[\tl{K}_0(\vec{l}\,,\lambda)+k]^{-1/2}\,.
\end{equation}
\subsection{The cosmological constant problem}
Presently I am merely interested in the ground state expectation value
\begin{equation}
  \label{eq:2063}
  \langle k, 0_{\otimes}|H_{em}(\vec{K})|k, 0_{\otimes}\rangle = \sum_{\vec{m}}\sum_{\lambda}
\langle k,0| H_{\vec{l}\,\lambda}(\vec{K})|k, 0 \rangle =2\, k\,\hbar
\sum_{\vec{m}} \om (\vec{l}\,)\,.
\end{equation}
The usual replacement
\begin{equation}
  \label{eq:2067}
  \sum_{\vec{m},\,\lambda} \to \frac{L^3}{\pi^2c^3}\,\int_{\om \geq 0}d\om\,\om^2
\end{equation}
leads to a strongly divergent ground state energy density
\begin{equation}
  \label{eq:2068}
  \frac{k \,\hbar}{\pi^2c^3}\,\int_{\om \geq 0}d\om\,\om^3\,.
\end{equation}
Cutting the infinite integral off at $\om = \hat{\om}$ yields the ``vacuum'' energy density
\begin{equation}
  \label{eq:2069}
  u_{em,\,0}(\hat{\om},\,k) = \frac{k\,\hbar}{4\pi^2c^3}\,\hat{\om}^4\,.
\end{equation}
Defining the effective length
\begin{equation}
  \label{eq:2070}
  \ell = \frac{2\pi\,c}{\hat{\om}} 
\end{equation}
finally gives
\begin{equation}
  \label{eq:2071}
   u_{em,\,0}(\ell,\,k) = \frac{4\pi^2\, k}{\ell^4}\,\hbar\,c \approx \frac{4\pi^2\, k}{\ell^4}\cdot(
2\cdot 10^{-11}\,\text{MeV\,cm}).
\end{equation}
We know from sec.\ 5 that the index $k$ may become arbitrarily small $>0$, perhaps in the course of time| So 
 the  $k$ in the  expression \eqref{eq:2071} may become so small - for a given value
 of the interaction length $\ell$ - 
 that the value of $ u_{em,\,0}(\ell,\,k)$ comes near  the order of magnitude of the observed dark energy
 density \cite{sper}
\begin{equation}
  \label{eq:2072}
  c^2\rho_{\Lambda} \approx 4\,\text{keV\,cm$^{-3}$}\,.
\end{equation}
Such a welcome adjustment of $k$ is, of course, here not proven at all, and one would like to have more sophisticated
 arguments
in the present framework for the desired appropriate value of the index $k$ in order to get a ``reasonable'' estimate
 for the cosmological constant.
Nevertheless, the mere existence of that index, originating from the non-trivial topology of the $(\vp,\,I)$-phase space
of the HO  and its related quantizing group $SO^{\uparrow}(1,2)$ (including its infinitely many covering groups),
 may be an important key for the solution of the cosmological constant problem!

I list a few of the many problems I leave open here: \begin{itemize}
\item The role of the index $k$ has to be examined for other matter fields, especially fermions and non-abelian gauge
 fields and associated interactions,
particularly for those with spontaneous symmetry breaking!
\item The compatibility with (local) Poincar\'{e} covariance and its concept of causality has to be analyzed.
\item Most of the prevailing discussions of
 the Casimir effect - with their by now quite sophisticated subtractions of two infinities - 
 (see the literature quoted above) and especially their experimental confirmations
 appear to contradict the introduction of an index $k$ different from $1/2$. There are different answers to such
an objection:

 First, it is evident from the discussions above, that the ground state of the HO Hamiltonian $H(Q,P)$ is
necessarily tight to $k=1/2$. In order to have $k\neq 1/2$ the basic quantum observables have to be the $\tl{K}_j$.
An analysis of the Casimir effect in terms of these new variables has not yet been done.

Second, there have been alternative proposals for deriving the Casimir effect (force) instead of subtracting infinite
vacuum energies \cite{schw}! 
\item As the number $k$ is a (dimensionless) measure for some energy, it may become time-dependent, i.e.\ dynamical,
 on a cosmic scale and might lead to a time-dependent cosmological constant. The index $k$ may also become a function
of the frequency $\om$ or (and) of space coordinates, like the dielectic constant $\epsilon$ from above.
\end{itemize}
\subsection{Birefringence and dichroism of the vacuum}
Comparing the expressions \eqref{eq:2057} and \eqref{eq:2063} suggests to preliminarily interpret the index $k$ here
as a kind of ``anomalous'' dielectric constant (or the square of an ``anomalous'' index of refraction, cf.\ Eq.\ 
\eqref{eq:1941}) of the vacuum. This interpretation leads (tentatively) to the following possible quantum 
optical application:

Lets assume we have in vacuum initially just two  photon modes  of the same frequency $\om$, the same initial wave
 number $\vec{l}$,  but orthogonal linear  polarizations. Both should initially
belong to the same index $k$. If one lets these photons pass through strong electric or (and) static magnetic fields 
$\vec{E}_0\,,\,\vec{B}_0$, these ``perturbations'' add constant terms proportional to $\epsilon_0\,\vec{E}_0^2$ or (and)
$\vec{B}^2_0/\mu_0$ to the free Hamiltonians $H_{\vec{l}\,\lambda}(\vec{K})\,,\,\lambda = 1,2\,$, (see also the discussion
around Eq.\ \eqref{eq:1928}). The energy of the static fields  may change the index $k$ of at least one of the fields by
a small amount $\delta\,k$ which could lead to the following possible effects: 
\begin{itemize} \item Compared to the photon the vacuum energy of which is ``lifted'' by an amount $\delta\,k > 0$
 the other photon
with the orthogonal polarization ``lost'' energy, leading to an effective ``dichroism''! \item 
If the energetically lifted photon returns to its original index $k$ after passing the external fields, then we have
an effective ``birefringence''! \end{itemize}
As to the conventional optical phenomena of this type in materials (electro-optical ``Kerr-effect'' or 
magneto-optical ``Cotton-Mouton-'' and ``Voigt-'' effects and related dichroisms etc.)  see Refs.\ \cite{dichr}.

  The effects mentioned should be proportional to the square of the external electric or (and) magnetic fields.

 Possibly the recent PVLAS experiment \cite{zav,cant} with its observation of vacuum dichroism induced by an external
 magnetic field can be understood in this framework!
\subsection{``Dark'' normal matter?} Let me dare to add a very speculative remark: As the quantum spectra 
\eqref{eq:925} and
\eqref{eq:926} of the two HO classical models \eqref{eq:921}  can be different, the index $k >0 $ possibly being very
small. So (radiation) energy may get ``stuck'' in the interval $1/2 > k > 0$ or even at higher excited levels which perhaps
 can decay by higher order electromagnetic transitions only. In such a speculation dark matter would be just ``normal''
 matter prevented from radiating normally (e.g., the abundance of diatomic molecular hydrogen \cite{h2} provides an
 abundance of
effective HOs). This could ``explain'' why visible and dark matter are of the same order of
 magnitude! In such a speculative scenario dark matter could have been formed only after the formation of atoms and
 molecules. All this has, of course, to be evaluated much more critically.
 
\section{Charged particles in external electric and magnetic fields}
\subsection{Charged harmonic oscillator  in an external electric field}
If one puts a harmonically vibrating particle of mass $M$ and charge $Z e_0\,,\,Z \in \mathbb{Z}\,,$ in an external 
electric field $E_0$ in
 $q$-direction then the potential term
 \begin{equation}
   \label{eq:39}
    -Ze_0\,E_0\,q
 \end{equation}
has to be added to the Hamiltonian of the HO:
\begin{equation}
  \label{eq:40}
  H= \frac{1}{2M}\,p^2 +\frac{M}{2}\om^2\,q^2 -Ze_0\,E_0\,q = \frac{1}{2M}\,p^2 +\frac{M}{2}\om^2\,(q-
\frac{Ze_0\,E_0}{\om^2\,M})^2 -
\frac{Z^2e_0^2\,E^2_0}{2\,\om^2\,M}\,.
\end{equation}
Defining
\begin{equation}
  \label{eq:41}
  \xi = q-\frac{Ze_0\,E_0}{\om^2\,M}
\end{equation}
we again have an effective HO with coordinate $\xi$ and the ground state energy shifted by the amount
\begin{equation}
  \label{eq:42}
  V_0 = -\frac{Z^2e_0^2\,E^2_0}{2\,\om^2\,M} \leq 0\,.
\end{equation}
Replacing $q$ in Eq.\ \eqref{eq:920} by $\xi$ yields
\begin{equation}
  \label{eq:43}
  H(\xi,p)=  \frac{1}{2M}\,p^2 +\frac{M}{2}\om^2\,\xi^2+V_0 = \om\,I +V_0\,.
\end{equation} The fine structure constant
\begin{equation}
  \label{eq:77}
  \alpha = \frac{e_0^2}{4\pi\,\epsilon_0\,c\,\hbar} \approx 7.3 \cdot 10^{-3}
\end{equation}
allows $V_0$ to be rewritten as
\begin{equation}
  \label{eq:78}
  -V_0= \hbar\,\om\,\frac{2\pi\,\alpha Z^2 \,\epsilon_0E^2_0}{(\om/c)^3\, M c^2}= \hbar\,\om \,\frac{\alpha Z^2}{4\pi^2}
\frac{
\epsilon_0 E^2_0\,\lambda^3}{M c^2}\,,~~\lambda = \frac{2\pi\,c}{\om} = \frac{c}{\nu}\,.
\end{equation}
Comparing with Eq.\ (19) suggest to introduce an effective Bargmann index
\begin{equation}
  \label{eq:79}
  k \to k_{eff} = k- \delta\,,~~~\delta = \frac{\alpha Z^2}{4\pi^2}\frac{\epsilon_0E^2_0\,\lambda^3}{M 
c^2}\,.
\end{equation}
In order to get an impression of the order of magnitude for $\delta$ in experiments consider an ion of rest energy
 $M\,c^2 \approx 
100 \,\text{GeV} \approx 10^{-8}\,$J and charge $e_0$ in a 1-dimensional harmonic Paul trap \cite{ion}.  With $E_0 
\approx 10^3\, \text{V/m}\,$ along the longitudinal  direction, 
$ \nu \approx 10^8\, \text{Hz}$ one gets approximately the  value $\delta \approx 10\,$, which makes
$k_{eff}$   negative! One further has to reduce the energy $\epsilon_0E^2_0\,\lambda^3$ compared to $Mc^2$ in
order to have $k_{eff}$ positive.
\subsection{Charged particle in an external magnetic field} 
It is well-known that the 3-dimensional motion of a particle with charge $q$ in a homogeneous magnetic field $\vec{B} =
 \text{curl}\vec{A}$ can be associated with an effective harmonic oscillator for the motion transversal to the magnetic
field \cite{land}: The Hamilton function is given by (here $m$ obviously means the mass, as opposed to previous Secs.)
\begin{equation}
  \label{eq:2124}
  H=\frac{1}{2m} \vec{\pi}^2\,,~~\vec{\pi}=m\,\dot{\vec{x}} = \vec{p} -q\,\vec{A}(\vec{x})\,,
\end{equation}
with the basic Poisson brackets
\begin{equation}
  \label{eq:2125}
  \{x_j,p_k\} = \delta_{j\,k}\,,~j,k =1,2,3\,.
\end{equation}
The Eqs.\ of motion are
\begin{equation}
  \label{eq:2126}
  \dot{x}_j = \{x_j,\,H\}= \frac{1}{m}(p_j-q\,A_j),,~~~\dot{p}_j= \{p_j,H\}= q\,\sum_{k=1}^{3}\dot{x}_k\partial_jA_k\,.
\end{equation}
It follows from the Poisson brackets \eqref{eq:2125} that
\begin{equation}
  \label{eq:2127}
  \{\pi_j,\pi_k\} = q\,(\partial_jA_k-\partial_kA_j) = q\,B_l\,,~~(j,k,l) = \text{ cycl.\ } (1,2,3)\,.
\end{equation}
For $\vec{B} = (0,0,B)$ we have
\begin{equation}
  \label{eq:2128}
  \{\pi_1,\pi_2\} = q\,B\,,~~\{\pi_1,\pi_3\} =0\,,~~\{\pi_2,\pi_3\} = 0\,.
\end{equation}
The last relations imply
\begin{equation}
  \label{eq:2129}
  \dot{\pi}_3 = \{\pi_3,H\} = 0\,,
\end{equation}
i.e.\ $\pi_3$ is a constant of motion. 

Of special interest here is the remaining ``transversal'' Hamilton function
\begin{equation}
  \label{eq:2130}
  H_{\perp}=\frac{1}{2m}(\pi_1^2 + \pi_2^2)\,.
\end{equation}
Defining 
\begin{equation}
  \label{eq:2131}
  \om = q\,B/m\,,~~\pi_1 = m\,\om\,\xi\,,~~\pi_2 =\pi_{\xi}\,,
\end{equation}
and assuming $q\,B >0$  we get
\begin{equation}
  \label{eq:2132}
   H_{\perp}=\frac{1}{2m}\pi_{\xi}^2 + \frac{1}{2}m\,\omega^2\,\xi^2\,,~~\{\xi,\pi_{\xi}\} = 1\,.
\end{equation}
This is an effective HO Hamilton function for the transversal motion of a particle with charge $q$ in a magnetic field
$\vec{B} = (0,0,B)$. As the ``canonical coordinate'' $\xi$ actually is proportinal to a time derivative of the
 original coordinates,
one needs another integration. This is provided by the quantities
\begin{equation}
  \label{eq:2133}
  b_1= x_1 +\frac{1}{m\,\om}\pi_{\xi}\,,~~b_2 = x_2 - \xi= x_2-\frac{1}{m\,\om}\,\pi_1\,;~~(x_1-b_1)^2+(x_2-b_2)^2 =
\frac{2}{m\,\om^2}\,H_{\perp}\,,
\end{equation}
which obey
\begin{equation}
  \label{eq:2134}
  \{b_j,\pi_k\} = 0\,,~j,k =1,2\,;~~~~\{b_2,b_1\} = \frac{1}{m\,\om}\,,
\end{equation}
 implying
\begin{equation}
  \label{eq:2135}
  \{b_j,H_{\perp}\} = 0\,,j=1,2\,,
\end{equation}
i.e.\ the $b_j$ are constants of motion. According to their definition they are the coordinates of the center of the
circle on which the particle moves in the tansversal $(1,2)$-plane.

If $q\,B < 0$ one just has to interchange the roles of $\pi_1$ and $\pi_2$ in the relations \eqref{eq:2130} and
 \eqref{eq:2131}
and defines $\om = |q\,B|\,$.

For the HO Hamilton function \eqref{eq:2132} one can introduce angle and action variables as usual:
With $\om > 0$ and defining
\begin{equation}
  \label{eq:2136}
  \xi = \sqrt{\frac{2I}{m\om}}\,\cos \vp\,,~~\pi_{\xi} = -\sqrt{2m\,\om I}\,\sin\vp\,,
\end{equation}
we get
\begin{equation}
  \label{eq:2137}
  H_{\perp}(\vp,I) = \om\,I\,,
\end{equation}
which can be dealt with as previously:

The usual quantization procedure for the Hamilton functions \eqref{eq:2130} or \eqref{eq:2132} is the standard one,
yielding the (Landau) energy levels
\begin{equation}
  \label{eq:2138}
  E_{\perp,n} = \hbar \om (n+1/2)\,,~~n=0,1,\ldots\,.
\end{equation}
However, quantizing the Hamilton function \eqref{eq:2137} in the spirit of the present paper yields the Hamilton operator 
\begin{equation}
  \label{eq:2139}
  \hat{H}_{\perp} = \hbar\om \tl{K}_0\,,
\end{equation}
with the possible energy levels
\begin{equation}
  \label{eq:2140}
  E_{k,n} = \hbar \om (n+k)\,,~n=0,1,2,\ldots\,.
\end{equation}
If $k \neq 1/2$ the usual Landau energy levels are being shifted to lower or higher values. Whether this really happens has,
of course, to be found out experimentally!
\section{Thermodynamics}
Next I collect some thermodynamical properties of a system with energy levels $E_n = \hbar\,\om\, (
n+k)$ in a heat bath of temperature $k_B\,T \equiv 1/\beta$ in order to see which quantity depends on the index $k$,
and which not! That index $k$ is here, of
 course, not to be confused with Boltzmann's constant $k_B$. \\
The following simple formulae should be of interest for the interpretation of experiments in preparation for measuring the
ground state energy of the HO by means of the AC Josephson effect \cite{beck}. \\
From the partition function
\begin{equation}
  \label{eq:2073}
  Z(\beta; k) = \sum_{n=0}^{\infty} e^{-\beta\,\hbar\,\om\,(n+k)} = \frac{e^{-\beta\,\hbar\,\om\,k}}{1-e^{-\beta\,\hbar\,\om}}
\end{equation}we get the probability to find the system in the $n$th state as
\begin{equation}
  \label{eq:2074}
  p_n(\beta) = e^{-\beta\,\hbar\,\om\,(n+k)}/Z(\beta; k) = e^{-\beta\,\hbar\,\om\,n}\,(1-e^{-\beta\,\hbar\,\om})\,,
\end{equation}
which is independent of $k$.

Furthermore we have \\
Free energy:
\begin{equation}
\beta\,F(\beta;k) = -\ln Z(\beta;k) = \beta\,\hbar\,\omega\,k +
\ln (1-e^{-\beta\,\hbar\,\om})\,.
 \label{eq:2075}\end{equation}
Internal energy: \begin{equation}
  U(\beta; k) = \langle E \rangle (\beta; k) = -\partial_{\beta}Z(\beta; k) = \hbar\,\om\,
\left(k +
\frac{1}{e^{\beta\,\hbar\,\om}-1}\right)\,. \label{eq:2077} \end{equation}
Energy mean square fluctuations: \begin{equation}
 (\Delta E)^2(\beta) = \partial_{\beta}^2\ln Z(\beta;k)= (\hbar\,\om)^2\,
\frac{e^{\beta\,
\hbar\,\om}}{(e^{\beta\,\hbar\,\om}-1)^2} 
= k_BT^2C_V\,.\label{eq:2078} \end{equation}
Entropy: \begin{equation}
  S(\beta)/k_B = \ln Z(\beta; k) + \beta\,U = -\ln (1-e^{-\beta\,\hbar\,\om}) + \frac{\beta\,
\hbar\,\om}{
e^{\beta\,\hbar\,\om}-1}\,.\label{eq:2079} 
\end{equation}
Here $C_V$ is the heat capacity of the system at constant volume.

We see that energy fluctuations (heat capacities) and entropy are independent of the index $k$.

\section*{Acknowledgements}
\addcontentsline{toc}{section}{\protect\numberline{}{Acknowledgements}}

I am indebted to a number of people for stimulating correspondences and discussions: I thank C.\ Beck, T.\ H\"{a}nsch,
W.\ Ketterle, K.-P.\ Marzlin and W.\ 
Schleich for correspondences concerning possible experimental determinations of the ground state energy of the HO, G.\
Agarwal, V.\ Bu\v{z}ek and B.\ Sanders for correspondences on the experimental production of Perelomov and Barut-Girardello
coherent states, and  W.\ Schleich for a correspondence on the experimental generation of Schr\"odinger-Glauber coherent
 states. I thank P.\ Toschek for discussions on the different harmonic traps in quantum optics and A.\ Ringwald for 
discussions on the PVLAS experiment. 

As before I thank DESY Hamburg, especially the Theory Group,  for its generous hospitality after my retirement from the
Institute for Theoretical Physics of the RWTH Aachen. The DESY Library has always been very helpful in providing the
necessary literature.

Most thanks go to my wife Dorothea who had to endure  my many outer and inner absences while I was preoccupied with the
present paper! 
\addcontentsline{toc}{section}{\protect\numberline{}{Appendices}}
\section*{Appendices}
\begin{appendix}
\section{Calculating the action variables for certain potentials of 1-dimensional systems}
The calculations of the action variable \eqref{eq:21} of subsec.\ 2.3.3 for the different potentials discussed there 
can all be reduced to that of the integral
\begin{equation}
  \label{eq:38}
 f(b) = \int_{-b}^{+b}du\, \frac{(b^2-u^2)^{1/2}}{1+u}\,,~~0 < b < 1\,,
\end{equation}
which may be transformed into \cite{grary3}
\begin{align}
  \label{eq:46}
  f(b) =& -\int_{-b}^{+b}du\, \frac{u}{(b^2-u^2)^{1/2}}+\int_{-b}^{+b}du\, \frac{1}{(b^2-u^2)^{1/2}}\\ & +(b^2-1)
\int_{-b}^{+b}du\, \frac{1}{(1+u)(b^2-u^2)^{1/2}}\,. \nonumber
\end{align}
Here the first term vanishes (replace $u$ by $-u$), the second gives $\pi$ \cite{grary5}, and the
 last $-(1-b^2)^{1/2}\,\pi\,$
\cite{grary6}, so that
\begin{equation}
  \label{eq:47}
  f(b) = [1-(1-b^2)^{1/2}]\,\pi\,.
\end{equation}
In the case of the Morse potential $V_{Mo}$ one puts in Eq.\ \eqref{eq:23}
\begin{equation}
  \label{eq:48}
  b^2 = \tl{E}\,,~~ u = e^{-\tl{q}}-1\,.
\end{equation}
In the case of the potential $V_{sMo}$ the substitution
\begin{equation}
  \label{eq:49}
  u = \tanh \tl{q}\,,
\end{equation}
combined with the observation that
\begin{equation}
  \label{eq:50}
  \int_{-b}^{+b}du\, \frac{(b^2-u^2)^{1/2}}{1-u^2} = \frac{1}{2} \int_{-b}^{+b}du\,(b^2-u^2)^{1/2}\left(\frac{1}{1+u} +
\frac{1}{1-u}\right) = 
\int_{-b}^{+b}du\, \frac{(b^2-u^2)^{1/2}}{1+u}
\end{equation}
works.

 For the potential $V_{PT}$ one substitutes $u = \sin \tl{q}$ and for $V_c$ one puts $u = \tl{q}^2 + const.\ $ 

\section{The covering groups of $\mathbf{SO^{\uparrow}(1.2)}$ and the positive discrete series of their 
irreducible unitary representations} 
I have stressed repeatedly in the Sections above that the irreducible unitary representations of those covering groups
of $SO^{\uparrow}(1,2)$ or $Sp(2,\mathbb{R})$ with a very small Bargmann index $k <1/2$ may be of special interest. In this
Appendix, therefore, I collect some here relevant  properties of those groups and the associated unitary representations.
 A rather complete list of the literature on the irreducible unitary representations of the group $SO^{\uparrow}(1,2)$ and
its covering groups is contained in the Refs.\ to Appendix B of Ref.\ \cite{ka1}. That Appendix contains also a summary
of essential properties of those groups.
\subsection{The universal covering group of $SO^{\uparrow}(1,2)$}
According to Bargmann \cite{barg4} the universal covering group $\tl{G} \equiv \widetilde{SO^{\uparrow}(1,2)}$ can be
 parametrized conveniently by starting from a modified parametrization of the group $SU(1,1)$ as given by the matrices
\eqref{eq:981}, namely by defining
\begin{eqnarray} \gamma& =& \beta/\alpha\,,~~ |\gamma|<1)\,, \label{eq:2096}\\
  \omega &=& \arg(\alpha)\,, \label{eq:2097}
 \end{eqnarray}with the inverse relations \begin{eqnarray}
\alpha & =& e^{\ds i\omega}(1-|\gamma|^2)^{-1/2}\,,~
 |\gamma|<1\,, \label{eq:2098} \\ \beta &=&
e^{\ds i\omega}\gamma(1-|\gamma|^2)^{-1/2}~. \label{eq:2099}\end{eqnarray} 
The inequality $|\gamma| <1$ in Eq.\ \eqref{eq:2096} follows from the relation $|\alpha|^2 - |\beta|^2 =1\,.$

With
\begin{equation}
  \label{eq:2104}
  SO^{\uparrow}(1,2)_{[m]}\,,~~\text{ $m$-fold covering of } SO^{\uparrow}(1,2)\,,
\end{equation}
 we have the following relations
\begin{eqnarray} SO^{\uparrow}(1,2)\,: && \{(\gamma,\,\omega)\,,~|\gamma| <1\,,~\om \in (-\pi/2,\,\pi/2]\,\}\,,
\label{eq:2100} \\
Sp(2,\mathbb{R}) \cong SU(1,1)=SO^{\uparrow}(1,2)_{[2]}\,:& & \{(\gamma,\,\omega)\,,~|\gamma| <1\,,~\om 
\in (-\pi,\,\pi]\,\}\,,\label{eq:2101} \\
SO^{\uparrow}(1,2)_{[m]}\,: &&\{(\gamma,\,\omega)\,,~|\gamma| <1\,,~\om \in (-m\,\pi/2,\,\,m\pi/2]\,\}\,,\label{eq:2102}\\
\tl{G} \equiv SO^{\uparrow}(1,2)_{[\infty]}\,:&& \tl{g}\equiv \{(\gamma,\,\omega)\,,~|\gamma| <1\,,~\om \in \mathbb{R}\,\}\,.
\label{eq:2103} 
 \end{eqnarray} From the multiplication of the matrices \eqref{eq:981} one deduces the group composition law
 \begin{equation}
   \label{eq:2105}
   (\gamma_3,\om_3)= (\gamma_2,\om_2)\circ (\gamma_1,\om_1)\,,
 \end{equation}
where
 \begin{eqnarray} \label{eq:2106}
\gamma_3 & = &
 (\gamma_1 +\gamma_2 e^{\ds
 -2i\omega_1})(1+\gamma^*_1\gamma_2e^{\ds -2i\omega_1})^{-1}\,, \\
 \label{eq:2107}\omega_3&=& \omega_1 + \omega_2 +\frac{1}{2i}\ln[(
1+\gamma^*_1\gamma_2e^{\ds - 2i\omega_1})(
1+\gamma_1\gamma^*_2e^{\ds 2i\omega_1})^{-1}]\,. \end{eqnarray}
For the four subgroups \eqref{eq:1437} - \eqref{eq:1453} the new parametrization means
\begin{eqnarray} R_0: && r_0= (\gamma =0,\,\om = \theta/2)\label{eq:2108} \\ 
 &&~~~~~~~ (0,\om_3) = (0,\, \om_2+\om_1)\,; \nonumber \\
A_0: && a_0 = (\gamma = i\,\tanh (\tau/2),\,\om = 0)\,,~ \tau \in \mathbb{R}\,,\label{eq:2109} \\
&&~~~~~~~ (\gamma_3,0) = (i\,\tanh[(\tau_2+\tau_1)/2],\,0)\,; \nonumber \\
B_0: && b_0 = (\gamma =\,\tanh (s/2),\,\om = 0)\,, ~s \in \mathbb{R}\,, \label{eq:2110} \\ 
&&~~~~~~~(\gamma_3,0) = (\tanh[(s_2+s_1)/2],\,0)\,; \nonumber \\
 N_0: && n_0 = (\gamma = \xi(\xi^2 +4)^{-1/2}\,e^{-i\,\om},\,\om = \arctan (\xi/2)\,)\,,~ \xi \in \mathbb{R}\,. 
\label{eq:2111}
\end{eqnarray} 

 For the universal covering group $\tl{G}$ the transformations
 \eqref{eq:1426} and \eqref{eq:1429} now take \\ the form
\begin{eqnarray}
 \label{eq:2112}\tl{I}^{\prime}&=& \rho(\tilde{g}, \varphi)\, \tl{I}\,,~\rho
 (\tilde{g}, \varphi)=
 |1 +e^{\ds i\varphi}\, \gamma|^2\, (1-|\gamma|^2)^{-1}~,  \\
 \label{eq:2113}e^{\ds i\varphi^{\prime}}&=& e^{\ds -2i\omega}\,
 \frac{e^{\ds i\varphi} +
  \gamma^*}{1+e^{\ds i\varphi}\gamma}~. \end{eqnarray}
  As $\partial\varphi^{\prime}/\partial\varphi= 1/\rho(\tilde{g},\varphi)$, the
  equality \eqref{eq:1431} holds again.

The transformations \eqref{eq:2113} act, however, not effectively on $\mathcal{S}_{\vp,\tl{I}}$ because the (infinite)
 discrete center 
\begin{equation}
  \label{eq:2114}
  C_{[\infty]} = (0,\,\om \in \pi\,\mathbb{Z}\,) \subset \tl{G}
\end{equation}
leaves all points $\sigma =(\vp,\tl{I})$ invariant. 
Correspondingly the center
\begin{equation}
  \label{eq:2115}
  C_{[m]} = (0, \om = 0,\,\pm \pi,\,\ldots,\,\pm m\,\pi) \subset SO^{\uparrow}_{[m]}(1,2)
\end{equation}
of an $m$-fold covering group leaves the points $\sigma$ invariant, too.

  With the elements of the group $SU(1,1)$  given by the restriction\\
  $ -\pi < \omega \leq +\pi,\, \alpha = \exp(i\omega)(1-|\gamma|^2)^{-1/2},\,
  \beta = \gamma\, \alpha~,$ the homomorphisms \begin{eqnarray}\label{eq:2116}
   \Phi_{[\infty]/2} :~~ \tl{G}\equiv SO^{\uparrow}_{[\infty]}(1,2) &\rightarrow& SU(1,1)\cong
 Sp(2,\mathbb{R})\,, \\
  \Phi_{[2]} :~~ SU(1,1)\cong Sp(2,\mathbb{R}) &\rightarrow& SO^{\uparrow}(1,2)\,,
 \end{eqnarray}
  have the kernels
  \begin{equation}
    \label{eq:2117}
    \text{ker}(\Phi_{[\infty]/2})=2\pi\,\mathbb{Z}\,,~\text{ker}(\Phi_{[2]})=Z_2\,, 
  \end{equation}
  respectively,
   and the composite
  homomorphism $ \Phi_{[\infty]}=  \Phi_{[2]}\cdot \Phi_{[\infty]/2}  $ has the kernel
  \begin{equation}
    \label{eq:2118}
    \text{ker}(\Phi_{[\infty]}= \Phi_{[2]}\cdot \Phi_{[\infty]/2}) = \pi\,\mathbb{Z}\,.
  \end{equation}

  As the space $\mathcal{S}^{2}_{\vp,\tl{I}}$ 
is homeomorphic to $\mathbb{R}^2-\{0\}
  =\mathbb{C}-\{0\}$,
  its universal covering space is given by $\varphi \in \mathbb{R},~ \tl{I} \in
  \mathbb{R}^+$, which is
  the infinitely sheeted Riemann surface of the logarithm.

 The transformations
  \eqref{eq:2112} and \eqref{eq:2113} may be reinterpreted as
 acting transitively and effectively
  on that universal covering space.
\subsection{Irreducible unitary representations of the positive discrete series for $k>0$}
I have already mentioned in subsec.\ 6.4.3 that in the Hilbert space of holomorphic functions on the unit disc
$\mathbb{D} = \{z \in \mathbb{C},\,|z| <1 \}$ with the scalar product
\begin{equation}
   \label{eq:2119}
  (f,g)_k=\frac{2k-1}{\pi}\int_{\mathbb{D}}f^*(z)g(z)
(1-|z|^2)^{2k-2}dxdy~. 
 \end{equation}
  one can define irreducible unitary representations for any $k>0$\,.

 The unitary operators representing the universal covering group in that space
are given by \begin{eqnarray}
 \label{eq:1425}[U(\tilde{g},k)f](z)&=&e^{\ds 2ik\omega}(1-|\gamma|^2)^{\ds k}
 (1+\gamma^*\,z)^{\ds -2k}
 f\left( \frac{\alpha z+\beta}{\beta^*\,z+\alpha^*}\right)\,,
 \\
  \tilde{g}&=&(\gamma,\om)\,,~~ \left( \begin{array}{cc} \alpha & \beta \\
 \beta^*& \alpha^* \end{array} \right)~=~\Phi_{[\infty]/2}(\tilde{g})~
 \in SU(1,1)\,.\label{eq:2120} \end{eqnarray}
Because $|\gamma\,z|<1$, the function $(1+\gamma^*\,z)^{-2k}$ is, for $k>0$,
 defined
in terms of the series expansion
\begin{equation}
  \label{eq:2121}
(1+\gamma^*z)^{-2k} = \sum_{n=0}^{\infty} \frac{(2k)_n}{n!}\,(-\gamma^*\,z)^n\,.  
\end{equation}
The phase factor
\begin{equation}
  \label{eq:2122}
  e^{\ds 2ik\om}
\end{equation}
in Eq.\ \eqref{eq:1425} determines the possible values of $k$ for a given covering group:

For $SO^{\uparrow}(1,2)$ itself we
have (see Eq.\ \eqref{eq:2100}) $\omega \in \mathbb{R} \bmod \pi$. Requiring the phase factor \eqref{eq:2122} to be unique
 implies $k=1,2,\cdots$.

 For $SU(1,1)$ we have $\omega \in
\mathbb{R} \bmod 2\pi$. Uniqueness of the phase factor then requires
$k=1/2,1,3/2,\cdots$.

Uniqness of the same phase factor as to the covering group $SO^{\uparrow}(1,2)_{[m]}$ for which $\omega \in \mathbb{R}
 \bmod m\pi$ requires
\begin{equation}
  \label{eq:2123}
  k= \frac{n}{m}\,,~n=1,2,\ldots. 
\end{equation}

For any irrational $k>0$ we get an irreducible representation of the universal covering group
 $SO^{\uparrow}(1,2)_{[\infty]}$. 
\section{Estimates for  the ratios $\mathbf{I_{2k}(2|z|)/I_{2k-1}(2|z|)}$ of modified Bessel functions of
 the first kind 
for $\mathbf{k>0}$}
In Appendix D.1 of Ref.\ \cite{ka1} I deduced the inequality
\begin{equation}\label{eq:62}
  \rho_k(|z|) = I_{2k}(2|z|)/I_{2k-1}(2|z|) <1
\end{equation}
for the ratio \eqref{eq:1771} which occurs frequently in expectation values with respect to Barut-Girardello coherent
states. The argments were:

It follows from the relation \cite{wat1}
\begin{equation}
  \label{eq:63}
  x\, \frac{dI_{\nu}}{dx}(x) = \nu\,I_{\nu}(x) + x\,I_{\nu +1}
\end{equation}
 that
 \begin{equation}
   \label{eq:64}
  I_{\nu +1}(x)/I_{\nu}(x) = \frac{d}{dx} \ln (I_{\nu}(x)/x^{\nu}) \,.
 \end{equation}
As \cite{wat2}
\begin{equation}
  \label{eq:65}
  I_{\nu}(x) = \frac{x^{\nu}}{2^{\nu}\,\sqrt{\pi}\,\Gamma (\nu + 1/2)}\,
\int_0^{\pi} d\theta\, e^{x\, \cos \theta}\, \sin^{2 \nu} \theta\,,
\end{equation}
we get for the ratio \eqref{eq:64}
\begin{equation}
 \label{eq:66}
  I_{\nu +1}(x)/I_{\nu}(x) =\frac{\int_0^{\pi} d\theta\,(\cos \theta)\, e^{x\, 
\cos \theta}\, \sin^{2 \nu} \theta}{
\int_0^{\pi} d\theta\, e^{x\, \cos \theta}\, \sin^{2 \nu} \theta} < 1\,.
\end{equation}
The argument is, however, only valid for $\nu > - 1/2\,$, i.e.\ for $k > 1/4\,$, because otherwise the integrals 
\eqref{eq:65} become singular. Thus, the interval $ k \in (0,1/4]$ has to be treated differently:

For $k=1/4$ we have \cite{wat3}
\begin{equation}
  \label{eq:67}
  \rho_{k=1/4}(|z|) = \frac{I_{1/2}(2|z|)}{I_{-1/2}(2|z|)} = \frac{\sinh 2|z|)}{\cosh 2|z|)} =\tanh 2|z| <1\,.
\end{equation}
For $k \in (0,1/4)\,$, however, the ratio $\rho_k$ may become larger than 1! This can already be seen from the asymptotic
expression \eqref{eq:55}: If we put $k= 1/4 - \delta,\,\delta \in (0,1/4)$ it takes the form
\begin{equation}
  \label{eq:68}
  \rho_{(0<k<1/4)}(|z|) \asymp 1 + \frac{\delta}{|z|} + \frac{\delta (1 + 2\delta)}{4|z|^2} + O(|z|^{-3})\,,~\delta \in
 (0,1/4)\,,~~ |z| \to \infty\,.
\end{equation}
Now the second and third non-leading terms in the expansion are positive, making the right-hand side larger than 1. 
\end{appendix}
The same feature may also been seen in the following way: Because of the relation \cite{wat4}
\begin{equation}
  \label{eq:69}
  I_{2k-1}(2|z|) = \frac{2k}{|z|}\,I_{2k}(2|z|) + I_{2k+1}(2|z|)
\end{equation}
we have 
\begin{equation}
  \label{eq:70}
  \rho_k(|z|) = \frac{|z|}{2k +|z|\,\rho_{k+1/2}(|z|)}\,,~~\rho_{k+1/2}(|z|) = \frac{I_{2k+1}(2|z|)}{I_{2k}(2|z|)}\,,
\end{equation}
which has the limit
\begin{equation}
  \label{eq:71}
  \lim_{k \to 0} \rho_k(|z|) = \frac{I_0(2|z|)}{I_1(2|z|)} > 1\,.
\end{equation}
Here the right-hand side even diverges for $|z| \to 0\,$! That the expression \eqref{eq:70} can become large for fixed
small $|z|$ and decreasing $k$ may also be seen from the approximation \eqref{eq:54} which yields
\begin{equation}
  \label{eq:72}
\rho_{k+1/2}(|z|) \to \frac{|z|}{2k+1}\left(1-\frac{|z|^2}{(2k+1)(2k+2)}\right) \text{ for } |z| \to 0\,.  
\end{equation}
If $|z|$ is so small that we can neglect the term of order $|z|^2$ in the bracket compared to $1$, we get for  the
relation \eqref{eq:70}
\begin{equation}
  \label{eq:73}
  \rho_k(|z|) \approx \frac{|z|}{2k + |z|^2/(2k+1)}\,.
\end{equation}
For $k \ll 1/2$ and $|z| > 2k + |z|^2$ the right-hand side of the expression \eqref{eq:73} becomes larger than $1\,$.

The possibility that $\rho_k(|z|) > 1$ for $k \in (0,1/4)$ can also be seen from the graphs in Figure 50-1 of Ref.\
\cite{span}.


\begin{thebibliography}{300}
\bibitem{sadd2} Here one puts $\zeta^n /(n!)^{1/\rho} = \exp {[} n\, \ln \zeta -(1/\rho)\, \ln \Gamma (n+1) {]} $,
 takes for $\ln
\Gamma (x+1)$ the usual asymptotic expansion (see, e.g.\ Ref.\ \cite{grary1}, subsec.\ 8.34) and then determines
 the saddle point.
\bibitem{bo4} Ref.\ \cite{bo1}, chap\ V, secs.\ D-F; in Eq.\ (103) there one has $2t=u$.
\end{thebibliography}
 \end{document}